\documentclass[preprint,superscriptaddress,nofootinbib,preprintnumbers,amsmath,amssymb,11pt]{revtex4-1}
\usepackage[T1]{fontenc}

\usepackage[titles]{tocloft}
\cftsetindents{section}{0em}{2.5em}
\cftsetindents{subsection}{2.5em}{2.5em}

\usepackage[lofdepth,lotdepth,caption=false]{subfig}
\usepackage{amsfonts}
\usepackage{mathrsfs}
\usepackage{leftidx}
\usepackage{amssymb}
\usepackage{amsmath}
\usepackage{placeins}
\usepackage{relsize}
\usepackage{slashed}
\usepackage{multirow}
\usepackage[dvipsnames]{xcolor}
\usepackage{tikz}
\usepackage{array}
\definecolor{red}{rgb}{0.9, 0,0}
\definecolor{cerulean}{rgb}{0., 0.62,0.9}
\definecolor{navy}{rgb}{0.05, 0.05,0.8}

\usepackage[colorlinks]{hyperref}
\hypersetup{
     colorlinks   = true,
     citecolor    = red,
	linkcolor = navy
}

\usepackage{soul}
\usepackage{epsfig}
\usepackage{graphicx}               % Standard graphics package
\usepackage{url}
\usepackage{float}
\usepackage{animate}  % animation

\newcommand{\be}{\begin{equation}}
\newcommand{\ee}{\end{equation}}
\newcommand{\bea}{\begin{eqnarray}}
\newcommand{\eea}{\end{eqnarray}}
\newcommand{\beq}{\begin{eqnarray}}
\newcommand{\eeq}{\end{eqnarray}}
\def\bit{\begin{itemize}}
\def\eit{\end{itemize}}
\def\ben{\begin{enumerate}}
\def\een{\end{enumerate}}

\newcommand{\bfq}{{\bf q}}
\newcommand{\bfu}{{\bf u}}

\newcommand{\bfp}{{\bf p}}
\newcommand{\bfk}{{\bf k}}
\newcommand{\bfl}{{\bf l}}
\newcommand{\bfv}{{\bf v}}
\newcommand{\bfr}{{\bf r}}

\newcommand{\bfD}{{\bf D}}
\newcommand{\bfE}{{\bf E}}
\newcommand{\bfH}{{\bf H}}
\newcommand{\bfG}{{\bf G}}
\newcommand{\bfP}{{\bf P}}

\newcommand{\bfZ}{{\bf Z}}

\newcommand{\bftau}{\boldsymbol{\tau}}

\newcommand{\bfe}{{\bf e}}
\newcommand{\bfeps}{\boldsymbol\epsilon}
\newcommand{\angstrom}{\mbox{\normalfont\AA}}

\newcommand{\bra}{\langle}
\newcommand{\ket}{\rangle}

\newcommand{\unitcell}{\Omega}
\newcommand{\Ncells}{N}
\newcommand{\Nlattice}{N\times n}
\newcommand{\Nunit}{n}
\newcommand{\oLO}{\omega_{\text{LO}}}
\newcommand{\dipmu}{\mu} % dipole reduced mass
\newcommand{\sapphire}{\mathrm{Al}_2\mathrm{O}_3}

\newcommand\DN[1][\relax]{%
\ifx\relax#1\relax\else{}^{#1}\fi \!X}

\DeclareMathAlphabet\mathbfcal{OMS}{cmsy}{b}{n} %boldface mathcal

\graphicspath{{./figures/}}

\def\brill{Brillouin}

\begin{document}

\count\footins = 1000 % this fixes a bug in Revtex package relating to footnote lengths

%--------------------   Previous format   ----------------------------

\title{Directional Detection of Light Dark Matter with Polar Materials}
\author{Sinead Griffin}
\affiliation{Department of Physics, University of California, Berkeley, CA 94720, USA} 
\affiliation{Molecular Foundry, Lawrence Berkeley National Laboratory, Berkeley, CA 94720, USA}
\author{Simon Knapen}
\affiliation{Department of Physics, University of California, Berkeley, CA 94720, USA}
\affiliation{Theoretical Physics Group, Lawrence Berkeley National Laboratory, Berkeley, CA 94720, USA}
\author{Tongyan Lin}
\affiliation{Department of Physics, University of California, Berkeley, CA 94720, USA}
\affiliation{Theoretical Physics Group, Lawrence Berkeley National Laboratory, Berkeley, CA 94720, USA}
\affiliation{Department of Physics, University of California, San Diego, CA 92093, USA }
\author{Kathryn M. Zurek}
\affiliation{Department of Physics, University of California, Berkeley, CA 94720, USA}
\affiliation{Theoretical Physics Group, Lawrence Berkeley National Laboratory, Berkeley, CA 94720, USA}\affiliation{Theoretical Physics Department, CERN, Geneva, Switzerland}

\begin{abstract} 
We consider the direct detection of dark matter (DM) with polar materials, where single production of optical or acoustic phonons gives excellent reach to scattering of sub-MeV DM for both scalar and vector mediators.  
Using Density Functional Theory (DFT), we calculate the material-specific matrix elements, focusing on GaAs and sapphire, and show that DM scattering in an anisotropic crystal such as sapphire features a strong directional dependence.
For example, for a DM candidate with mass 40 keV and relic abundance set by freeze-in, the daily modulation in the interaction rate can be established at 90\% C.L. with a gram-year of exposure.
Non-thermal dark photon DM in the meV -- eV mass range can also be effectively absorbed in polar materials.

\end{abstract}

\maketitle
\tableofcontents
%%%%%%%%%% INTRODUCTION %%%%%%%%%%

\clearpage

\section{Introduction}

Sub-GeV dark matter (DM) has become an important direction in dark matter searches in recent years. While low mass DM models have long been recognized to be viable theoretically, only recently have they come within experimental reach for direct searches. The correct dark matter abundance can be obtained in a variety of models, such as  Hidden Valleys~ \cite{Strassler:2006im}, secluded DM \cite{Boehm:2003hm,Pospelov:2007mp}, Asymmetric Dark Matter \cite{Kaplan:2009ag,Petraki:2013wwa,Zurek:2013wia}, freeze-in dark matter \cite{Hall:2009bx}, supersymmetric hidden sectors \cite{Hooper:2008im,Feng:2008ya,Cohen:2010kn}, and strongly interacting massive particles~\cite{Hochberg:2014dra}. All of these models can contain light mediators which couple the DM to the Standard Model, and in some cases such a mediator is crucial to set the relic abundance. The presence of light, weakly coupled mediators has made detection in high energy collisions challenging, while the sensitivity of direct detection experiments has been limited by the small kinetic energy of light DM in the Milky Way.  Recent advances in low threshold techniques have put direct detection within reach, however.

There is now a wide range of upcoming experimental probes for DM with mass between 1 MeV and 1 GeV (see Ref.~\cite{Battaglieri:2017aum} for a comprehensive overview).  For DM-electron couplings, currently the best sensitivity in this mass range is achieved by semiconductor target experiments such as SuperCDMS \cite{Agnese:2018col}, SENSEI \cite{Crisler:2018gci} and DAMIC \cite{Aguilar-Arevalo:2016ndq}, as well as noble liquid experiments such as DarkSide~\cite{Agnes:2018oej} and XENON10/100~\cite{Essig:2017kqs}.   A graphene target in a prototype of the PTOLEMY experiment also has directional sensitivity to sub-GeV DM with electron couplings~\cite{Hochberg:2016ntt}.  For DM-nucleon couplings, the CRESST collaboration \cite{Angloher:2015ewa} has obtained sensitivity to masses as low as 0.5 GeV. Other experiments targeting nuclear recoils in this mass range include DAMIC \cite{Aguilar-Arevalo:2016ndq}, NEWS-G \cite{Arnaud:2017bjh} and SuperCDMS SNOLAB \cite{Agnese:2016cpb}, while there are proposals to use liquid helium~\cite{Guo:2013dt}, molecules~\cite{Essig:2016crl}, or crystal defects~\cite{Kadribasic:2017obi,Budnik:2017sbu} as detection targets. 

Extending the reach to keV-MeV DM particles presents greater challenges, but also new opportunities.  The main challenge is to detect the extremely small energy depositions in ultra pure targets. Superconducting targets \cite{Hochberg:2015pha,Hochberg:2015fth} (with a meV electronic band gap), and Dirac materials \cite{Hochberg:2017wce} (with an arbitrarily small gap) have been shown to be promising low threshold targets for both scattering and absorption of low mass dark matter, while molecular magnets~\cite{Bunting:2017net} have been considered for absorption.  Furthermore, a DM particle with mass less than $\sim 1$ MeV has a deBroglie wavelength that is longer than the inter-particle spacing in typical materials, implying that the DM effectively couples to the collective excitations of atoms (phonons) in the target. Such DM-phonon scattering processes have different kinematics and allow for a greater amount of energy to be extracted from the DM than for scattering off a single free nucleus. This was the idea of Ref.~\cite{Schutz:2016tid}, where it was shown that a DM collision can produce multiple phonon modes in superfluid helium, extending the reach of such a target by 2-3 orders of magnitude in DM mass compared to ordinary nuclear recoils. (Note that there is a phase space penalty for emitting multiple states in a restricted configuration, as discussed in detail in Ref.~\cite{Knapen:2016cue}.) The emission of single or multiple phonon modes also allows for absorption of meV-eV mass bosonic DM in both superconductors~\cite{Hochberg:2016ajh}  and semiconductors~\cite{Hochberg:2016sqx}.

In this paper we consider the production of a single excitation --- an optical phonon  in a polar material --- from the interaction of a sub-MeV DM particle.  Optical phonons occur in all materials with more than one atom in their primitive cell, including {\em e.g.}~germanium and diamond crystals.  Unlike acoustic phonons, where the atoms (or rather, the ions) in the primitive cell oscillate in phase, optical phonons arise when the inequivalent atoms in the primitive cell oscillate out of phase. The optical phonons are gapped at low momentum with typical energies of $10$ - $100$ meV, which is well-matched to the total kinetic energy of sub-MeV DM. In a polar material such as  GaAs or sapphire (Al$_2$O$_3$), the inequivalent atoms in the primitive cell also have different effective charges, such that the coherent, out-of-phase motion of the ions in the optical modes generates a strong oscillating dipole. In Ref.~\cite{Knapen:2017ekk}, we argued that this dipole is particularly advantageous for DM interactions through a dark photon mediator, which can couple directly to the dipole. Polar materials moreover tend to be semiconductors or insulators, which means that the dark photon field does not experience the strong screening effects inherent to conductors (e.g.~superconductors \cite{Hochberg:2015pha}). GaAs and sapphire are also well-understood materials, where the technology already exists to make ultra pure single-crystal targets.  This is in contrast to Dirac and Weyl materials, which show similar theoretical promise for sub-MeV DM~\cite{Hochberg:2017wce} but are not yet feasible for large-scale high-quality synthesis.

In our previous analysis~\cite{Knapen:2017ekk}, we used several analytic approximations that allowed us to obtain the DM scattering rate in the isotropic limit for a relatively simple polar material, GaAs, which has only 6 phonon modes. In the present paper, we study a more complex material, sapphire, which we argue is better suited for direct detection.   To this end we employ more advanced numerical condensed matter techniques, notably Density Functional Theory (DFT), which allows us to accurately compute the scattering rate in sapphire and to validate the analytic treatment of GaAs used in Ref.~\cite{Knapen:2017ekk}. 
A key reason why sapphire is a more promising target is that its crystal structure is anisotropic, implying that the DM scattering rate will depend on the angle of the DM momentum with the primary crystal axis. This manifests itself as a modulation in the rate with the sidereal day, a smoking gun signature for DM that can be used to test the origin of any signal.\footnote{Other proposals with directional sensitivity to low mass DM scattering include using graphene \cite{Hochberg:2016ntt} or Dirac material targets~\cite{Hochberg:2017wce} for DM-electron scattering, and taking advantage of direction-dependent thresholds for defect production in crystals~\cite{Budnik:2017sbu} or nuclear recoils in semiconductors~\cite{Kadribasic:PRL18_ERmodulation}. } 

The outline of this paper is as follows. In Section~\ref{subsec:principle}, we begin by presenting the theoretical case for polar materials and elaborate on their benefits, focusing specifically on GaAs and sapphire. In the remainder of Section~\ref{sec:CM}, we describe in detail the crystal structures, the method for computing phonon band structures, and set up the formalism for calculating the direction-dependent DM scattering rate.
Next, we consider a few specific model benchmarks: in Section~\ref{sec:darkphoton}, we present the reach and daily modulation for DM scattering via an ultralight dark photon mediator, or  millicharged DM with photon-mediated scattering. 
Section~\ref{sec:nucleonint} considers nucleon scattering mediated by a light scalar, where we find modulation rates as large 30\% for $m_X \sim 50$ keV. We lastly consider absorption of dark photon DM with mass of 10-100 meV into optical phonons and multiphonons in Section~\ref{sec:absorption}, and conclude in Section~\ref{sec:conclusions}.  More details on the derivation of DM scattering rates are provided in Appendices \ref{app:phonons}-\ref{app:neutronscat}, while some results on scalar-mediated electron scattering are contained in Appendix~\ref{app:electrons}. Finally, some additional details on our method to estimate the statistical discrimination for the daily modulation signal are given in Appendix~\ref{app:statistics}.
A detailed description of the experimental setup as well as the calculation of the most important backgrounds will be presented in an upcoming paper. In the meanwhile, for a brief description of the experimental setup and estimates of the backgrounds we refer the reader to Ref.~\cite{Knapen:2017ekk}.

%%%%%%%%%%%%%%%%%%%%%%%%%%%%%%%%%%%
\section{Polar materials}
\label{sec:CM}
%%%%%%%%%%%%%%%%%%%%%%%%%%%%%%%%%%%

In this section we first lay out the qualitative features that make polar materials excellent targets for sub-MeV dark matter. For the purposes of this paper, we focus on the examples of GaAs and sapphire\footnote{Note that we will use sapphire or $\sapphire$ here interchangeably to mean crystalline aluminum oxide (Al$_2$O$_3$); this is also sometimes called corundum ($\alpha-\sapphire$) in the literature.} since they are commonly used and well-characterized materials, including in some existing or proposed direct detection experiments\footnote{Notably, the CRESST collaboration recently published results on $\sim$GeV DM with a sapphire target~\cite{Angloher:2017sxg}. Here the deposited energy is measured in phonons, but the initial scattering is a nuclear recoil. In contrast, for sub-MeV mass scales, we are considering the process where single phonons are directly excited by the DM.}.  We then review the crystallographic properties of GaAs and sapphire, setting up the theoretical framework necessary for performing DM scattering calculations.

%%%%%%%%%%%%%%%%%%%%%%%%%%%%%%%%%%%
\subsection{Advantages for direct detection}
\label{subsec:principle}
%%%%%%%%%%%%%%%%%%%%%%%%%%%%%%%%%%%

As briefly discussed in the introduction, polar materials have several features that make them attractive as targets for the scattering and absorption of light DM.  These properties are:
\begin{enumerate}
\item Even in the limit of low momentum transfer, a relatively large energy deposition is possible when scattering into optical phonons;
\item Sapphire has an anisotropic crystal structure, allowing for directional detection;
\item The optical response allows for dark photon interactions, but is still sufficiently weak that screening effects do not hinder the sensitivity.
\end{enumerate}  
We detail each of these features in turn.

\subsubsection{Acoustic vs.~Optical Phonons}

\begin{figure}[t]
\centering
\subfloat[][LO mode]{
 \begin{tikzpicture}
    \node[anchor=south west,inner sep=0] at (0,0) {\includegraphics[width=0.4\linewidth]{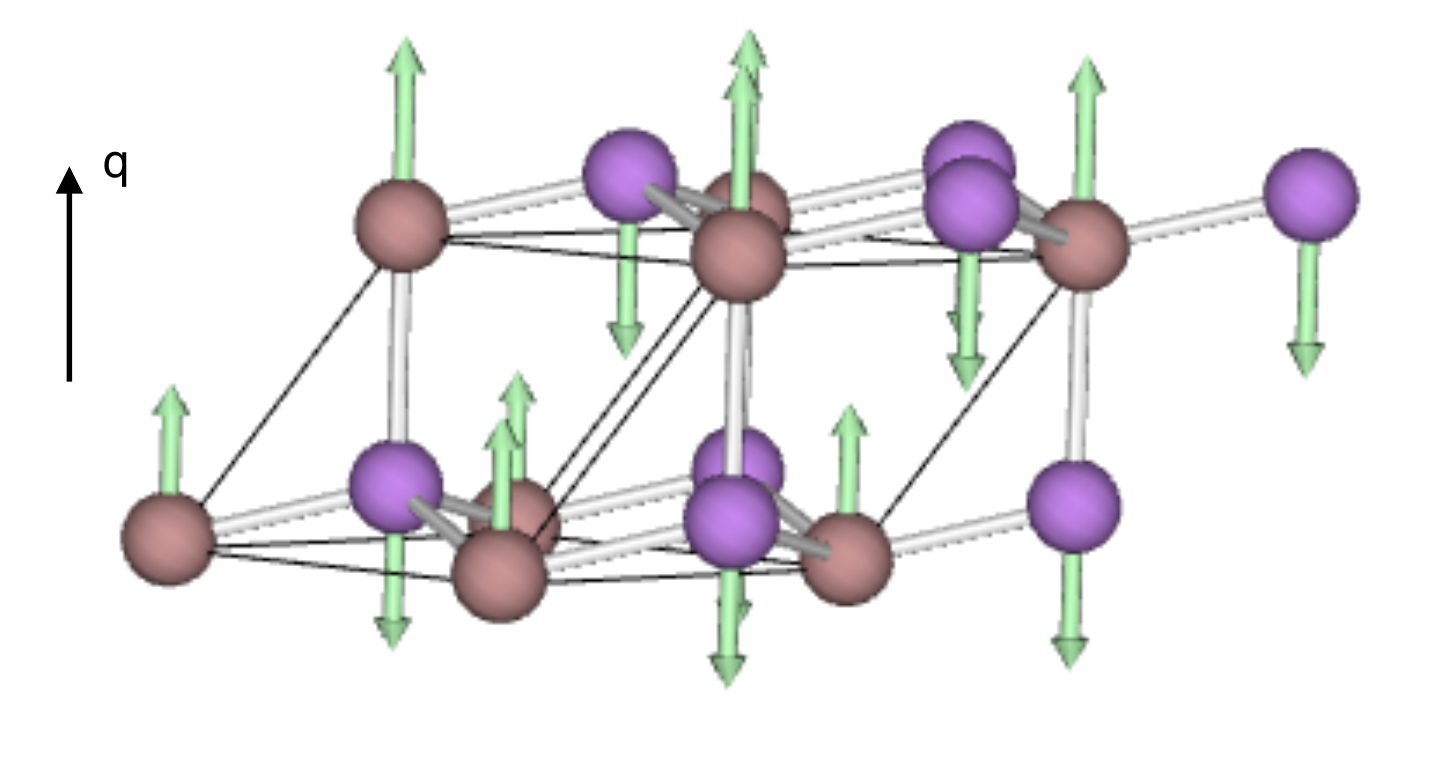}};
    \draw [black,thick, ->] (-0.25,1.0) -- (-0.25,2.0)  node [right] {$q$};
\end{tikzpicture}
}
\hfill
\subfloat[][TO mode]{
\begin{tikzpicture}
    \node[anchor=south west,inner sep=0] at (0,0) {\includegraphics[width=0.4\linewidth]{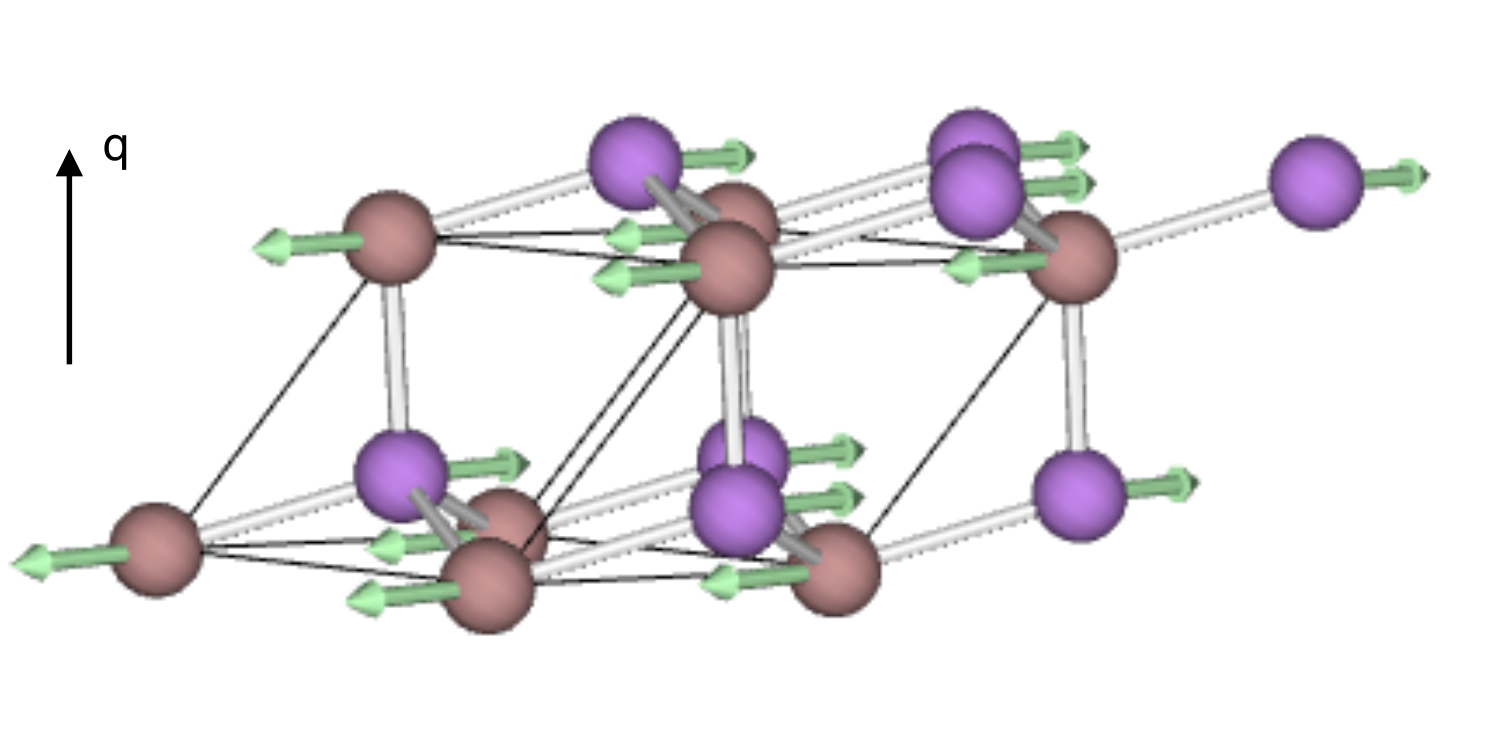}};
    \draw [black,thick, ->] (-0.25,1.0) -- (-0.25,2.0)  node [right] {$q$};
\end{tikzpicture}
}
\caption{ Visual representation of the optical modes in GaAs, for a lattice containing two primitive cells in each direction. The black lines outline a single primitive cell, containing one As atom (purple) and 8 times 1/8 of a Ga atom (brown). The green arrows indicate the atomic motions  at a snapshot in time, while the black arrow is the phonon propagation direction. Figures generated with \cite{phononwebsite}. 
\label{fig:GaAsphonons}}
\end{figure}

\begin{figure}[t]\centering
\includegraphics[height=5cm]{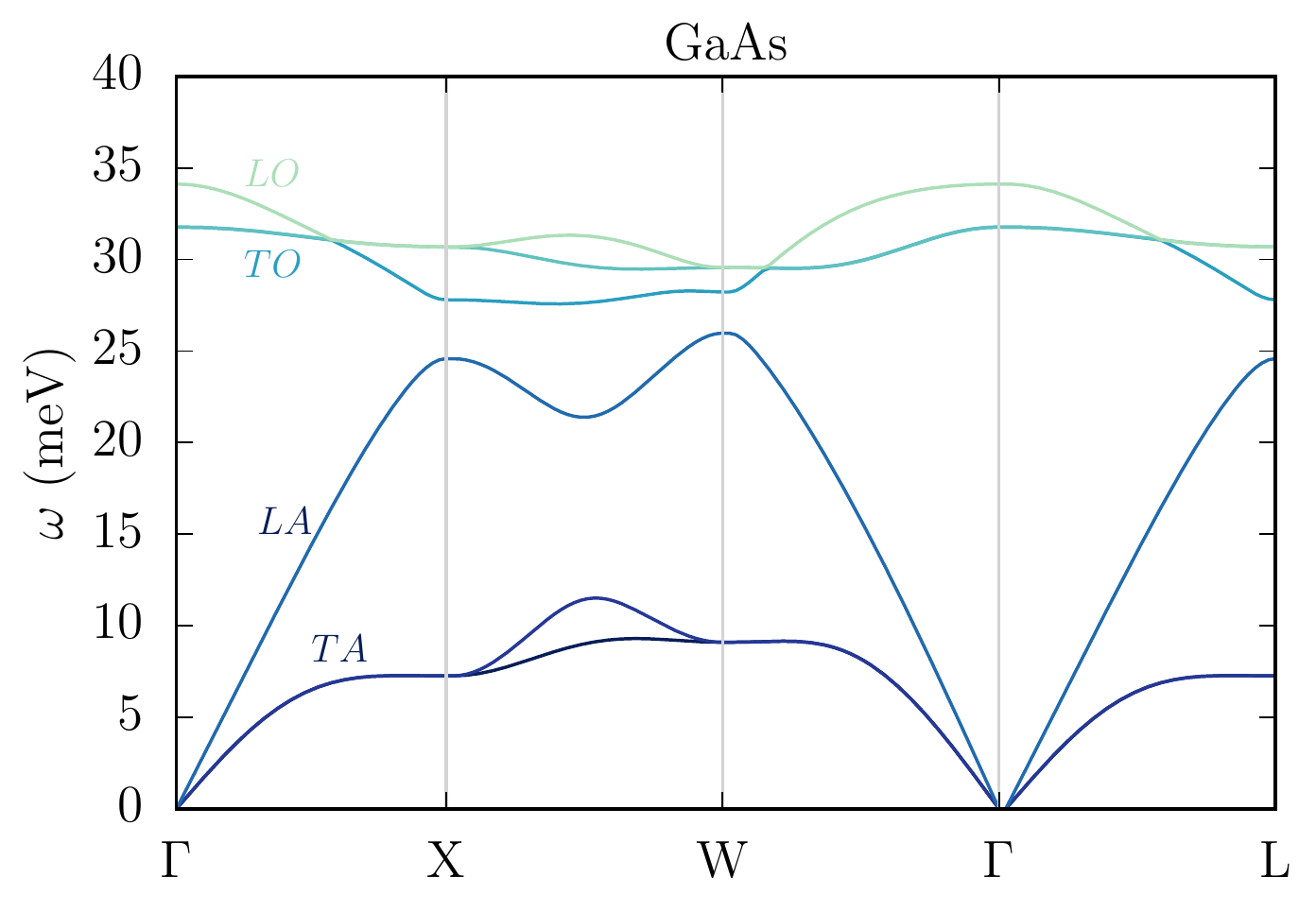}\hspace{1.5cm}
\includegraphics[height=5cm]{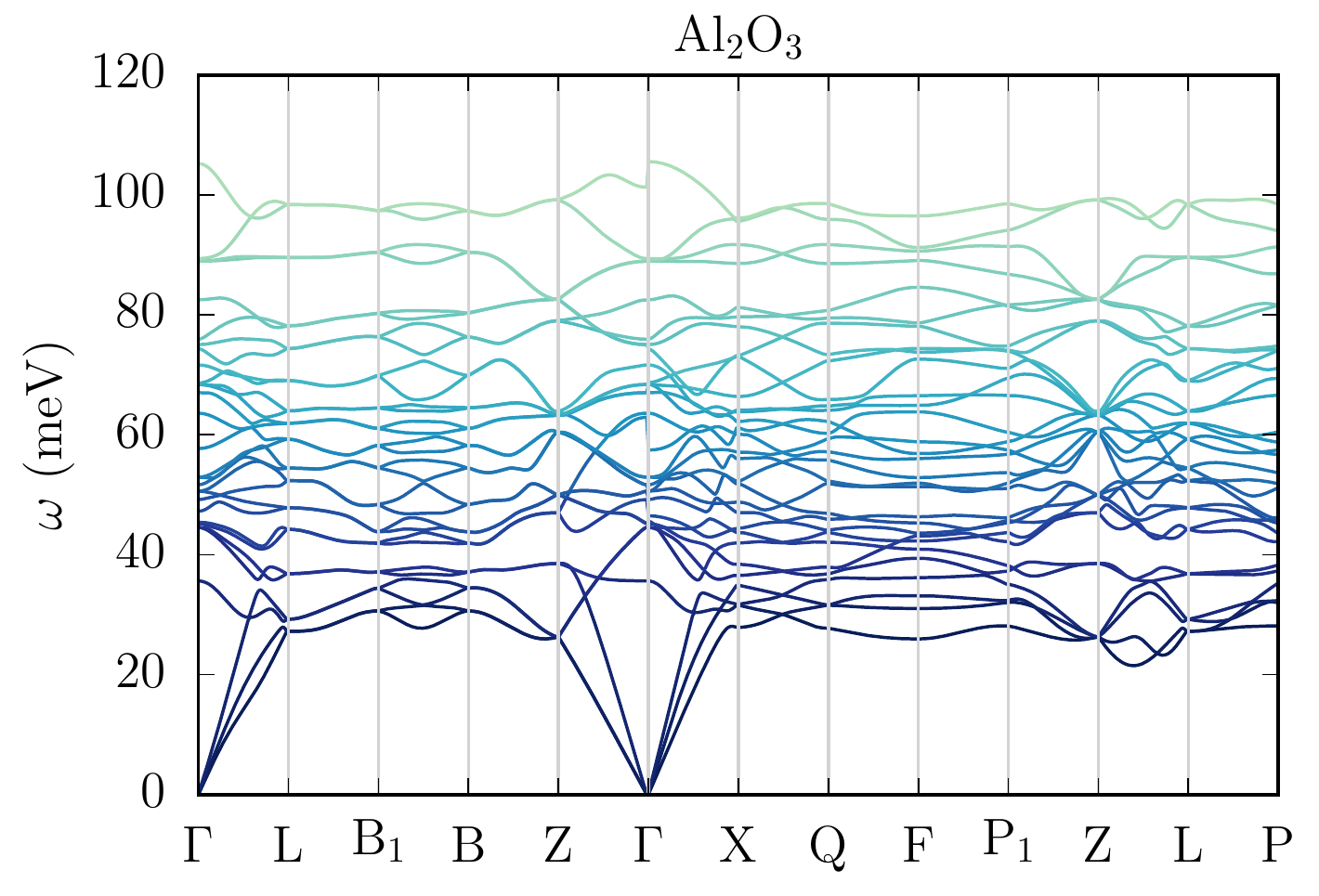}
\caption{Phonon band structures for GaAs (left) and sapphire (right) as computed with \textsf{phonopy} \cite{phonopy}. The x-axis traces out a path in the \brill\ zone. As is conventional in the condensed matter literature, the points in  the Brillouin zone with high symmetry are indicated with Roman and Greek characters (see Fig.~\ref{crystal_conventional} in Appendix~\ref{app:phonons}), where $\Gamma$ always refers to the origin of the \brill\ zone $\bfq=(0,0,0)$. 
\label{fig:GaAsphononbands} }
\end{figure}

Polar crystals have  a primitive unit cell with more than one type of atom. The total number of phonon modes is given by the number of atoms in the primitive cell multiplied by a factor of three due to three spatial degrees of freedom.  Three of the phonon modes are always acoustic modes, where atoms in the primitive cell oscillate exactly in phase in the long-wavelength limit.  The rest of the modes are optical phonon modes, where the atoms oscillate out of phase.  For GaAs there are three optical modes, two transverse (TO) modes and one longitudinal (LO) mode, as visualized in Fig.~\ref{fig:GaAsphonons}.  For sapphire, there are 10 atoms in the primitive cell and as a result 30 phonon modes, divided into 3 acoustic branches and 27 optical branches. The dispersion of the phonons (energy of the mode as a function of the momentum) for each material is shown in Fig.~\ref{fig:GaAsphononbands}. Note that we have shown the band structure along a high-symmetry  path within the \brill\ zone; for reference, the typical size of the \brill\ zone in physical units is on the order of $\sim$ keV.

The acoustic modes (labeled transverse (TA) and longitudinal (LA) acoustic) have the standard gapless, linear dispersion at $|\bfq|\approx0$ (``$\Gamma$'' point in Fig.~\ref{fig:GaAsphononbands}.) The slope is given by the speed of sound $c_s=\omega / |\bfq|$ near $|\bfq| \to 0$, where the longitudinal sound speed is $c_s \sim 4000$ m/s in GaAs and $c_s \sim 10^4$ m/s in sapphire, though for sapphire the sound speed is somewhat direction dependent~\cite{2001JAP....90.3109W}.
In the long wavelength limit, these modes carry no energy, as they correspond to translations of the lattice as whole. In this sense, the acoustic modes can be considered as the Goldstone modes\footnote{There are no Goldstone modes associated with the spontaneous breaking of the rotation invariance: in the presence of a broken translational symmetry, rotations do not give rise to a linearly independent set of Goldstone modes (see e.g.~\cite{Low:2001bw,Watanabe:2013iia}).}
of the spontaneous breaking of the translation invariance by the crystal. The optical phonons are not protected by Goldstone's theorem, and at $\bfq\approx 0$ they can be thought of as a standing, non-propagating wave which stores a finite amount of energy.

{\it A priori}, the dark matter can excite both the optical and acoustic modes, but the energy deposited in the acoustic modes is much smaller and is only detectable in the most optimistic circumstances. Concretely, for $m_X\lesssim$ MeV, the DM momentum $m_X v\, \lesssim$ keV is sufficiently small that it is only possible to excite a phonon mode within the first \brill\ zone. Consider a DM scattering with momentum transfer $\bf q$ and energy deposition $\omega$, which excites a single acoustic phonon; the phonon must absorb all of the energy and momentum transferred. This leads to the scaling
\begin{equation}\label{eq:acousticscaling}
\omega=c_s\, |\bfq|  \lesssim 2\, c_s\, v\, m_X\sim 7\,\mathrm{meV}\times \frac{m_X}{\mathrm{100\,keV}}.
\end{equation}
with $v\sim 10^{-3}$ the DM velocity and assuming the speed of sound for sapphire. The threshold for near future devices will be at best in the $10 -100$ meV range, which means that single acoustic phonon excitations from light DM will be difficult or impossible to detect, depending on $m_X$. However, the scaling in \eqref{eq:acousticscaling} does not apply for the optical modes since they have an energy of $\omega\sim 30$ meV or more as $|\bfq| \to 0$, as is evident from Fig.~\ref{fig:GaAsphononbands}. 

The gapped dispersion of optical phonons is a particularly appealing feature, as it allows nearly the maximum amount of DM kinetic energy to be extracted in the scattering, even when the momentum transfer is much less than a keV. This is in contrast to recoils off free nuclei, where the energy deposited from light DM is much less than the initial DM kinetic energy. The presence of optical phonons is also advantageous compared to a material such as superfluid helium. Superfluid helium does have gapped quasiparticle excitations (rotons), but they only occur at high $\bfq$ and are much lower energy that the optical phonons in a solid. Since single phonon production in superfluid helium is undetectable in the foreseeable future, one must resort to multi-phonon production to break the relation in \eqref{eq:acousticscaling}, as was demonstrated in Refs.~\cite{Schutz:2016tid,Knapen:2016cue}. However, the rate is suppressed since this is a higher order process relying on anharmonic phonon couplings. On the other hand, in Sec.~\ref{sec:nucleonint} we will see that for scalar-mediated DM the rate for producing optical phonons in GaAs and sapphire is also somewhat suppressed due to destructive interference between the different atoms in the primitive cell. The end result is that, for the scalar-mediated model, the reach for polar materials is comparable  to that of superfluid helium.

\subsubsection{Crystal Properties and Directional Detection}

The crystal structure of polar materials can be anisotropic. This is the case for sapphire, which has a rhombohedral lattice structure and therefore a  primary crystal axis.  This anisotropy manifests itself both in the spectrum of the phonons (see Fig.~\ref{fig:GaAsphononbands}) and in the strength of their coupling to the dark matter. The latter feature depends on the type of mediator, and will be discussed in detail in Secs.~\ref{sec:darkphoton} and \ref{sec:nucleonint}. Regardless of the type of mediator, this means that the scattering rate will modulate with the sidereal day as the angle between the primary crystal axis and the DM wind changes due to the rotation of the Earth. In other words, different regions of the \brill\ zone are sampled at different times of the day, which results in a changing rate due to the $\bfq$-dependence of the  phonon energies and DM-phonon coupling. We set up our notation and conventions for the directionally dependent rate in Sec.~\ref{subsec:modulation}.

\subsubsection{Optical Properties and Dark-Photon Mediated Scattering}

The optical response is particularly important if the DM scattering is mediated by a dark photon that is kinetically mixed with the SM photon. Since the dark photon has a coupling to the electromagnetic current, naively the best target for direct detection would be a population of free charges, such as the conduction electrons in a metal. However, the same free charges screen the dark photon field at the low frequencies that are of interest. The ideal material therefore has few conduction electrons but a large polarizability. Superfluid helium fails the latter criterion, as the polarizability of a He atom is very small, rendering helium transparent to both SM and dark photon fields \cite{Knapen:2016cue}. 

To compare the characteristics of polar materials with other candidate targets like superconductors and Dirac/Weyl materials, it is useful to express the electromagnetic response in terms of the longitudinal and transverse in-medium polarization tensor $\Pi_{T,L}$,
\beq
\Pi_L = q^2(1-\epsilon), ~~~ \Pi_T = \omega^2(1-\epsilon),
\eeq
which is expressed in terms of the relative permittivity $\epsilon$ of the material, and where $q^2 = \omega^2 - |\bfq|^2$. Accounting for the in-medium effects, the matrix element for scattering can be written as
\beq
\langle J^\mu_{\rm EM} J^{\nu}_{\rm DM} \rangle = \sum_{T,L} \frac{  e g_X }{ q^2 - m_{A'}^2} \frac{\kappa q^2 \, P^{\mu \nu}_{T,L}}{q^2 - \Pi_{T,L}},
\eeq
where ${P}^{\mu \nu}_{T,L}$ are the projection operators for transverse and longitudinal polarizations, $\kappa$ is the vacuum mixing parameter between the dark and SM photon, $m_{A'}$ is the dark photon mass and $g_X$ is the gauge coupling of DM with the dark photon. Thus, as demonstrated by previous work on dark photon mediated interactions \cite{An:2013yua,Hochberg:2015fth}, the in-medium polarization tensor is essential in determining the reach. 

Superconductors were the first targets to be considered for direct detection of sub-MeV DM \cite{Hochberg:2015fth}. In the limit of $| \bfq| \gg \omega$, the dielectric function in metals and superconductors displays Thomas-Fermi screening:
\begin{align}\label{eq:supercon}
	\epsilon_{\rm metal} \sim 1 + \frac{ \lambda_{\rm TF}^2 } {|\bfq|^2}
\end{align}
with $\lambda_{\rm TF}^2 = 3 e^2 n_e/(2 E_F) \simeq (\rm few\ keV)^2$.  The maximum momentum transfer for sub-MeV DM-electron scattering is $|\bfq| = 2 m_X v_X \lesssim \textrm{keV}$, such that $\epsilon_{\rm metal}$ tends to be very large.  This screening severely limits the sensitivity of superconductors to dark photon mediated DM.   

This screening can be lifted in Dirac and Weyl materials. These have an arbitrarily small gap for excitations of electrons to the conduction band, but a gauge symmetry in the material protects the photon from obtaining a large in-medium polarization \cite{Hochberg:2017wce}. This effect can be understood as a result of gauge invariance, where charge is renormalized but the photon obtains no mass.  Calculating a simple one-loop polarization graph with the linear dispersion typical of Dirac materials near the Dirac point, one obtains the dielectric response 
 \begin{align}\label{eq:epsdirac}
	\epsilon_{\rm Dirac} \sim 1 + \frac{e^2}{12 \pi^2 \epsilon_b v_F},
\end{align}
where $\epsilon_b$ is the background dielectric constant supplied by the nuclei and bound electrons, and $v_F$ is the Fermi velocity, which is typically $10^{-3} - 10^{-2}$.  The resulting dielectric constant is typically an ${\cal O}(1)$ number, such that excellent reach to dark photon mediated DM can be obtained.  However, Dirac/Weyl materials are still the subject of intense research, and it is not yet known how to fabricate large, ultra pure samples needed for DM detection. 

In polar materials, there is a gap for electronic excitations on the order of 1-10 eV, so there is no screening by conduction electrons. And while electron excitations are forbidden, dark photon mediated DM can instead couple to the dipole moment of the optical phonons. The interaction is only screened by the valance electrons, an effect which is encoded in the high frequency dielectric constant ($\epsilon_\infty $).  Its value for GaAs and sapphire is in the ${\cal O}(1$ - $10)$ range, and we derive the screening of the dipole interaction in detail in Appendix~\ref{app:frolich}. Thus polar materials satisfy the criteria of large polarizability with little screening by free charges. Compared to Dirac/Weyl materials, there is some penalty to coupling though a dipole moment but the phase space for the scattering process is larger, such that the projected reach is comparable. Polar materials moreover have the advantage that the technology already exists to fabricate the bulk, ultra pure targets needed for detection.

%%%%%%%%%%%%%%%%%%%%%%%%%%%%%%%%%%%
\subsection{Crystal properties \label{subsec:crystal} }

GaAs adopts a cubic zincblende crystal structure (space group \textit{F-43m}) with two atoms (Ga and As) in the primitive unit cell (left panel of Fig.~\ref{fig:lattice}). These two atoms in the primitive cell give a total of six degrees of freedom corresponding to the six phonon modes as was shown in the left panel of Fig.~\ref{fig:GaAsphononbands}. The covalent bond in GaAs is polar, like in other III-V compound semiconductors, owing to the moderate difference in electronegativity between the Ga and As ions. This results in the Ga and As carrying opposite net effective charges. The phonon branches corresponding to the out-of-phase motion of these net-charged ions will therefore couple to electric fields, hence the name ``optical phonons''. 

\definecolor{Alcolor}{RGB}{125, 152,181}
\definecolor{Ocolor}{RGB}{229, 16,7}
\definecolor{Gacolor}{RGB}{112, 17,174}
\definecolor{Ascolor}{RGB}{61, 156,50}
\begin{figure}

 \begin{tikzpicture}
    \node[anchor=south west,inner sep=0] at (0,0) {\includegraphics[width=0.45\textwidth]{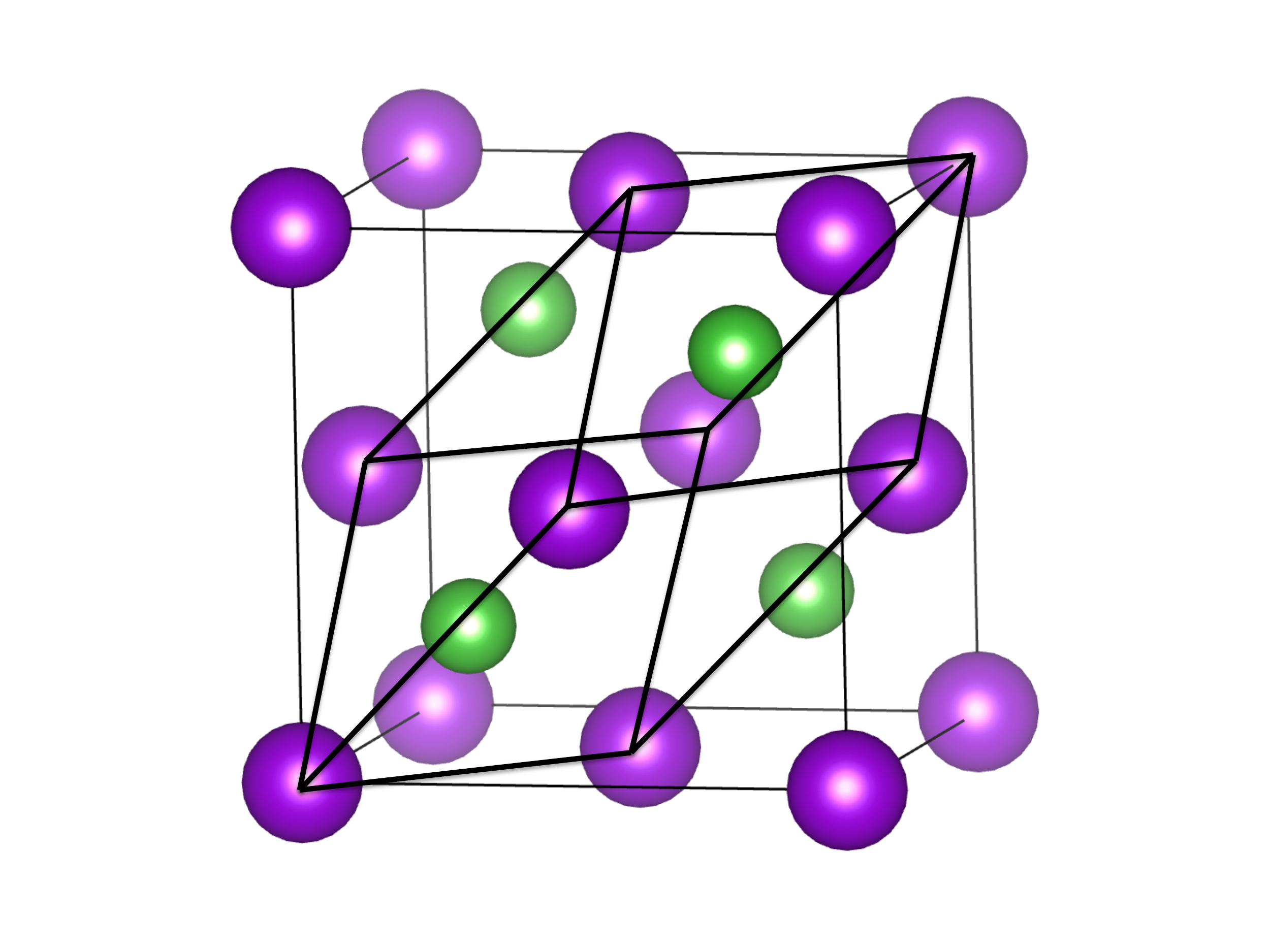}};
     \node[inner sep=0,color=Goldenrod] at (1.82,1.06) {{\Huge $\mathbf{\star}$}};
      \node[inner sep=0,color=Goldenrod] at (4.43,3.65) {{\Huge $\mathbf{\star}$}};
               \node[inner sep=0] at (1.1,1) {\textcolor{Gacolor}{{\bf \small Ga}}};
          \node[inner sep=0] at (2.25,2.1) {\textcolor{Ascolor}{{\bf \small As}}};
\end{tikzpicture}
\hfill
  \begin{tikzpicture}
    \node[anchor=south west,inner sep=0] at (0,0) {\includegraphics[width=0.45\textwidth]{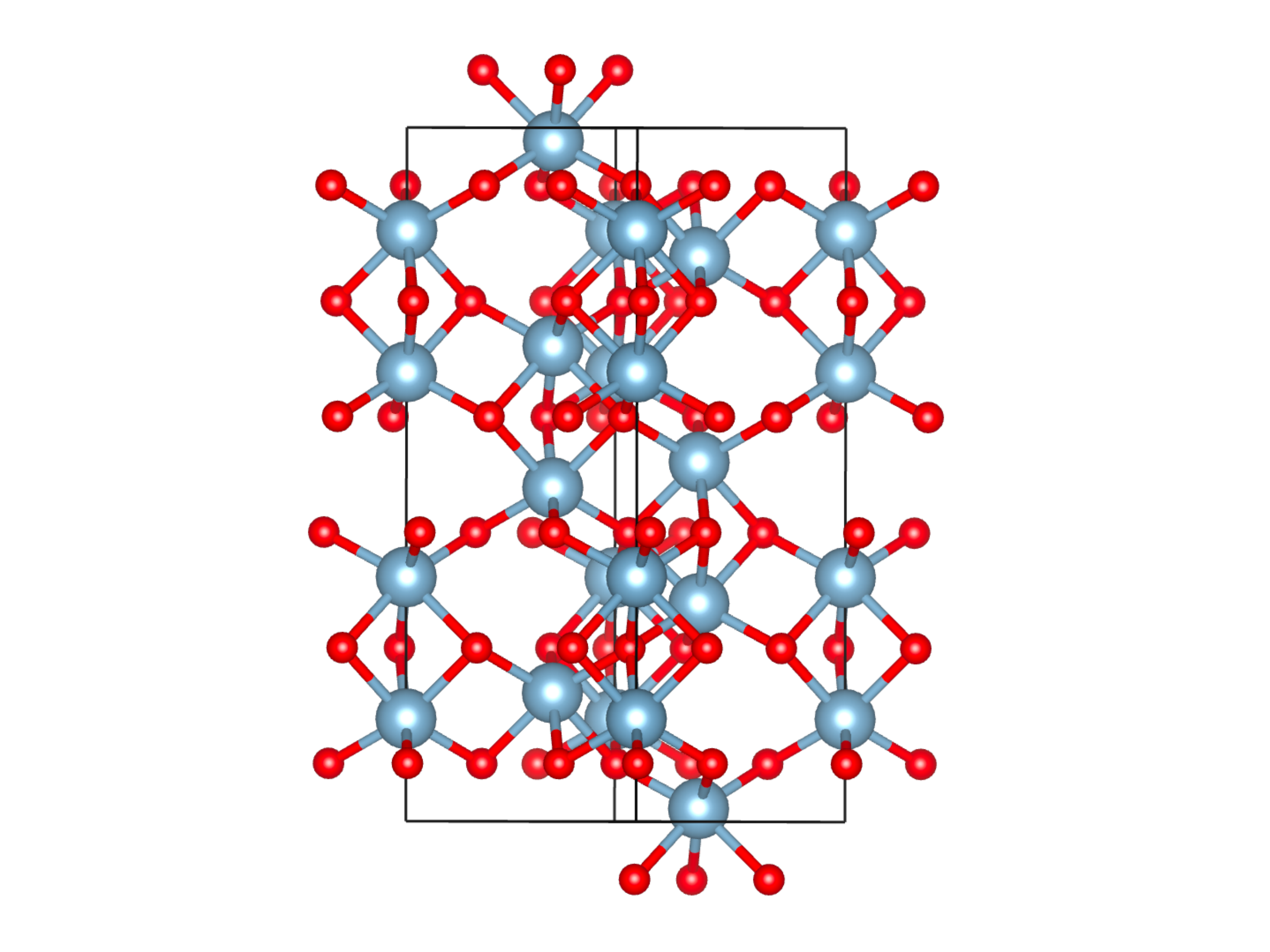}};
     \node[inner sep=0,color=Goldenrod] at (2.46,3.51) {{\large $\mathbf{\star}$}};
      \node[inner sep=0,color=Goldenrod] at (2.46,4.35) {{\large $\mathbf{\star}$}};
      \node[inner sep=0,color=Goldenrod] at (2.46,2.27) {{\large $\mathbf{\star}$}};
      \node[inner sep=0,color=Goldenrod] at (2.46,1.4) {{\large $\mathbf{\star}$}};
      
       \node[inner sep=0,color=Goldenrod] at (1.98,3.23) {{ \footnotesize$\mathbf{\star}$}};
        \node[inner sep=0,color=Goldenrod] at (2.31,3.23) {{ \footnotesize$\mathbf{\star}$}};
        \node[inner sep=0,color=Goldenrod] at (2.9,3.23) {{ \footnotesize$\mathbf{\star}$}};
        
          \node[inner sep=0,color=Goldenrod] at (1.9,2.53) {{ \footnotesize$\mathbf{\star}$}};
        \node[inner sep=0,color=Goldenrod] at (2.47,2.53) {{ \footnotesize$\mathbf{\star}$}};
        \node[inner sep=0,color=Goldenrod] at (2.88,2.53) {{\footnotesize $\mathbf{\star}$}};
        
         \node[inner sep=0] at (1.88,4.3) {\textcolor{Alcolor}{{\bf \small Al}}};
          \node[inner sep=0] at (1.88,3.7) {\textcolor{Ocolor}{{\bf \small O}}};

\end{tikzpicture}
\caption{Conventional unit cell for GaAs (left) and sapphire (right). The atoms belonging to a single primitive cell are labeled with a $\star$, and for GaAs, the primitive cell is represented by the thick black lines. In the case of GaAs, we have a cubic unit cell with the crystal structure having the same symmetry in all three principal crystallographic directions. For sapphire, the in-plane axes are equivalent, but differ from the out-of-plane (vertical) crystal axis, giving the more anisotropic crystal.  \label{fig:lattice} }
\end{figure}
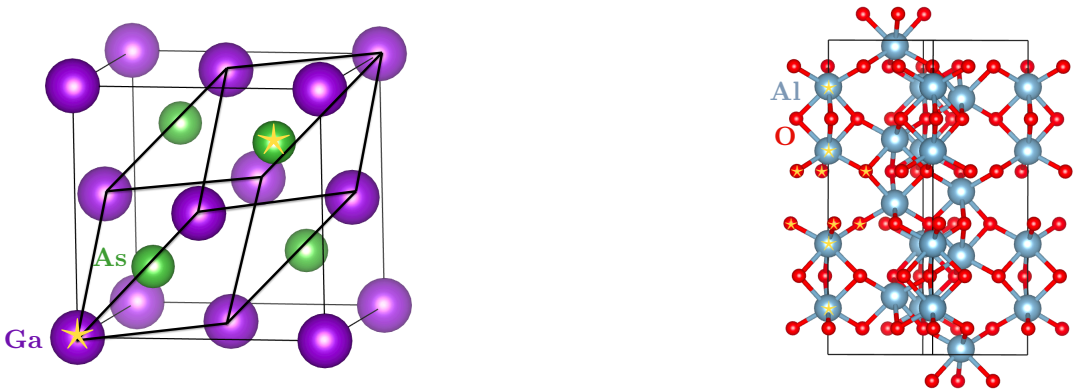

Sapphire with the chemical formula Al$_{2}$O$_{3}$ has a more complex rhombohedral crystal structure (space group \textit{R-3c}) with four Al and six O atoms in the primitive unit cell (right panel Fig.~\ref{fig:lattice}). Each Al ion is six-coordinated with oxygen, which form a corner-sharing network to make up the crystal lattice. The corresponding phonon spectrum was shown in the right panel of Fig.~\ref{fig:GaAsphononbands}. Because of the inequivalent in-plane and out-of-plane crystal directions, sapphire has a primary crystal axis, which implies that the scattering rate depends on the angle between the momentum transfer and the primary axis. It also means that the isotropic approximation used in Ref.~\cite{Knapen:2017ekk} does not hold for sapphire, though we will see that it works well for a more symmetric crystal like GaAs. For sapphire, the scattering rate must be computed numerically,  using first-principles methods that incorporate the crystalline and chemical specificity of the sapphire crystal.

Tab.~\ref{tab:usefulstuff} lists some useful quantities for both materials from the point of view of DM-phonon scattering. While the quantities are all temperature-dependent, the differences between room-temperature and liquid helium temperatures are typically percent-level or less~\cite{sapphireproperties}. Notably, the dielectric constants are ${\cal O}(1)$ quantities, which is relevant for dark photon mediated DM, as discussed in the previous subsection. 
We list both the low frequency ($\epsilon_0$) and high frequency ($\epsilon_\infty$) dielectric constants,  where the high frequency case refers to $\omega$ above the optical phonon frequencies, but still well below the electronic band gap of the material. At high frequencies,  the ions in the lattice have no time to respond to a rapidly changing electric field, and the dielectric function only receives contributions from the valence electrons. At low frequencies, the optical phonons contribute to the dielectric function as well, such that $\epsilon_0 > \epsilon_\infty$ in a polar material. We also note that our first-principles calculations are carried out at zero temperature, providing a close reference for liquid He temperatures.

\begin{table}
\begin{tabular}{|c|c|c|}\hline
&GaAs&   $\text{Al}_2 \text{O}_3$\\\hline
$\rho_T$ &5.32 $\mathrm{g}/\mathrm{cm}^3$&3.98 $\mathrm{g}/\mathrm{cm}^3$\\
$a$&5.76 $\angstrom$& 4.81 $\angstrom$, 13.11 $\angstrom$\\
$\epsilon_0$&12.9& 11.5 (parallel to $c$-axis)\\
&&  9.3 (perpendicular to $c$-axis)\\
$\epsilon_\infty$ &10.89&3.2\\
\hline
\end{tabular}
\caption{Values of the density, calculated lattice constants and dielectric constant for a primitive cell of GaAs and a conventional unit cell of sapphire. 
}
\label{tab:usefulstuff}
\end{table}

\subsection{Theoretical description of phonons\label{sec:theoreticaldesc}}

With advances in first-principles modeling of materials and in high-performance computing, it is routine to calculate the electronic and vibrational properties of crystals from first principles. For DM direct detection in particular we need the phonon spectrum over the whole Brillouin zone, since this is an input for the DM scattering (or absorption) rate calculation. Here we briefly discuss how these calculations are performed and establish the notation for the remainder of the paper. Readers familiar with the subject or only interested in the results can choose to skip the remainder of this section.

The positions of the atoms (or ions) in the crystal are denoted by $ \bfu_{j,\bfl}+\bfr _j^0 +\bfl$,
where $ \bfu_{j,\bfl}$ is displacement of the atom relative to its equilibrium position, $\bfr_j^0$ is the equilibrium position of the atom relative to the origin of the primitive cell and $\bfl$ is a lattice vector labeling the primitive cell. The index $j$ therefore runs over the atoms in the primitive cell. In what follows, a boldface symbol always refers to a tensor or vector in position or reciprocal space. The potential energy $\mathcal{V}$ is a function of the displacements and can be expanded as 
\begin{equation}
\mathcal{V}=\mathcal{V}^{(0)}+\sum_{\bfl,j} \mathbfcal{V}^{(1)}_{\bfl,j} \cdot \bfu_{j,\bfl}+\frac{1}{2}\sum_{\bfl,\bfl',j,j'} \bfu_{j,\bfl} \cdot \mathbfcal{V}^{(2)}_{\bfl,j,\bfl',j'}\cdot \bfu_{j',\bfl'}+\cdots
\end{equation}
where the $\mathbfcal{V}^{(2)}_{\bfl,\bfl',j,j'}$ etc are the force constants. The force constants can be calculated from \textit{ab initio} density functional theory methods. For this work we use density functional theory (DFT) as implemented in the VASP package~\cite{VASP1} with full calculation details given in Appendix~\ref{app:phonons}. Firstly, the equilibrium crystal lattice and atomic positions are found by minimizing the forces on the atoms and stresses in the crystal cell. From this optimized crystal structure, the force constants can be calculated using two different methods. The first displaces atoms in the cell in symmetry-inequivalent directions, calculates the resulting forces on the atoms in the unit cell, and from these builds up the force constant matrix. The second method, the linear response method, uses density functional perturbation theory (DFPT) to calculate the forces. In this work, we will use the former method, known as the `frozen-phonon' method, to calculate the full phonon spectrum as it is computationally less expensive. For the Born effective charges, we will use DFPT.

In the harmonic approximation, one only considers the leading non-vanishing order, keeping only $\mathbfcal{V}^{(2)}_{\bfl,j, \bfl',j'}$. The displacement operator is then quantized in terms of phonon modes:
\begin{align}\label{eq:displacementquant}
\bfu_{j,\bfl}(t) = \sum_{\nu}^{3 \Nunit}\sum_{\bfq}  \sqrt{\frac{1}{2 \Ncells m_j \omega_{\nu,\bfq}}} \left(\mathbf{e}_{\nu,j,\bfq} \hat a_{\nu,\bfq} e^{i\bfq \cdot (\mathbf{l}+ \bfr^0_j )-i \omega_{\nu,\bfq}t}+\mathbf{e}^\ast_{\nu,j,\bfq} \hat a^\dagger_{\nu,\bfq} e^{-i\bfq\cdot(\mathbf{l}+ \bfr^0_j )+i \omega_{\nu,\bfq}t}  \right)
\end{align}
where the $\hat a_{\nu,\bfq}$ ($\hat a^\dagger_{\nu,\bfq}$) are the creation (annihilation) operators of a phonon mode in branch $\nu$ with momentum $\bfq$. The total number of branches is $3n$, where  $n$ is the number of atoms in the primitive cell. $\omega_{\nu,\bfq}$ is the energy of phonon branch $\nu$ with momentum $\bfq$, and $\bfe_{\nu,j,\bfq}$ is the unit vector indicating the direction of oscillation of atom $j$ for phonon branch $\nu$. Finally, $m_j$ is the mass of atom $j$, and $\Ncells$ is the number of cells in the lattice. The $\bfq$ form an $\Ncells$-point discretization of the \brill\ zone, such that the variance of the displacement $\langle \bfu_{j,\bfl} \cdot \bfu_{j',\bfl'}\rangle$ is an intrinsic quantity under $\Ncells\to\infty$. 

In Fourier space, the equations of motion for the displacements then reduce to a standard eigenvalue problem for a given momentum vector,
\begin{equation}
\sum_{j'}\bfD_{\bfq,j,j'}\cdot\bfe_{\nu,j',\bfq}= \omega_{\nu,\bfq}^2 \bfe_{\nu,j,\bfq} \ ,
\end{equation}
where the eigenvectors are normalized such that $ \sum_j \bfe_{\nu,j,\bfq}^* \cdot \bfe_{\nu,j,\bfq} = 1$.
The dynamical matrix $\bfD_{\bfq,j,j'}$ is given by
\begin{equation}
	\bfD_{\bfq,j,j'}=\sum_{\bfl'}\frac{\mathbfcal{V}^{(2)}_{0,j,\bfl',j'}}{\sqrt{m_j m_{j'}}} e^{i\bfq\cdot(\bfr^0_{j'}+\bfl'-\bfr^0_{j})}.
	\label{eq:dynmatrix}
\end{equation}
For a rigorous derivation, see {\em e.g.}~\cite{Schober2014}. Once the force constants are known from a DFT calculation, the eigenvalue problem can be solved for $\omega_{\nu,\bfq}$ and $\bfe_{\nu,j,\bfq}$ using the post-processing software package \textsf{phonopy} \cite{phonopy}. From these the phonon-derived properties such as the phonon band structures shown in Fig.~\ref{fig:GaAsphononbands}, can be calculated.

The dynamical matrix receives an additional non-analytic contribution from the Born effective charges, which modifies the frequencies of the LO phonons.
The \textit{Born effective charge} is the electric charge that effectively contributes to the polarization induced during an atomic displacement, and is used to quantify the coupling between optical phonons and electric fields. Formally, the Born effective charge tensor $\bfZ^{*}$ is defined as the change in polarization $P$ resulting from a displacement $u$:
\begin{equation}
\mathbf{Z}^{*}_{ij} = \frac{\Omega}{e}\frac{\partial P_{i}}{\partial u_{j}} = \frac{1}{e}\frac{\partial F_{i}}{\partial E_{j}},  \quad i,j,k = x,y,z,
\end{equation}
where $\Omega$ is the unit cell volume, and $e$ is the electric charge. It can also be written in terms of the change in the force $F$ in a direction $i$ with respect to a homogeneous electric field $E$ in direction $j$. The Born effective charge tensor $\bf{Z^{*}}$ can be calculated using DFPT. This uses density functional theory to calculate the response of the system to a finite electric field as detailed in \cite{Gajdos_et_al:2006, Baroni/Resta:1986}. The Born effective charges are hence computed from the change in the Hellmann-Feynman forces and mechanical stress tensor due to the changes in the wavefunction. 

The calculated Born effective charges for GaAs and Al$_{2}$O$_{3}$ are  
\begin{align}
&\bf{Z}^{*}_{\text{Ga}}=\left(\!\begin{array}{ccc}2.27&&\\&2.27&\\&&2.27 \end{array}\!\!\right) 
&&\bf{Z}^{*}_{\text{As}}=\left(\!\begin{array}{ccc}-2.27&&\\&-2.27&\\&&-2.27 \end{array}\!\!\right)&\\
\intertext{and} 
&\bf{Z}^{*}_{\text{Al}}=\left(\!\begin{array}{ccc}2.980&&\\& 2.980&\\ &&2.951\end{array}\!\!\right) 
&&\bf{Z}^{*}_{\text{O}}=\left(\!\begin{array}{ccc}-1.937&&\\& -1.937&\\ &&-1.967\end{array}\!\!\right)&
\label{eq:bornsaph}
\end{align}
We note that due to the high symmetry of the GaAs crystal, the Born effective charges are a scalar quantity. For Al$_{2}$O$_{3}$, owing to the different anisotropic chemical environment surrounding Al and O atoms, the Born effective charges tensors are in general tensors that differ for inequivalent atoms, as listed in Appendix~\ref{app:crystallographic}. For our numerical calculations we use the diagonal, cell-averaged values for Al and O given above. 

The LO phonon modes correspond to ions with opposite effective charges moving in opposing directions along $\hat \bfq$, causing long-range macroscopic electric fields in a polar crystal. In contrast, TO phonons correspond to oppositely-charged ions moving in adjacent planes parallel to each other, resulting in no long-range Coulomb interaction (see Fig.~\ref{fig:GaAsphonons}). The additional force created by the electric field interaction with the LO phonon modes results in a frequency change in the LO mode as $\bfq\to0$. The lifting of the degeneracy between the LO and TO phonon modes at the Brillouin zone center -- the so-called LO-TO splitting -- can be calculated by including the non-analytic contribution to the dynamical matrix, given by
\begin{equation}
	{\bf{D}}^{NA}_{\bfq, j, j'}= \frac{e^2}{\Omega}\frac{(\bfq\cdot {\bfZ^\ast_j}) \, (\bfq\cdot {\bfZ^\ast_{j'}})}{ \bfq \cdot \epsilon^{\infty} \cdot \bfq}
\end{equation}
in Lorentz-Heaviside units. Hence calculating the $\bf{Z^{*}}$ allows one to determine the corrected LO modes. Here we use diagonal $\mathbf{\epsilon^\infty}$ tensors, as given in Table~\ref{tab:usefulstuff}.

The $\omega_{\nu,\bfq}$ and $\bfe_{\nu,j,\bfq}$ obtained from \textsf{phonopy} will be the most important inputs for the DM scattering rate calculations.  The next missing ingredient is the effective coupling of the dark matter to the displacement operator in \eqref{eq:displacementquant}. This coupling is model dependent and we treat it separately for dark photon and scalar mediator cases in Secs.~\ref{sec:darkphoton} and \ref{sec:nucleonint} respectively.

\subsection{Crystal alignment relative to dark matter flux}
\label{subsec:modulation}

Before turning to the scattering rate computation, we first establish our assumptions and conventions regarding the DM velocity distribution and the orientation of the DM wind in the frame of the crystal, which will determine the directional signal. The incoming DM velocity in the lab frame is modeled in a standard way, with a boosted Maxwell-Boltzmann distribution:
\begin{align}
	f(\bfv) &= \frac{1}{N_0}  \exp \left[ - \frac{(\bfv + \bfv_e)^2}{v_0^2} \right]   \ \ \Theta( v_{\text{esc}} - |\bfv + \bfv_e|), \label{eq:velodist} \\
	&N_0 = \pi^{3/2} v_0^3 \left[ {\rm erf} ( \tfrac{v_{\text{esc}}}{v_0} ) - \tfrac{2}{\sqrt{\pi}}\tfrac{v_{\text{esc}}}{v_0} \exp\left( -( \tfrac{v_{\text{esc}}}{v_0} )^2 \right) \right] 
\end{align}
with $v_0=220$ km/s, and truncated by the escape velocity $v_{\text{esc}}=500$ km/s. The velocity of the Earth with respect to the DM wind is indicated with  $\bfv_e$, with $|\bfv_e|\approx240$ km/s on average.

\begin{figure}
\includegraphics[width=0.5\textwidth]{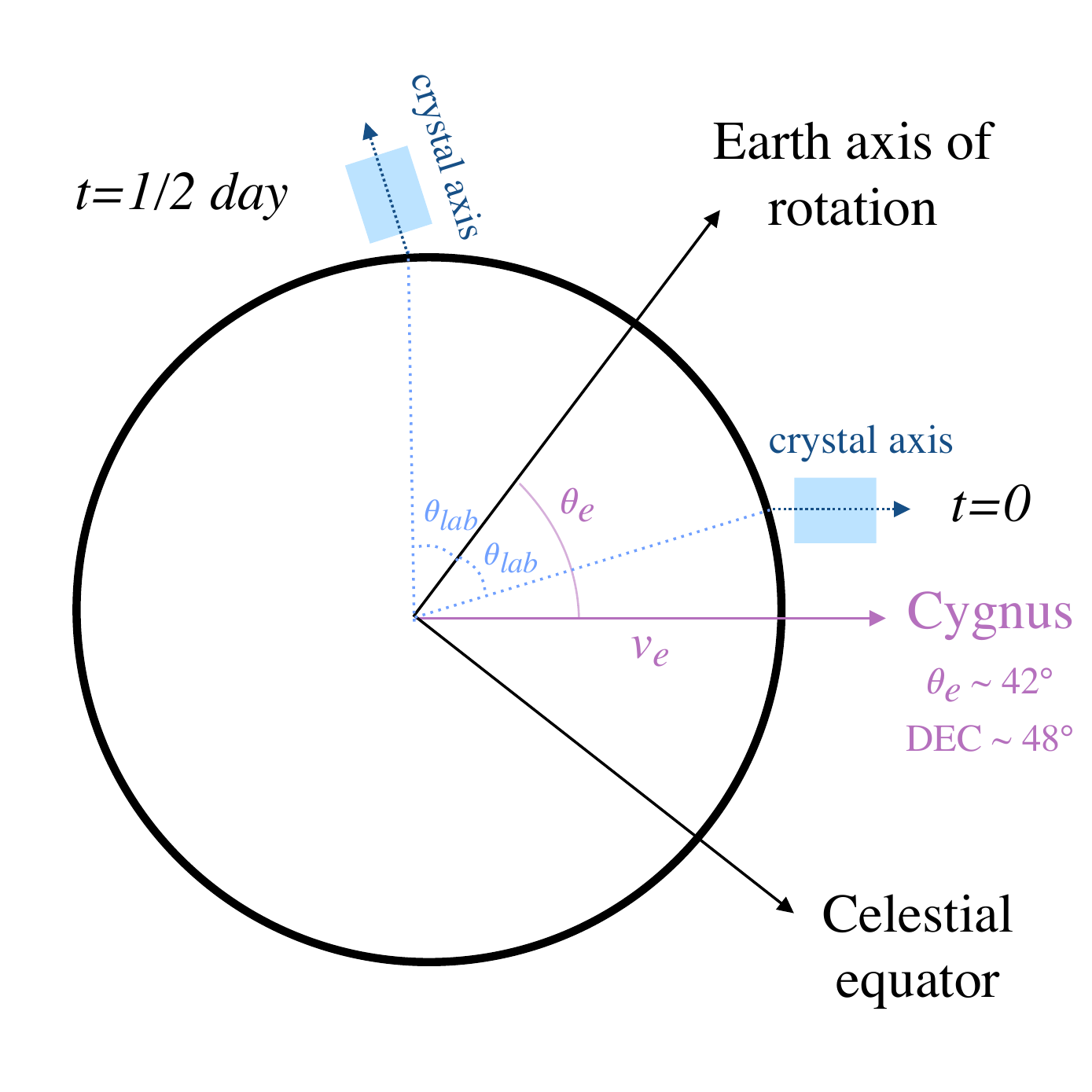}
\vspace{-0.5cm}
\caption{The setup assumed in our calculation of DM scattering rate. At $t=0$, the $z$-axis of the crystal coordinate system is aligned with the Earth's velocity ${\bf v_e}$. With this choice, the modulation is independent of the position of the lab, indicated by $\theta_{\rm lab}$. The Earth's velocity is approximately in the direction of Cygnus, which is at an angle of $\theta_e \approx 42^\circ$ relative to the Earth's axis of rotation. We also show the orientation of the crystal after a half-day rotation. \label{fig:angles}}
\end{figure}

The orientation of $\bfv_e$ relative to the crystal changes as the Earth rotates around its axis. Combined with the anisotropic crystal structure, this  sources a daily modulation of the scattering rate. We neglect the yearly modulation due to the Earth's orbit around the Sun. The orientation is illustrated in Fig.~\ref{fig:angles}, where $\theta_e \approx 42^\circ$ is the angle between the Earth's rotation axis and the direction of its velocity and $\theta_{\text{lab}}$ gives the latitude at which the experiment is constructed. We choose the crystal orientation and coordinate system such that the $z$-axis in the crystal frame is aligned with the Earth's velocity at $t=0$. For GaAs, we choose the $z$-axis in the crystal frame along one for the faces of the cubic lattice. For sapphire, the $z$-axis is taken to be aligned with the primary crystal axis, which is the axis defined by the Al atoms in Fig.~\ref{fig:lattice}. This implies that at $t=1/2$ day, the primary axis of the sapphire crystal is at an angle of  roughly $90^\circ$ with the DM wind. While we have not explicitly optimized for the crystal configuration, we expect that the choice here should (nearly) maximize the amplitude of the daily modulation since the biggest anisotropies in sapphire are that between the crystal axis and the axis perpendicular to it. 

Since we explicitly orient the crystal relative to the dark matter wind, there is no dependence of the DM flux or scattering rate on the latitude at which the experiment is located. As a function of time, the unit vector of $\bfv_e$ in the crystal coordinate frame is
\begin{equation}
\hat \bfv_e=\left(\!
\begin{array}{c}  
\sin\theta_e\sin\phi\\
\sin \theta_e \cos\theta_e (\cos\phi-1)\\
\cos^2\theta_e+\sin^2\theta_e\cos\phi
 \end{array}
  \!\right)
\end{equation}
with $\phi=2\pi\times t/24 \text{h}$ the angle parametrizing the rotation of the Earth around its axis. 

%%%%%%%%%%%%%%%%%%%%%%%%%%%%%%%%%%%
\section{Dark photon mediated scattering\label{sec:darkphoton}}
%%%%%%%%%%%%%%%%%%%%%%%%%%%%%%%%%%%

We begin by considering Dirac fermion DM that interacts with the SM via a kinetically mixed dark photon.  The model is defined by the vacuum Lagrangian
\begin{equation}
	\mathcal{L}=\mathcal{L}_{\text{SM}} -\overline X (\slashed{\partial} - i g_X \slashed{A}')X -\frac{1}{4}F'^{\mu\nu}F'_{\mu\nu}  -\frac{\kappa}{2} F^{\mu\nu}F'_{\mu\nu} -\frac{m_{A'}^2}{2} A'^\mu A'_\mu \, ,
\end{equation}
where $\kappa \ll 1$ is the kinetic mixing parameter, and $g_X$ and $m_{A'}$ are respectively the gauge coupling and mass of the $A_\mu'$ field. For finite $m_{A'}$, one can perform a diagonalization to the mass basis, where the electron picks up a small charge (in vacuum) of $\kappa e$ under the dark photon.   On the other hand, in the limit where $m_{A'}\rightarrow 0$, we can perform a field redefinition $A'\rightarrow A' - \kappa A$ to write the Lagrangian as
\begin{equation}
\mathcal{L}=\mathcal{L}_{\text{SM}} -\overline X (\slashed{\partial} - i g_X \slashed{A}'- i \kappa g_X \slashed{A})X -\frac{1}{4}F'^{\mu\nu}F'_{\mu\nu},
\label{eq:millicharge}
\end{equation}
where the dark matter $X$ has a small charge $e'\equiv \kappa g_X$ under the photon.  In either of these cases, there is a coupling of the DM current to the electromagnetic current that is proportional to $\kappa g_X e$.

For sub-MeV dark matter, the relic abundance can be explained by freeze-in~\cite{Essig:2011nj,Chu:2011be,Essig:2015cda} via the out-of-equilibrium process $e^+ e^- \to \overline X X$. Since this production rate is proportional to the coupling combination $\kappa^2 g_X^2$, requiring that $X$ is 100\% of the dark matter predicts also a compelling target for DM scattering off SM particles. Requiring that the DM-$A'$ coupling satisfies self-interaction bounds and that the kinetic mixing $\kappa$ is consistent with dark photon constraints, one finds that $m_{A'} \lesssim 10^{-11}$ eV~\cite{Knapen:2017xzo}. Since this mass is much smaller than a typical in-medium effective photon mass, we can take the massless $A'$ limit. We are then in the situation given by Eq.~\eqref{eq:millicharge} above, where we can treat the DM as a millicharged particle for the purposes of our calculations. One could also consider the model above with $m_{A'}=0$ as a specific model of millicharged DM\footnote{There have been claims in the literature that millicharged particles are effectively ejected from the disk by SN shocks, and that they cannot re-enter the disk due to the Milky Way's magnetic fields~\cite{Chuzhoy:2008zy,SanchezSalcedo:2008zd}. We will not consider such bounds further for several reasons: first, in the dark photon model, whether the DM behaves as a millicharged particle depends on $m_{A'}$ and the in-medium photon mass in the disk ($\sim 10^{-11}$ eV). Furthermore, even in the $m_{A'} \to 0$ limit, the millicharges considered here are extremely small, $e' \lesssim 10^{-10}$, implying a significantly reduced efficiency for injecting particles in the SN shock.  The arguments raised in Refs.~\cite{Chuzhoy:2008zy,SanchezSalcedo:2008zd} merit further study in the context of light kinetically mixed dark photons, but are beyond the scope of this work. }.

For interactions mediated by an ultralight dark photon, the long-range coupling of DM with a phonon in the crystal is then similar to that of electrons with phonons, but with an amplitude suppressed by $e'/e$. Here, we specifically mean the interaction associated with a $1/r^2$ Coulomb field. (There are also short-range electron-phonon interactions in a material, but in a polar material these interactions are typically much smaller.) The long-range interaction between electrons and phonons in semiconductors and insulators is described by the Fr\"ohlich Hamiltonian~\cite{frolich1954,YuCardona}. Physically, an electron injected into the crystal sources a small electric field, which induces an oscillation of the ions via the Born effective charges. This oscillation can then be identified with a phonon mode. 

Since the DM scattering in polar materials behaves similarly to millicharged dark matter, we can directly use the Fr\"ohlich Hamiltonian as a description for this process. The main difference with the electron case is that the DM is a free particle, while for electrons the appropriate in-medium wavefunctions must be included. On a practical level, this is a simplification of the computation since the plane wave approximation is sufficient to describe the DM. In the following section we will summarize the most important formulas and numerical results, and provide the relevant derivations in Appendix~\ref{app:frolich}.

%%%%%%%%%%%%%%%%%%%%%%%%%%%%%%%%%%%
\subsection{Fr\"ohlich interaction}
%%%%%%%%%%%%%%%%%%%%%%%%%%%%%%%%%%%

Adapted for the DM case, the Fr\"ohlich matrix element\footnote{Note that \eqref{eq:frolichgeneral} differs from the expression in~\cite{PhysRevLett.115.176401} with a phase factor, as we have used a different convention for the phase of $\bfe^*_{j,\nu,\bfq}$. } for a periodic lattice is given by~\cite{PhysRevLett.115.176401}
\begin{equation}
	\label{eq:frolichgeneral}
	\mathcal{M}_{{\bfq + \bfG,\nu}}= i e e' \sum_{j}\frac{1}{\sqrt{2 \unitcell m_j \omega_{\nu,\bfq}}}\frac{(\bfq+\bfG)\cdot\bfZ^\ast_j \cdot \bfe^*_{j,\nu,\bfq}}{(\bfq+\bfG)\cdot\bfeps_\infty \cdot(\bfq+\bfG)}.
\end{equation}
This result is derived in Appendix~\ref{app:frolich}. 
Here $\nu$, $j$ and $\bfG$ are the phonon branches, the atoms in the primitive cell and the reciprocal lattice vectors, respectively. The momentum transfer is given by $\bfq + \bfG$, while $\bfq$ is the momentum of the phonon restricted to the first \brill\ zone. $e$ is the electron charge\footnote{We adopt Lorentz-Heaviside units where the vacuum permittivity $\varepsilon_0=1$ and $e = \sqrt{4 \pi \alpha_{em}}$, while \cite{PhysRevLett.115.176401} uses SI units. } and $\unitcell$ is the volume of the primitive unit cell. In general $\bfeps_{\infty}$ is a tensor, though it is well approximated by the scalar quantity $\epsilon_\infty$ times the unit tensor.   The phonon eigenvectors $\bfe^*_{j,\nu,\bfq}$, energies  $\omega_{\nu,\bfq}$, and the Born effective charges $\bfZ^\ast_j$ are all computed from first principles, as described in Sec.~\ref{sec:theoreticaldesc}. A similar formulation is often used in the literature to describe the coupling of electrons with optical phonons, albeit with the inclusion of the electron wavefunctions in the medium. 

The expression above can be understood more intuitively by taking the isotropic and long-wavelength approximation. In this limit, and assuming a single phonon branch $\nu$ contributes, the matrix element simplifies to
\begin{align}
	\mathcal{M}^{\text{iso}}_{{\bfq}}&=i  \frac{e'}{\epsilon_\infty}\sum_{j}  \frac{e Z^\ast_j\ \bfq \cdot \bfe^*_{j,\nu,\bfq}}{ \sqrt{2 \unitcell m_j \omega_{LO}}}\frac{1}{|\bfq|^2}, 
\end{align}
where we have dropped the reciprocal lattice vector (since the result is dominated by $\bfG = 0$). For a given phonon eigenmode, the physical displacements of atom $j$ are proportional to $ \bfe^*_{j,\nu,\bfq}/\sqrt{2m_j \omega_{LO}}$; weighted by $e Z^\ast_j$, this is simply the dipole moment corresponding to the lattice displacements. The eigenvector is dotted into $\bfq$, selecting for the longitudinal mode, while the overall $1/|\bfq|$ scaling is that expected for a dipole-charge coupling. Finally, the field generated by the dipole is screened by the valence electrons, which accounts for the $1/\epsilon_\infty$ factor.

The above expression can be further simplified for certain crystals. In Ref.~\cite{Knapen:2017ekk}, we considered GaAs, which has a single LO phonon branch. As discussed in the previous section, GaAs has a cubic symmetry with $Z^\ast_{\rm Ga} = - Z^\ast_{\rm As}$ and $m_{\rm Ga} \approx m_{\rm As}$. We can then make the additional approximation in the long-wavelength limit,
\begin{align}
	\mathcal{M}^{\text{iso}}_{{\bfq}}&\approx i\frac{ee'}{\epsilon_\infty } \frac{Z^\ast}{\sqrt{2 \unitcell \mu \omega_{LO}}}\frac{1}{|\bfq|}\\
	&=ie' \Bigg[\frac{\omega_{LO}}{2}\left(\frac{1}{\epsilon_\infty}-\frac{1}{\epsilon_0}\right)\Bigg]^{1/2}\frac{1}{|\bfq|}\label{eq:frolichiso}
\end{align}
with $\mu$ the reduced mass of the Ga and As atoms.
In the second equality, we expressed the Born effective charge $Z^*$ in terms of the measured quantities $\oLO$ (the frequency of the LO phonon as $\bfq \to 0$),  $\epsilon_{\infty}$, and $\epsilon_{0}$. The derivation for this identity is given in Appendix~\ref{app:frolich}. 
None of these simplifications apply for sapphire, however, and there we must numerically sum over all phonon eigenmodes.

%%%%%%%%%%%%%%%%%%%%%%%%%%%%%%%%%%%
\subsection{Reach}
%%%%%%%%%%%%%%%%%%%%%%%%%%%%%%%%%%%

The scattering rate for an incoming DM particle with velocity $v_i$ is obtained from Fermi's golden rule, 
\begin{align}
\Gamma = 2\pi \sum_{\nu} \int_{\text{BZ}}\! \frac{d^3\bfq}{(2\pi)^3}|\mathcal{M}_{{\bfq,\nu}}|^2\delta(E_i-E_f-\omega_{\nu,\bfq}),\label{eq:Gammafrolich}
\end{align}
where the momentum integral is over the first \brill\ zone. 
For scattering for sub-MeV dark matter,  we simplify the matrix element in Eq.~\eqref{eq:frolichgeneral} by observing that  the momentum transfer $q\sim v_X m_X$ is small compared to the size of the \brill\ zone, except for $m_X$ approaching 1 MeV. In addition, the matrix element is proportional to $1/|\bfq|$ and therefore the rate is dominated by those phonon modes well within the first \brill\ zone. We can therefore neglect Umklapp processes where phonons are created with wavevectors outside the first \brill\ zone; this amounts to setting $\bfG = 0$ in Eq.~\eqref{eq:frolichgeneral}. (We expect that the reach does extend to higher masses via such processes, but will not consider this regime further here.) 

The integral above can be performed analytically if we take the isotropic limit  for the matrix element in Eq.~\eqref{eq:frolichiso}:
\begin{equation}
\Gamma^{\text{iso}}(v_i)= \frac{|e'|^2}{4\pi}\frac{\omega_{LO}}{v_i}\left(\frac{1}{\epsilon_\infty}-\frac{1}{\epsilon_0}\right)\log\frac{1+ \sqrt{1-2\oLO/m_X v_i^2}}{1- \sqrt{1-2\oLO/m_X v_i^2}}\Theta\left(m_X v_i^2/2-\oLO\right)
\end{equation}
where $v_i$ is the initial velocity of the DM and the Heaviside $\Theta$-function enforces energy conservation. 

 \begin{figure}[t]
\includegraphics[width=0.6\textwidth]{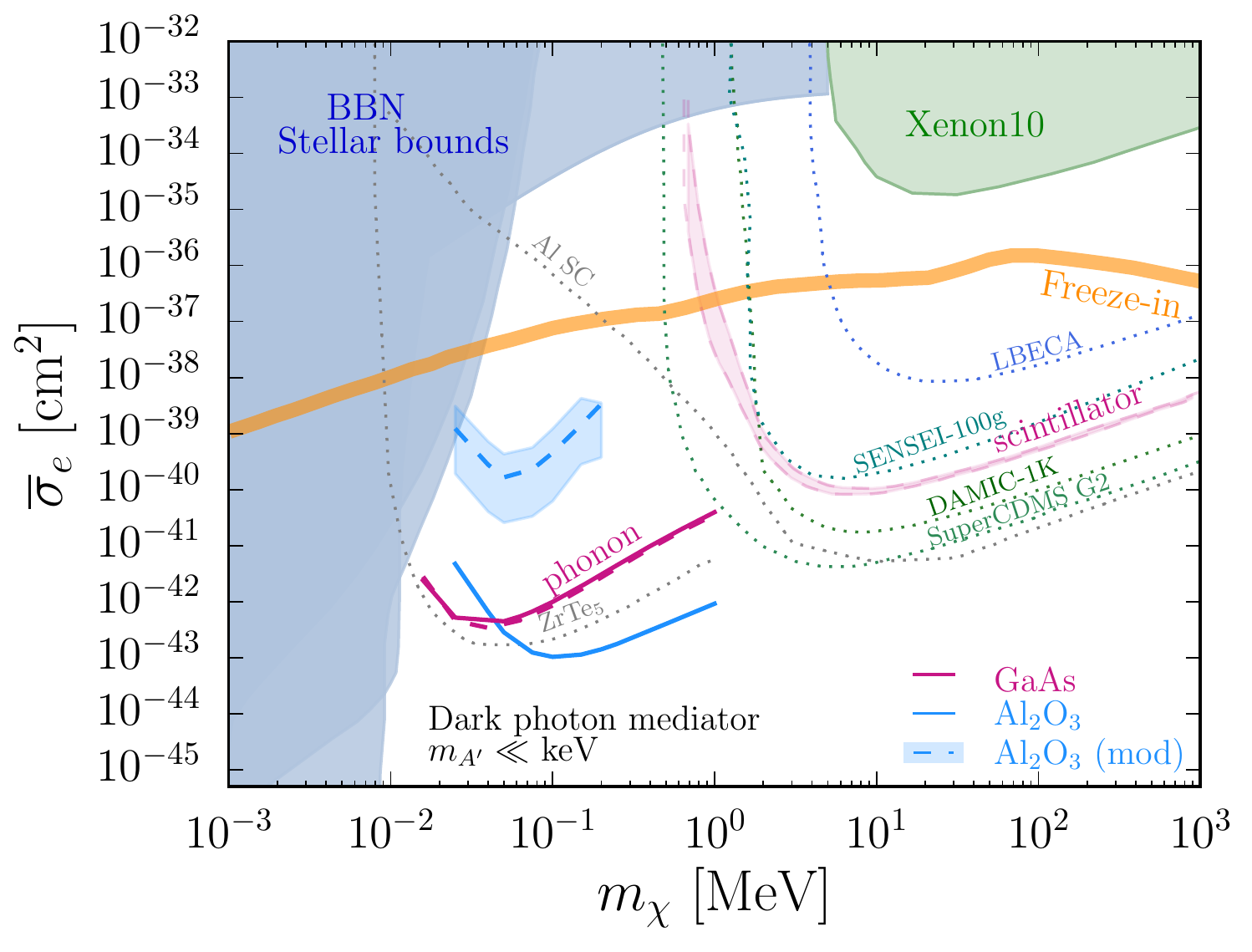}
\caption{Reach for dark photon mediated scattering in GaAs and sapphire, assuming one kg-year exposure. For sapphire, we indicate the sensitivity if one requires a $2\sigma$ observation of the daily modulation (see Sec.~\ref{sec:dailymodfrolich}). For GaAs, we also show the result using the analytic approximations in \cite{Knapen:2017ekk} (dashed line), which is nearly identical to the full numerical result. GaAs can also be operated as scintillator for dark matter masses above 1 MeV~\cite{Derenzo:2016fse}, as indicated by the dashed purple lines.  Existing constraints and other proposed experiments are described further in the text. \label{fig:darkphotonlimit}}
\end{figure}

For the full numerical result as well as for the analytic approximation, the scattering rate for the target is obtained by integrating over the initial DM velocities:
\begin{equation}
R=\frac{1}{\rho_T}\frac{\rho_X}{m_X}\int\! d \bfv_i^3 f(\bfv_i) \Gamma(\bfv_i)
\end{equation}
 with $f(\bfv_i)$ the dark matter velocity distribution in Eq.~\eqref{eq:velodist} and $\rho_T$ the mass density of the target material. 

To estimate the reach, we compute the expected 90\% exclusion on $e'$ assuming zero events observed with no expected background.\footnote{Backgrounds from coherent photon and coherent neutrino scattering are estimated to be no higher than a few events / kg year exposure \cite{Knapen:2017ekk}.} To compare with existing constraints and other proposed experiments, we express the result in terms of 
\begin{equation}
	\bar \sigma_e = \frac{4 e'^2 \alpha \mu^2_{X e}}{(\alpha m_e)^4}.
\end{equation}
which corresponds to the typical cross section of dark matter with a bound electron, {\em e.g.}~in a semiconductor-based experiment.  $\mu_{X e}$ is the DM-electron reduced mass, $\alpha$ is the fine-structure constant and $m_e$ is the electron mass. The result is shown in Fig.~\ref{fig:darkphotonlimit} for both GaAs and sapphire. For GaAs, we compare the isotropic limit with the numerical result including phonon eigenmodes and find excellent agreement. Also shown are existing stellar cooling~\cite{Vogel:2013raa}, BBN~\cite{Davidson:2000hf} and Xenon10~\cite{Essig:2012yx} constraints, as well as the projected reach of other experimental proposals~\cite{Essig:2015cda,Battaglieri:2017aum,Hochberg:2015pha,Hochberg:2017wce}. Interestingly, we find that as little as a gram-month exposure would suffice to reach the freeze-in benchmark. In the sub-MeV range, an experiment based on a Dirac material \cite{Hochberg:2017wce} is currently the only other proposal which could compete with polar materials. Given that Dirac materials have not yet been fabricated in the quantities needed for a dark matter detector, we expect that the polar material concept could be realized on a substantially shorter timescale. Also shown in Fig.~\ref{fig:darkphotonlimit} (dashed blue) is the expected sensitivity for sapphire if one requires a daily modulation signal at 2$\sigma$. We elaborate on the daily modulation in the next section.

%%%%%%%%%%%%%%%%%%%%%%%%%%%%%
\subsection{Daily modulation\label{sec:dailymodfrolich}}
%%%%%%%%%%%%%%%%%%%%%%%%%%%%%

The anisotropy in the crystal structure induces a dependence of the scattering rate on the crystal orientation, which translates to a modulation over the sidereal day. Here there are two effects that lead to modulation: the directional dependence in the phonon couplings to the DM model, and in the phonon energies. For GaAs, which has a high degree of symmetry, we find that this modulation is negligible. Instead, for sapphire, there is a sizable anisotropy in the DM scattering rate. In the rest of this section, we discuss the dominant effects and present results on the modulation.

\begin{figure}
        \centering
        \includegraphics[width=0.5\textwidth]{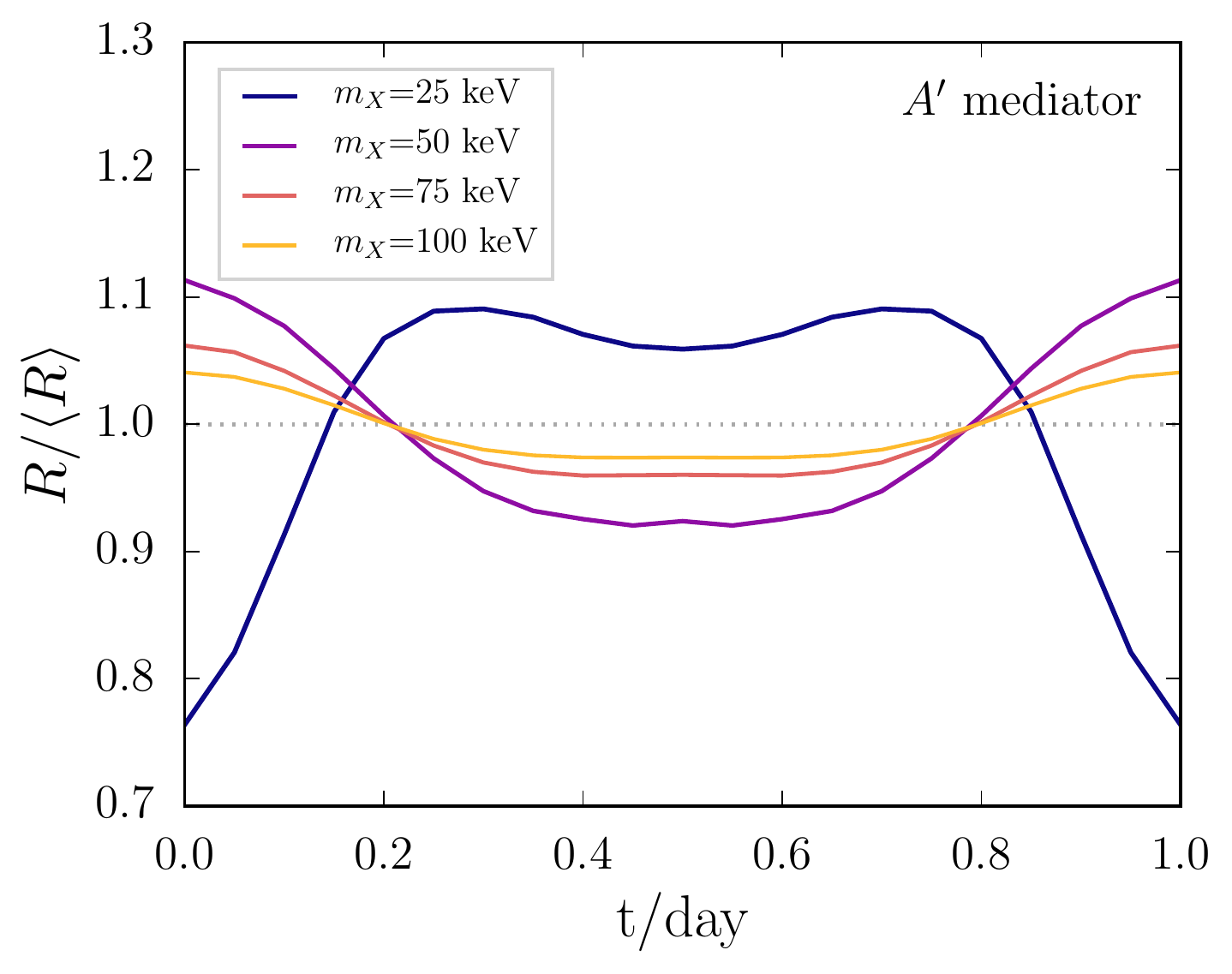} 
        \caption{Modulation of the scattering rate for sapphire over a sidereal day, assuming a 25 meV threshold.  \label{fig:darkphotondirection} }
\end{figure}

The daily modulation in sapphire for several DM masses is shown in Fig.~\ref{fig:darkphotondirection}. Here we assumed a threshold of 25 meV, well below the energies of the optical phonons. Since any backgrounds are expected to be either flat in time, or at least out of phase with the sidereal day over many periods, this can be used as an additional indicator of a DM signal. Assuming a kg-year exposure, we can estimate the cross section needed to reject the null hypothesis of non-modulating scattering at the $2\sigma$ level. This is given by the dashed blue line in Fig.~\ref{fig:darkphotonlimit}, which requires that in 50\% of the simulated signal datasets, the null hypothesis can be rejected. The shaded band indicates the $\pm 1\sigma$ band around the mean: specifically, the cross section needed if we instead require this to be true in 16\% or 84\% of the simulated datasets  (assuming only statistical fluctuations).  We refer to Appendix~\ref{app:statistics} for details on our statistical treatment.

\begin{figure}
        \centering
\subfloat[][Matrix element by mode\label{fig:frolichband}]{\includegraphics[height=6cm]{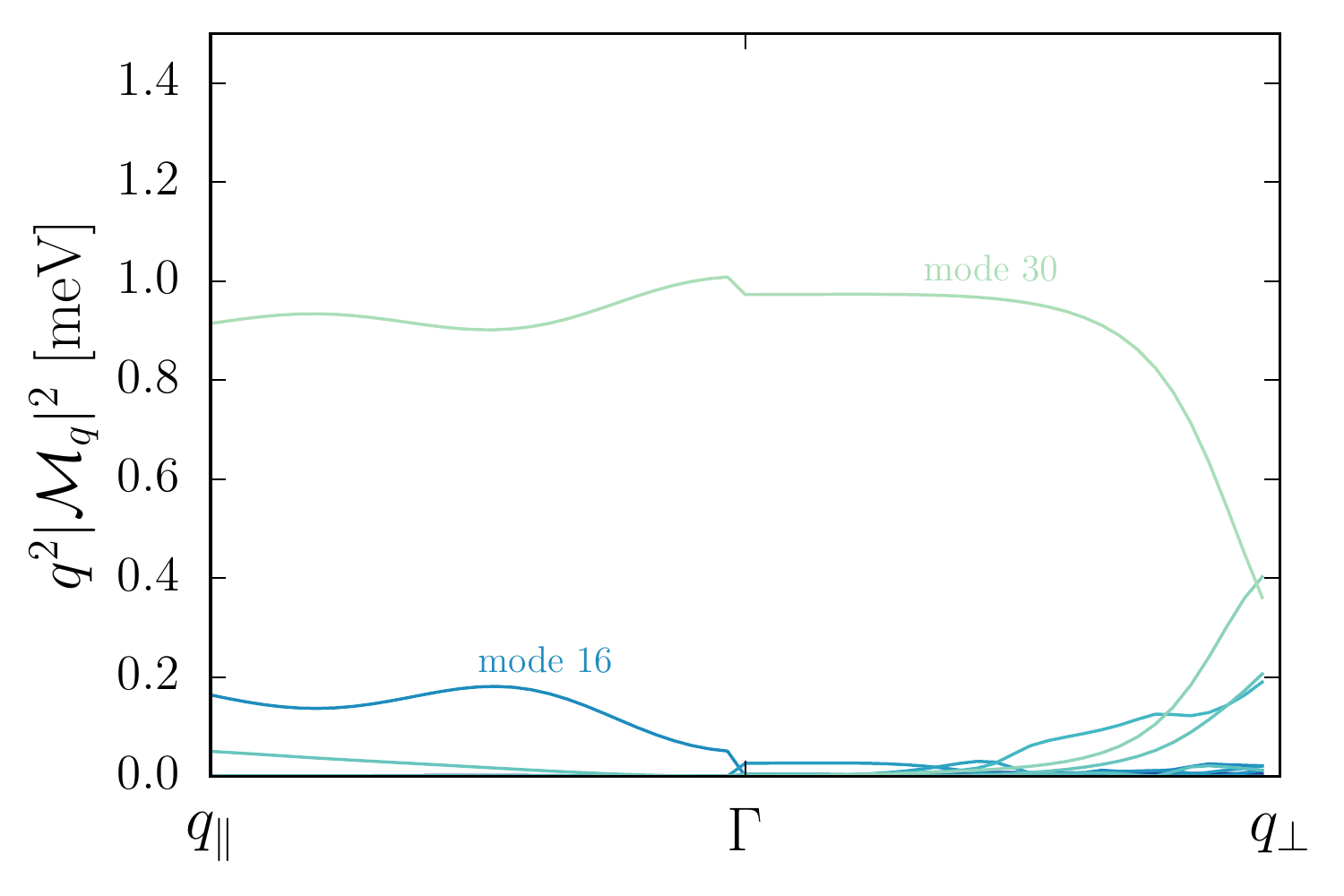}}
\subfloat[][Rate breakdown by mode\label{fig:bymodeall}]{\includegraphics[height=6cm]{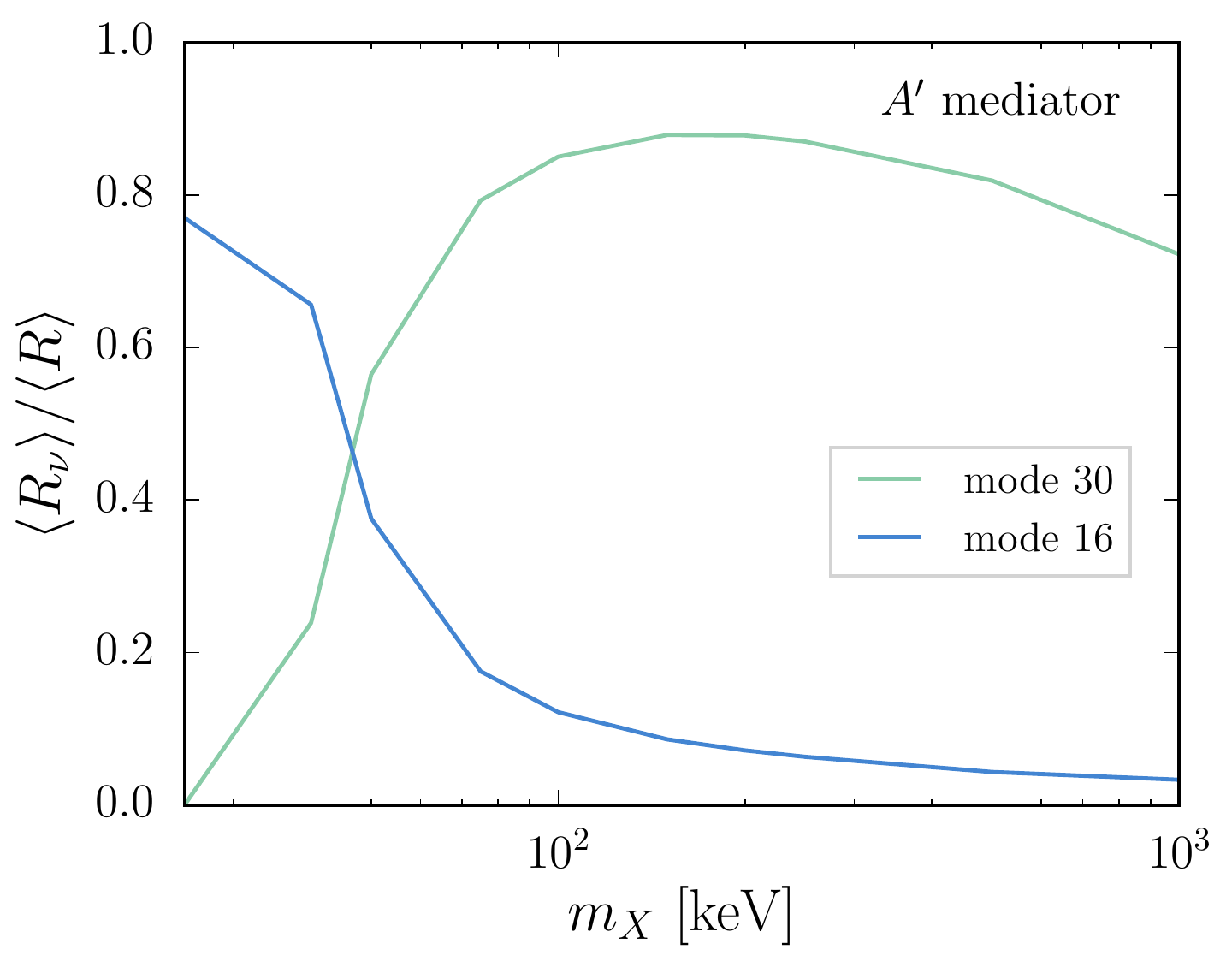}}
  \caption{ (a) Fr\"ohlich matrix element squared for each phonon mode in sapphire, along a path parallel ($q_\parallel$) and orthogonal ($q_\perp$) to the primary crystal axis. We label only the dominant modes, as the other modes contribute negligibly to the scattering rate. (b) Fraction of the total rate contributed by the two most important modes, as a function of $m_X$.  }
\end{figure}

To understand the origin of the modulation in Fig.~\ref{fig:darkphotondirection}, it is useful to deconstruct the total scattering rate in terms of the rate from individual phonon branches. In Fig.~\ref{fig:frolichband}, the squared Fr\"ohlich matrix element from Eq.~\eqref{eq:frolichgeneral} is plotted separately for all phonon modes; here we show a band from the direction parallel to the crystal axis ($q_\parallel$) to the origin of the \brill~zone $(\Gamma)$, and then from $\Gamma$ to the direction orthogonal to the crystal axis ($q_\perp$). The most striking feature is that   the contribution of a single mode appears to dominate the matrix element. This is also the most energetic mode in the spectrum, mode 30, with $\omega\approx 104$ meV. (We label the 30 phonon modes according to their energy in the vicinity of the origin of the \brill\ zone, from least energetic ``mode 1'' to most energetic ``mode 30''.) Fig.~\ref{fig:frolichband} also highlights the directional-dependence and the $q$-dependence of the phonon couplings, which enters directly into the scattering rate. As can be seen from Fig.~\ref{fig:angles},  we have assumed that the crystal axis is aligned with the DM wind at $t=0$, so that the scattering is preferentially along the crystal axis (the degree to which this is true depends on the DM mass, of course). Meanwhile, the crystal axis is nearly perpendicular to the DM wind at $t=0.5$ day, with the dominant scattering into those modes which have large dipole along the $q_\perp$ directions. Fig.~\ref{fig:frolichband} thus suggests that the highest rate occurs along the $q_\parallel$ direction, corresponding with $t=0$, which is consistent with the location of the maximum in Fig.~\ref{fig:darkphotondirection} for $m_X\gtrsim 50$ keV. 

\begin{figure}[t]
\centering
\subfloat[][Mode 30]{
\includegraphics[width=0.45\textwidth]{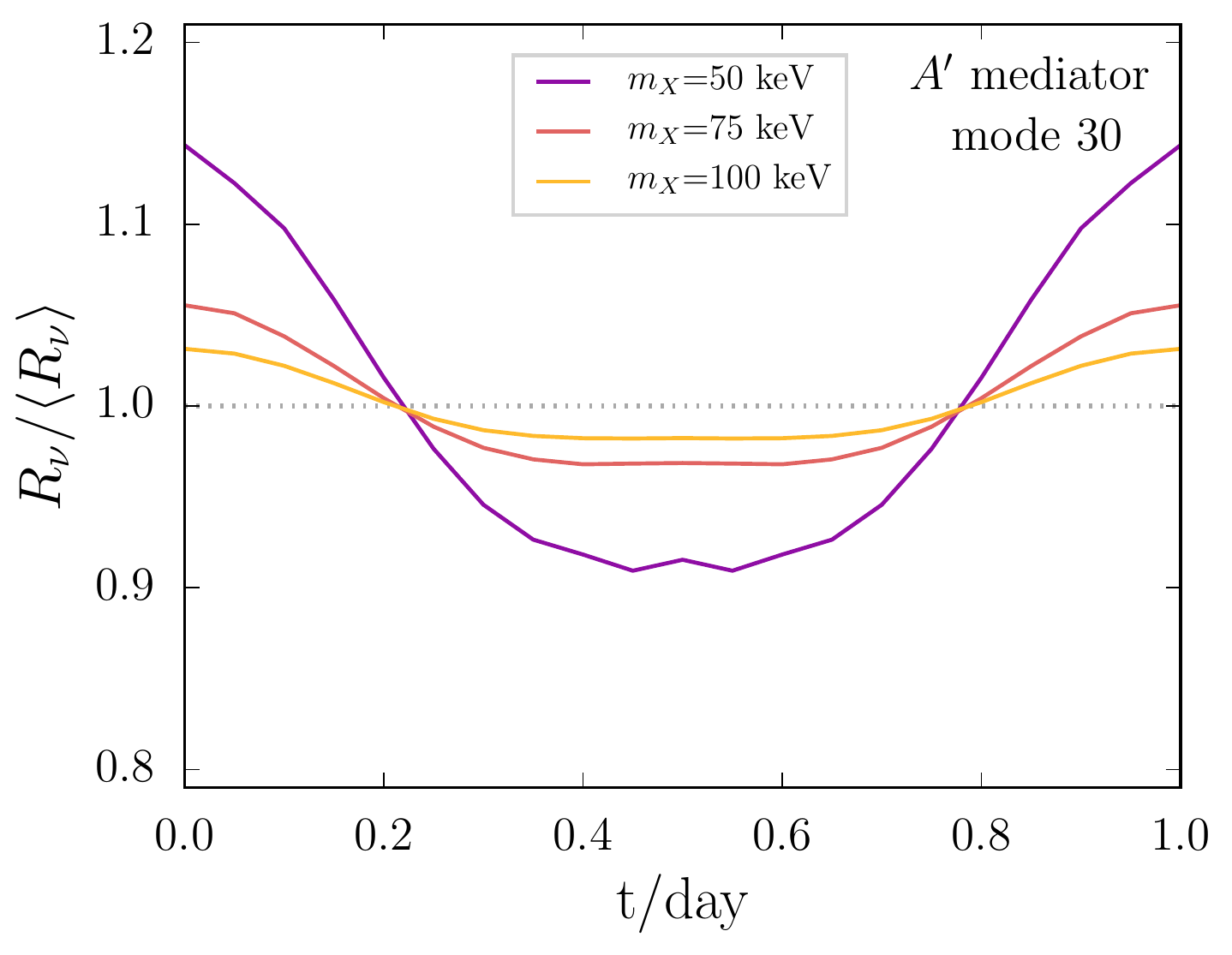}
\label{fig:bymodeall30}}\hfill
\subfloat[][Mode 16]{
\includegraphics[width=0.45\textwidth]{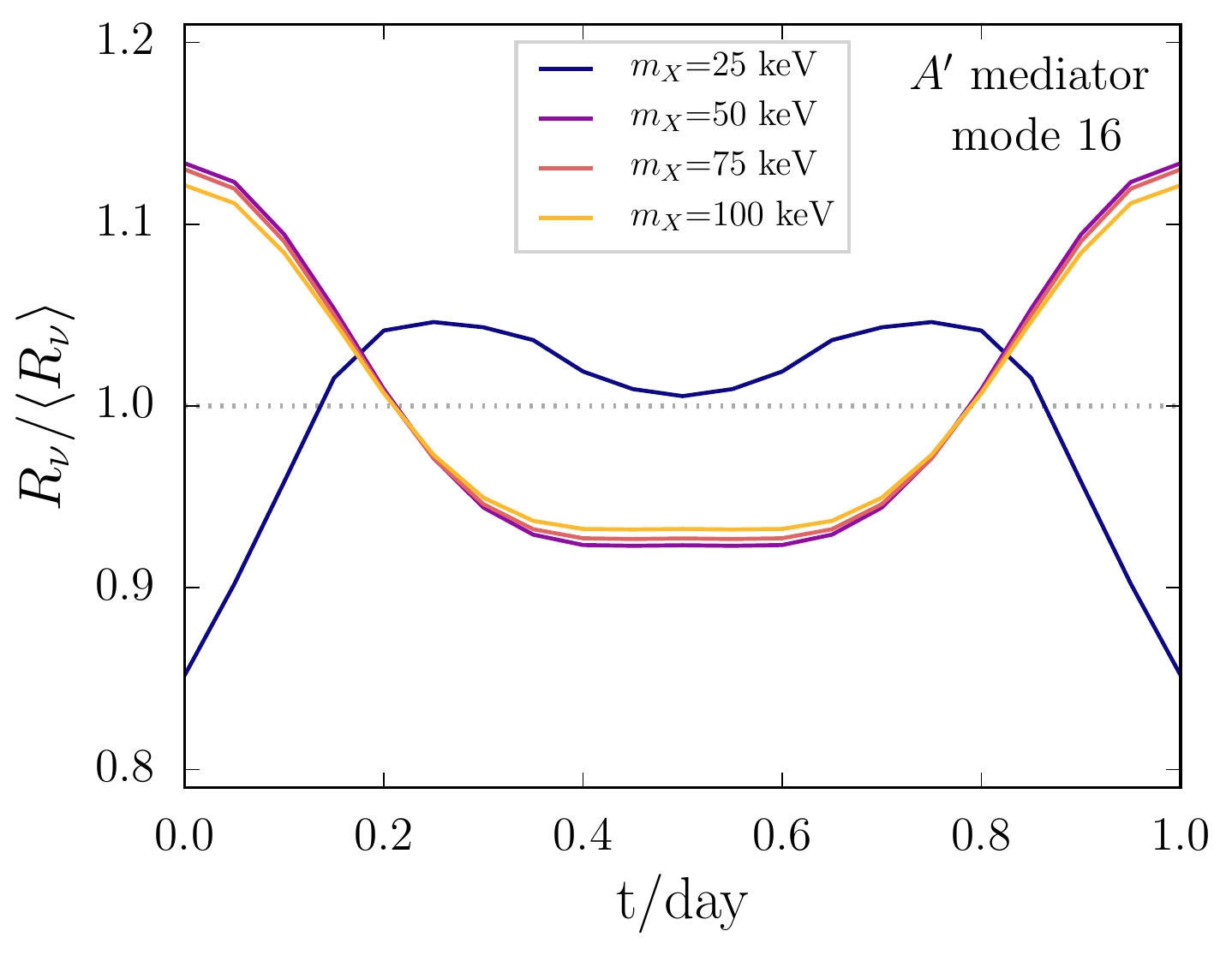}
\label{fig:bymodeall16}}
\caption{ The modulation for individual phonon modes in sapphire over a sidereal day. Mode 30 does not contribute for $m_X=25$ keV, since it is kinematically inaccessible.
\label{fig:bymode}}

\end{figure}

For $m_X\lesssim50$ keV, mode 30 is kinematically forbidden, and the modulation pattern therefore changes. In this case, the lower-lying mode 16 ($\omega\approx $ 60 meV) takes over, as shown in Fig.~\ref{fig:bymodeall}. For such low $m_X$, threshold effects from the  directionally-dependent phonon energies dominate the modulation, which is the reason for the large difference in the modulation pattern of mode 16 between $m_X=25$ keV and $m_X=50$ keV. Mode 30 and mode 16 are visualized in the low $\bfq$ regime in the animation in panels (a-b) of Fig~\ref{fig:animation}. In mode 30, all Al atoms are exactly in phase with each other but in anti-phase with the O atoms, and it therefore has the largest dipole of all the phonon modes. Because of its large dipole, this mode also represents the largest disturbance in the electrostatic potential of the system, which explains why it is the most energetic. In mode 16, the Al atoms also move coherently but with a lower amplitude along the crystal axis as compared to mode 30. As such it is subdominant, unless mode 30 is kinematically inaccessible. It is worthwhile to inspect the modulation patterns of the contributions from mode 16 and mode 30 separately, which we present in Fig.~\ref{fig:bymode} for several DM masses. One can see that mode 16 gives rise to a much larger amplitude, and that its phase is shifted with respect to that of mode 30, especially at low mass. This explains the dramatic change in the modulation pattern in Fig.~\ref{fig:darkphotondirection} for $m_X=25$ keV, for which mode 30 is forbidden.

\begin{figure}[t]
\centering
\subfloat[][Mode 30, low $\bfq$]{
\animategraphics[loop,autoplay,poster=first,width=0.3\linewidth]{20}{figures/mode30}{}{}
\label{fig:animation30}}
\hfill
\subfloat[][Mode 16, low $\bfq$]{
\animategraphics[loop,autoplay,poster=first,width=0.3\linewidth]{20}{figures/mode16}{}{}
\label{fig:animation16}}
\hfill
\subfloat[][Mode 30, high $\bfq$]{
\animategraphics[loop,autoplay,poster=first,width=0.3\linewidth]{20}{figures/mode30_highq}{}{}
\label{fig:animation30_highq}}
\caption{Animation of the atoms in the primitive cell, where we show the phonon modes in sapphire that dominate the scattering for dark photon mediated processes. Both  modes (30 and 16) are characterized by a large oscillating dipole of the Al (gray) and O (red) atoms. At high momentum, the relative motion between the atoms is less coherent, illustrated in panel (c). Adobe Acrobat reader is required to view these animations. Animations were generated with \cite{phononwebsite}.\label{fig:animation}}

\end{figure}

The amplitude of the modulation decreases for higher $m_X$, where larger $\bfq$ values are sampled in the \brill~zone. We expect that in this limit, scattering starts to transition towards scattering with a single nucleus, which is isotropic. In other words, at high momentum transfer the DM is blind to the long-range crystal structure. In practice, this effect manifests itself in a gradual randomization of the eigenvectors as $\bfq$ is increased on a particular phonon branch. To illustrate this effect, the animation in Fig.~\ref{fig:animation30_highq} shows mode 30 for a point near the edge of \brill~zone with $|\bfq|\sim1$ keV, which displays less coherent oscillations within the unit cell.

Finally, we comment on the theoretical uncertainties of our first principles calculation.  As a proxy for the uncertainty, we have also calculated total scattering rates and modulation patterns using inequivalent rather than averaged Born effective charges for each of the different Al and O atoms in the primitive cell (see discussion in App.~\ref{app:crystallographic}). We find small differences in the total rate at the level of a few percent for $m_X\gtrsim 75$ keV, though the difference grows for lighter DM, up to a factor of $\sim 4$ for $m_X\approx 25$ keV. The modulation amplitude differs by roughly a factor of two between the two assumptions for $m_X\approx 25$ keV, but this difference reduces to roughly 50\% for $m_X\gtrsim 50$ keV. We find that the overall modulation pattern remains unchanged.

%%%%%%%%%%%%%%%%%%%%%%%%%%%%%%%%%%%
\section{DM-nucleon scattering\label{sec:nucleonint}}
%%%%%%%%%%%%%%%%%%%%%%%%%%%%%%%%%%%

In this section, we consider the benchmark where DM couples primarily to nuclei through a light scalar mediator. The underlying interactions are
\begin{align}
	\label{eq:nucleonmodel}
	\mathcal{L}\supset - \frac{1}{2}m_X^2X^2 - \frac{1}{2}m_\phi^2 \phi^2 - \frac{1}{2}y_X m_X \phi X^2 - y_n \phi (\overline n n + \overline p p) , 
\end{align}
where we assume a scalar DM particle, $X$, and an identical coupling $y_n$ to both neutrons and protons. We further take $m_\phi$ small compared to the typical momentum transfer; the so-called light-mediator regime. Similarly to the dark photon mediator, this model is already subject to a number of astrophysical and terrestrial constraints, and we refer the reader to Ref.~\cite{Knapen:2017xzo} for a detailed discussion.\footnote{In particular, this model is most motivated for sub-component DM. In our figures we will however assume that $X$ is $100\%$ of the DM, as the reach is easily rescaled to a particular subcomponent fraction. } 

\subsection{DM-phonon form factor}

The scattering of DM off the nuclei of a lattice is similar to the scattering of cold neutrons, except for an additional form factor associated with the light mediator. There is extensive literature on the scattering of cold neutrons (for a review, see for example Ref.~\cite{Schober2014}), as this process is important to experimentally measure phonon dispersion relations. In the cold neutron case, the cross section for a neutron to scatter off a single nucleus $N$ is written as $\sigma = 4 \pi \overline b_N^2$, where $\overline b_N$ is the average neutron-nucleus scattering length. Accounting for the lattice structure requires summing over nuclei, weighted appropriately by the phonon wavefunctions. For DM scattering via a light scalar, the techniques for cold neutron scattering in the lattice can then be directly applied. 

For a nearly massless mediator, the differential cross section diverges as $1/|\bfq|^4$, though the divergence is cut off by the experimental threshold. Since this threshold varies for different experiments, it is conventional to introduce an effective DM-nucleon cross section,
\begin{align}
\label{eq:neucleoneffectivecrosssec}
	\overline \sigma_n \equiv \frac{y_n^2 y_X^2}{4\pi}\frac{\mu_{X n}^2}{q_0^4} \approx  \frac{y_n^2 y_X^2}{4\pi}\frac{m_{X}^2}{q_0^4} ,
\end{align}
where $q_0\equiv v_0 m_X$ is a reference momentum and $\mu_{X n}$ is the DM-nucleon reduced mass. (The choice for $q_0$ is merely a convention, and drops out in the scattering rate.) We can similarly define an effective DM-nucleon scattering length from the relation $\overline \sigma_n = 4\pi \overline b_X^2$.

The primary quantity for describing the response of a crystal to an incident neutron or DM particle is the {\emph{dynamic structure factor}} $S(\bfq,\omega)$. Here we provide only the final expressions for $S(\bfq,\omega)$; we summarize their derivation in Appendix~\ref{app:neutronscat}.  In particular, if the momentum transfer is below the size of the \brill\ zone, $S(\bfq,\omega)$ can be written as:
\begin{align}
\label{eq:neutronsqo}
	S(\bfq,\omega)= \frac{1}{2}\sum_{\nu} \frac{|F_\nu(\bfq) |^2}{\omega_{\nu,\bfq}}\delta (\omega_{\nu,\bfq}-\omega)
\end{align}
where the sum runs over the phonon modes ($\nu$).  The phonon form factor $F_\nu({\bf q})$ is given by
\begin{equation}\label{eq:phononformfac}
	F_\nu({\bf q})=\sum_j \frac{A_j}{\sqrt{m_j}}e^{-W_j({\bf q})}\bfq\cdot\mathbf{e}_{\nu,j,\bfq}
\end{equation}
where the sum runs over the atoms $j$ in the primitive cell, and $A_j$ is the atomic mass number. Here we have used that for our benchmark model, the DM-nucleus scattering length is given by $\overline b_j = A_j \overline b_X$, since we have a coherent sum over all nucleons in the long-wavelength limit.  The Debye-Waller function, $W_j$, measures the average motions of the atoms\footnote{Here we use `atom' and `'nucleus' interchangeably to refer to the scattering center.} in a phonon excitation, and is given by 
\begin{align}
W_j (\bfq) =\frac{1}{4 \Ncells}\sum_{\nu,\bfk}\frac{1}{ m_j \omega_{\nu,\bfk}}|\bfq\cdot\mathbf{e}_{\nu,j,\bfk}|^2
\end{align}
Note the quantity is finite, since the $1/N$ factor is compensated by the sum over all phonon modes $\bfk$. For all our results, it is a good approximation to take $W_j\approx 0$, as the spread on the motions of the atoms is small compared to the inverse momentum transfer. Taking $m_j > 16 $ GeV since the lightest nucleus is O  and $\omega_{\nu, \bfk} > \textrm{meV}$ for the most optimistic experimental threshold, we still find $\sqrt{m_j \omega_{\nu, \bfk}} = 4$ keV, which is larger than the typical momentum transfer for scattering of sub-MeV DM. 

As derived in Appendix~\ref{app:neutronscat}, the integrated scattering rate per unit of target mass is given in terms of the dynamical structure factor,
\begin{equation}\label{eq:appneutronrate}
	R =\frac{\rho_X}{m_X} \frac{\bar b_X^2}{\rho_T\unitcell m_X^2}\int\!d^3\bfv f(\bfv)\int\!d^3\bfq\, \left(\frac{q_0}{|\bfq|}\right)^4S(\bfq,\omega) ,
\end{equation}
where $\rho_T$ is the mass density of the target and $\unitcell$ is the volume of the primitive unit cell. The $\left(q_0/|\bfq|\right)^4$ form factor is the result of the light mediator. The expressions for the massive mediator limit can be obtained by dropping this form factor and substituting $q_0^4$ with $m_\phi^4$ in equation \eqref{eq:neucleoneffectivecrosssec}.

\subsection{Reach}

Contrary to the case with a dark photon mediator, all atoms in the primitive cell contribute with the same sign to the form factor in \eqref{eq:phononformfac}. The modes which couple most strongly to the dark matter are those where all atoms move in the same direction, and thus interfere constructively in \eqref{eq:phononformfac}. In addition, the $\bfq\cdot\mathbf{e}_{\nu,j,\bfq}$ factor indicates that only the longitudinal modes with  motion of the atoms parallel to the momentum $\bfq$ contribute, to leading order in the small $\bfq$ expansion (see Appendix~\ref{app:neutronscat}). Thus, the DM coupling to the longitudinal acoustic mode will be the largest. The optical modes also contribute, but since at least some atoms move in opposite directions, there are inevitably cancellations (destructive interference) between the contributions of various atoms. These effects can be seen most easily for GaAs, where the form factor can be approximated by
\begin{equation}
	F_\nu({\bf q})\approx \frac{|\bfq| }{\sqrt{m_p}}\left( \sqrt{A_{\text{Ga}}}e^{-i\bfq\cdot \bfr_{\text{Ga}}} \pm \sqrt{A_{\text{As}}}e^{-i\bfq\cdot \bfr_{\text{As}}}\right),
	\label{eq:formfactor_approx}
\end{equation}
where the $+$ sign applies for the LA mode and the $-$ sign for the LO mode. We have included relative phases for the motion of the Ga and As, to account for the fact that the motion will not be perfectly in phase away from the long-wavelength limit (see also \eqref{eq:dynmatrix}, where the phases appear explicitly in the dynamical matrix). 

Since Ga and As have similar mass numbers, we see from the equation above that there is  destructive interference for the optical mode, which leads to a suppression of the rate by several orders of magnitude compared to the acoustic mode. 
For sapphire, the mass hierarchy between the two elements is slightly larger, but since there six O atoms as compared to four Al atoms in the primitive cell, both elements end up contributing a similar amount to the scattering rate. To fully remove the suppression due to the destructive interference, it would be interesting to consider a polar material with a large mass difference between the elements, such as PbS.

\begin{figure}[t]
\includegraphics[width=0.49\textwidth]{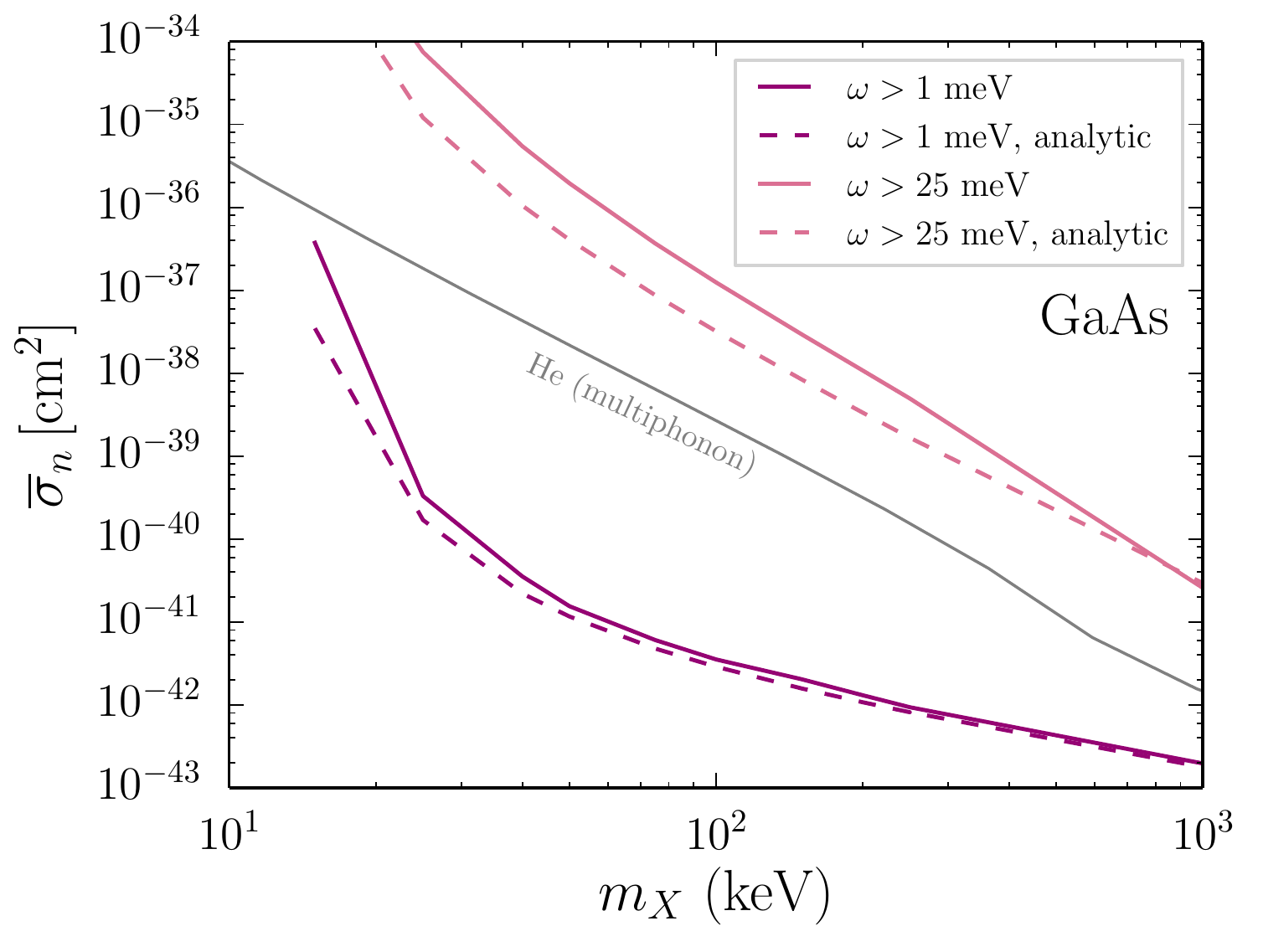}\hfill
\includegraphics[width=0.49\textwidth]{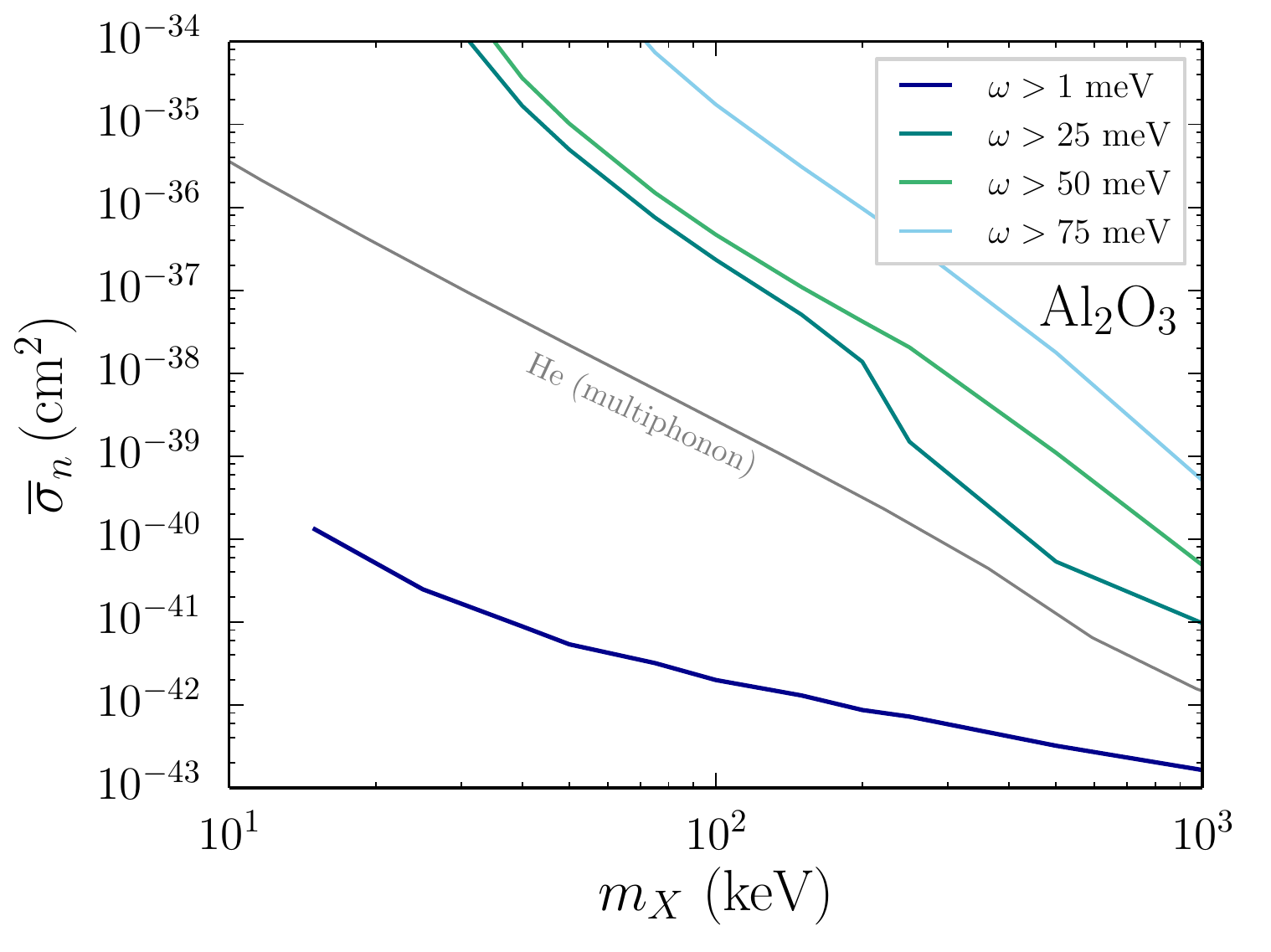}
\caption{The projected reach for GaAs (left) and $\text{Al}_2\text{O}_3$ (right) for a kg-year exposure and different experimental thresholds. The solid lines show the reach using the numerically computed phonon modes, while the dashed lines use the analytic approximation in \eqref{eq:formfactor_approx}. Also shown is a projection for a superfluid helium target that is sensitive to multiphonon production from DM, with kg-year exposure and meV threshold~\cite{Knapen:2016cue}.
\label{fig:neutronreach}}
\end{figure}

Here we use the numerically computed phonon eigenmodes to calculate the scattering rate, while we previously applied the analytic approximation of \eqref{eq:formfactor_approx} to GaAs in Ref.~\cite{Knapen:2017ekk}. As before, we estimate the reach by computing the projected 90\% CL limit under the assumption of no backgrounds and no events observed. The result for both GaAs and sapphire is shown in Figure~\ref{fig:neutronreach} for a kg-year exposure. The analytic approximation for GaAs matches the numerical result very well for the acoustic branch (dark purple), and for the optical branch (light purple) it reproduces the numerical result to within a factor of $\sim$ 3. As expected, the reach dramatically improves if the threshold is low enough to pick up the acoustic modes, and in this case substantially outperforms a superfluid helium detector in multiphonon mode \cite{Knapen:2016cue}.

If only the optical modes are accessible, the reach is comparable or somewhat weaker than that of superfluid helium. In this case only one LO mode contributes for GaAs, and it is imperative that the threshold is lower than 30 meV. For sapphire, there are several modes in the spectrum which contribute comparably to the total rate. The cross section and the reach therefore differ for different experimental thresholds in Fig.~\ref{fig:neutronreach}, as more phonon modes can be accessed for lower thresholds.  This is to be contrasted with the case of the dark photon mediator, where mode 30 alone was responsible for almost all of the rate, provided that it is kinematically accessible. The threshold dependence of the rate is thus not present for the dark photon mediator, and could be a discriminating variable between the models, should a signal be observed. As for GaAs, the sapphire reach would increase substantially if the acoustic phonons could be accessed.  In particular, the improved reach for the 25 meV threshold and $m_X > 200$ keV in sapphire is due to one of the acoustic modes: at this point, the momentum transfer becomes just large enough to access a portion of the acoustic branches (see Fig.~\ref{fig:GaAsphononbands}). This substantially enhances the rate, giving rise to the feature in Fig.~\ref{fig:neutronreach}.

%%%%%%%%%%%%%%%%%%%%%%%%%%%%%%%%%%%
\subsection{Daily modulation}
%%%%%%%%%%%%%%%%%%%%%%%%%%%%%%%%%%%

\begin{figure}[t]
\includegraphics[width=0.45\textwidth]{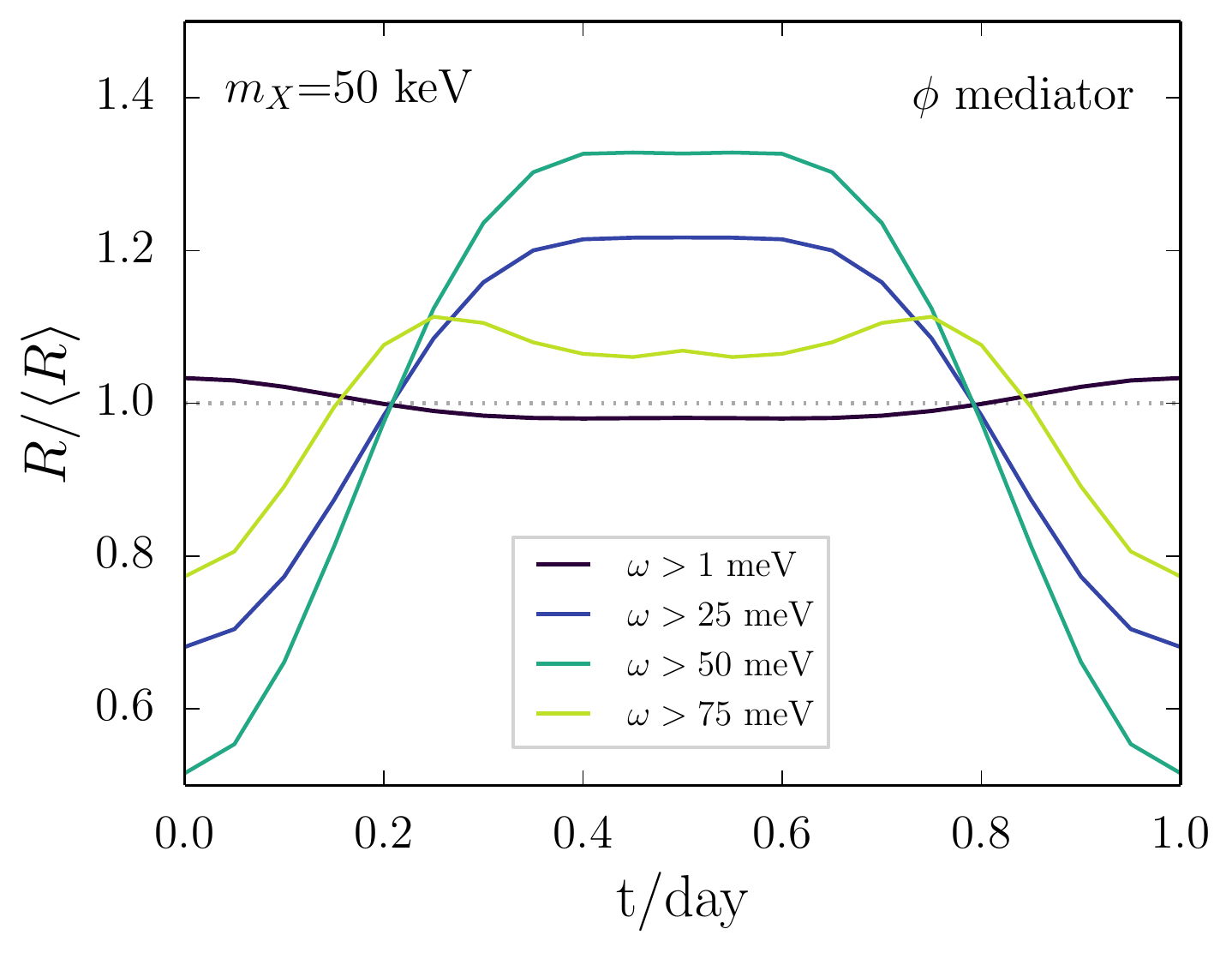}\hfill
\includegraphics[width=0.45\textwidth]{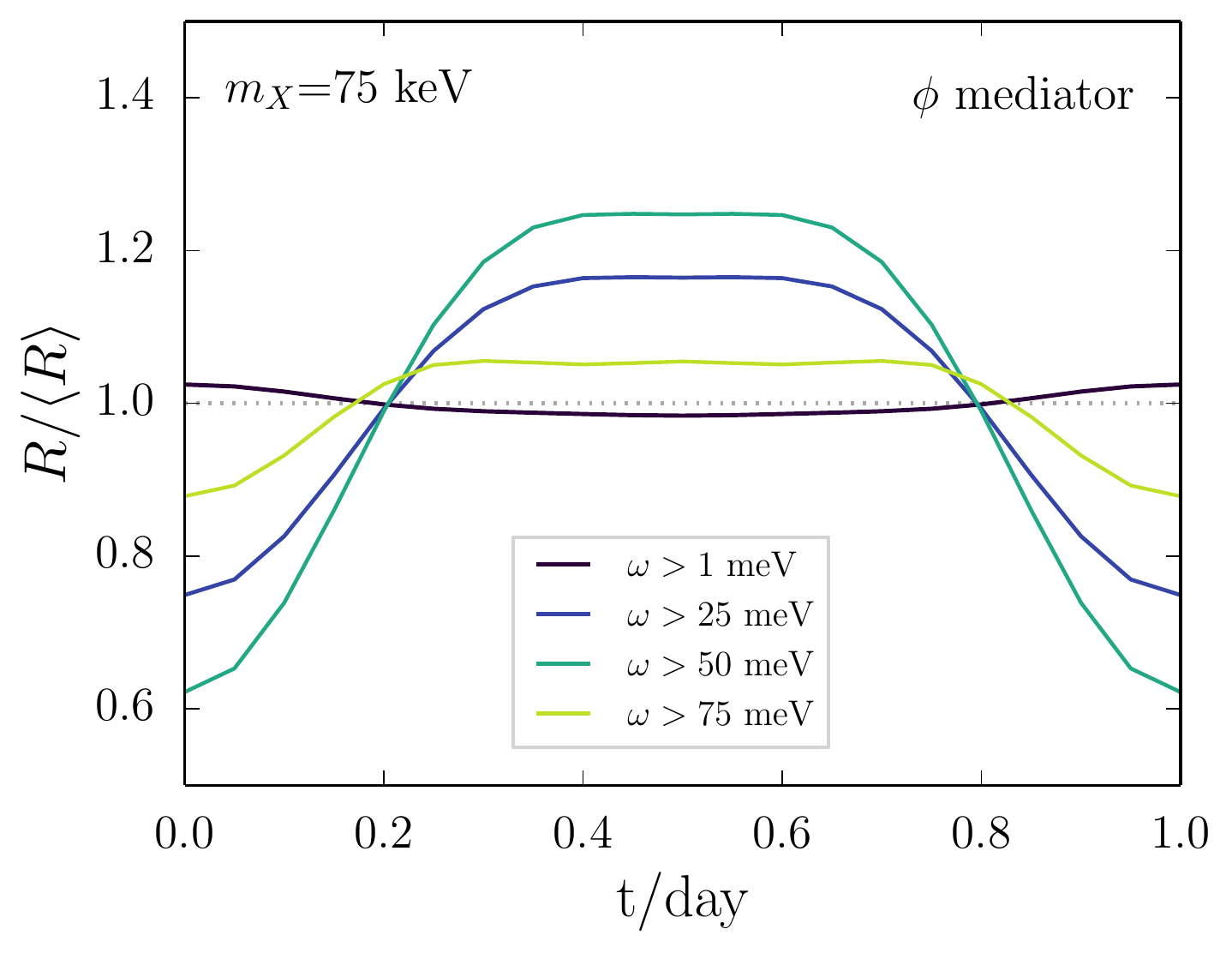}
\includegraphics[width=0.45\textwidth]{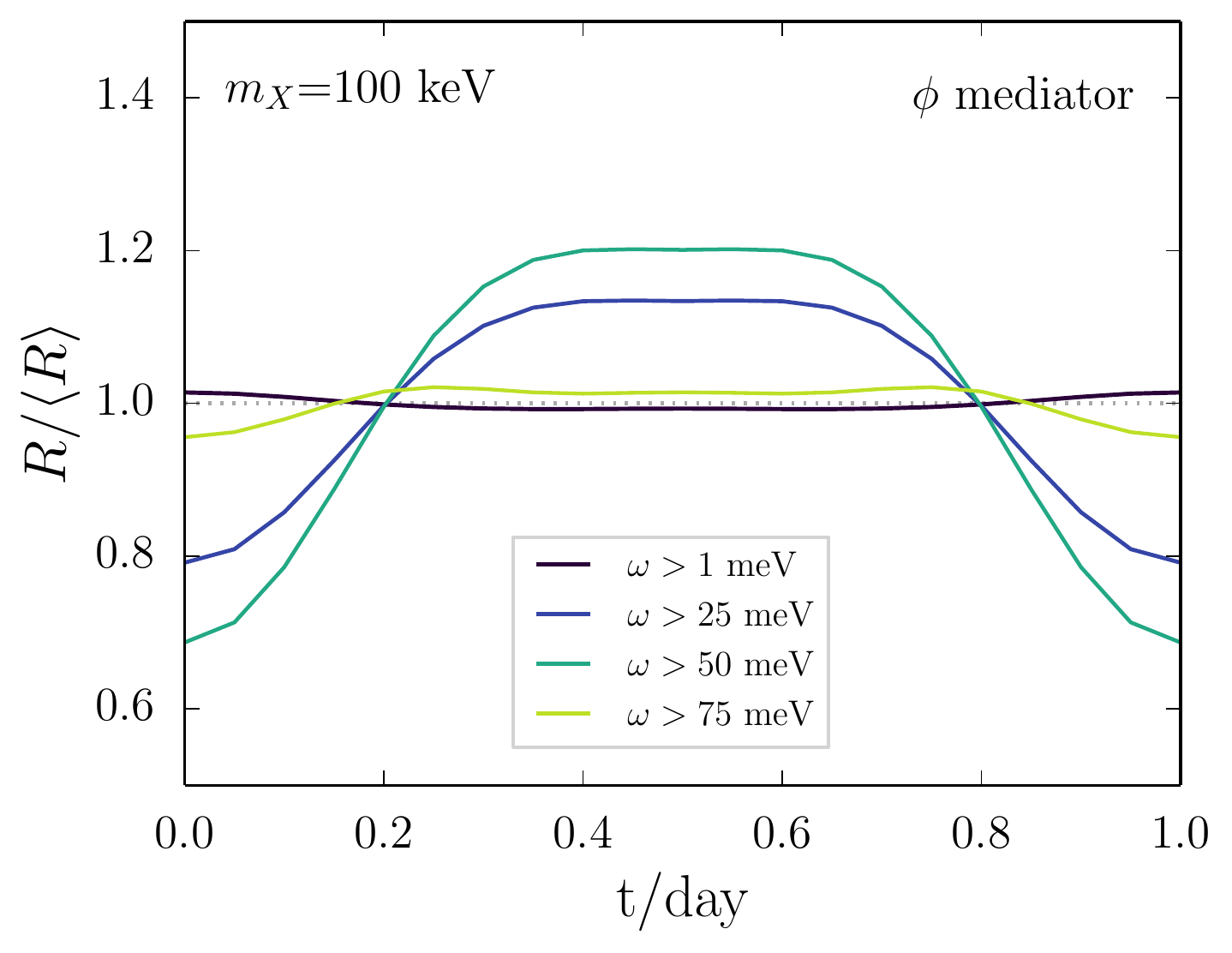}\hfill
\includegraphics[width=0.45\textwidth]{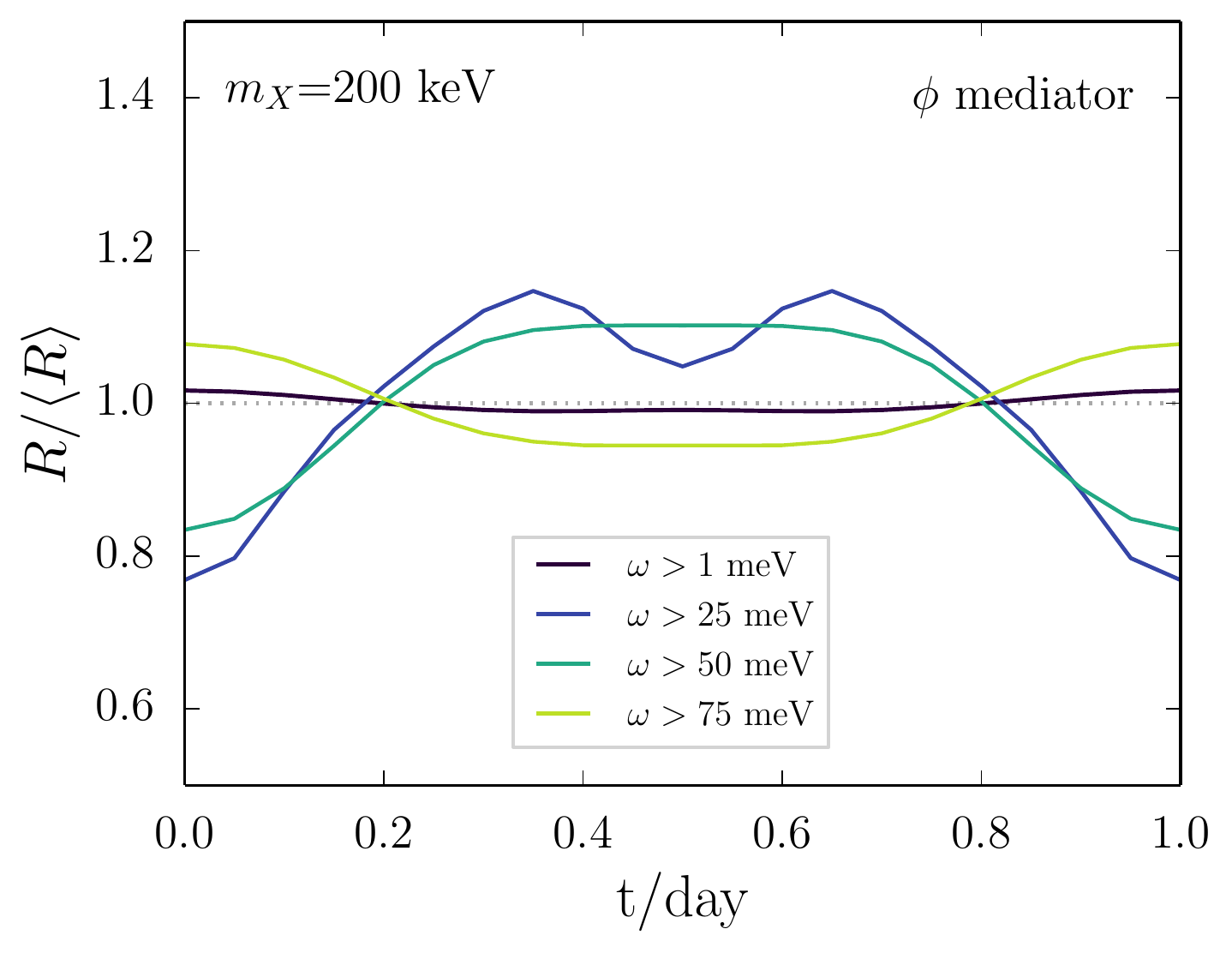}
\caption{Modulation of the scalar-mediated scattering rate in sapphire over a sidereal day, for different DM masses and experimental thresholds.    \label{fig:neutrondirection}}
\end{figure}

Similar to the case of dark photon mediated scattering, the rate modulates with sidereal day due to anisotropies in the phonon spectrum and the phonon form factor. Here the directional dependence of the form factor is encoded in the eigenvectors $\mathbf{e}_{\nu,j,\bfk}$ in Eq.~\eqref{eq:phononformfac}. The modulation rates for different DM masses and possible experimental thresholds are shown in Fig.~\ref{fig:neutrondirection} for sapphire; similar to before, we find much smaller modulation rates for GaAs, with sub-percent modulation except for $m_X \lesssim 30$ keV. As for the dark photon mediated scattering, the modulation decreases for larger DM masses, as the $\mathbf{e}_{\nu,j,\bfk}$ tend to be more randomized for higher $\bfq$. However, the modulation amplitude drops more slowly compared to dark photon mediated scattering, and even for $m_X \approx $~200 keV the modulation can still be as large as \mbox{$\sim$ 20\%.}

To understand the dependence of the modulation on the threshold, we first observe the lack of a substantial modulation for the lowest (1 meV) threshold. The reason is that the acoustic modes dominate in this case. Since all atoms move in phase on the acoustic branches, the primary modulation comes from the anisotropy of the sound speed, which is fairly small. For a higher threshold (>25 meV), we instead rely primarily on the optical modes. As explained in the previous section, in the $\bfq \to 0$ limit, the contributions from the different atoms tend to destructively interfere for the optical branches. The effect of finite $\bfq$ corrections is then to partially remove these cancellations; this effect will vary along different crystal directions, leading to a sizable directional dependence of the scattering rate. (One way of seeing this is to consider the effect of the phase factors in \eqref{eq:formfactor_approx}.) 
 As the threshold is further increased, fewer optical modes can contribute to the rate. Since each mode has a unique modulation pattern, this means that the total modulation pattern depends on the threshold.  In addition, different DM masses sample different regions in the \brill\ zone, which means that the relative weight of the phonon modes shifts as the DM mass is varied. This too has an effect on modulation pattern, as can be seen most clearly by comparing the curves for $m_X=50$ keV and $m_X=200$ keV benchmarks in Fig.~\ref{fig:neutrondirection}. Both features may help with characterizing the DM mass, should a signal be observed.

\FloatBarrier

%%%%%%%%%%%%%%%%%%%%%%%%%%%%%%
\section{Absorption of dark photons \label{sec:absorption}}
%%%%%%%%%%%%%%%%%%%%%%%%%%%%%%

The presence of optical phonons in polar materials also makes it an excellent target for absorption of dark photon DM. In the sub-keV regime, dark photons are a viable DM candidate, and can be detected by an optical absorption signal if there is a small mixing with the SM photon. Similar to Section~\ref{sec:darkphoton}, we consider the Lagrangian
\begin{align}
	\mathcal{L} \supset -\frac{1}{4}F^{\mu\nu}F_{\mu\nu} + J_{\rm EM}^\mu A_\mu -\frac{1}{4}F'^{\mu\nu}F'_{\mu\nu}  -\frac{\kappa}{2} F^{\mu\nu}F'_{\mu\nu} -\frac{m_{A'}^2}{2} A'^\mu A'_\mu
	\label{eq:darkphotonabs}
\end{align}
with kinetic mixing $\kappa$ and Stuckelberg mass $m_{A'}$.  Polar materials are sensitive to dark photons in the mass range of $\sim$ meV up to a few hundred meV, due to the wide range of phonons coupling to EM fields and the possibility of multiphonon absorption. Electronic excitations also allow sensitivity to DM with eV or greater mass, although a number of existing experiments are already making progress in this regime.

We begin with a review of the absorption of dark photons in optically isotropic materials, such as GaAs. The mixing present in Eq.~\eqref{eq:darkphotonabs} is modified in the presence of an in-medium polarization for the photon, which can be written as~\cite{An:2014twa,Hochberg:2015fth}
\begin{align}
	\Pi_{\gamma \gamma}({\bf q}, \omega) = \omega^2 (1 - \hat n^2).	
	\label{eq:inmediumpol}
\end{align}
Note that the above result holds for both longitudinal and transverse polarizations, and we have taken the limit of $|{\bf q}| \to 0$, appropriate for absorption processes, such that we can write $\Pi_{\gamma \gamma}({\bf q}, \omega) \equiv \Pi(\omega)$.  $\hat n = n + ik $ is the frequency-dependent complex index of refraction, and is related to the permittivity $\hat \epsilon $ and to the optical conductivity $\hat \sigma$ of the material:
\begin{align}
	\hat \epsilon  = \hat n^2 = 1 + \frac{ i \hat \sigma}{\omega}.
\end{align}
Note that the real part of $\hat \sigma$, $\sigma_1$, appears in the imaginary part of the polarization tensor $\Pi(\omega)$. It can thus be seen that $\sigma_1$ is the absorption rate of SM photons. 
For energies near the LO and TO phonon frequencies, the permittivity of a polar material can be described analytically as~\cite{0022-3719-7-13-017}
\begin{align}
 	\hat \epsilon(\omega) = \epsilon_\infty \prod_\nu \frac{\omega_{{\rm LO},\nu}^2 - \omega^2 + i \omega \gamma_{{\rm LO},\nu}}{\omega^2_{{\rm TO},\nu} -\omega^2 +  i \omega \gamma_{{\rm TO},\nu}},
	\label{eq:permittivity}
\end{align}
where we have included a product over all optical branches $\nu$. Each branch is split into longitudinal and transverse modes with energies  $\omega_{\rm TO, LO}$, while  $\gamma_{\rm TO, LO}$ are the damping parameters. $\epsilon_\infty$ is the contribution of the electrons for energies below the electronic band gap. It is at the LO phonon frequencies where $\epsilon(\omega)$ becomes suppressed. For GaAs, there is one active branch and data on the parameters at low temperatures can be found in Ref.~\cite{PSSB:PSSB2221950110}. However, note that the permittivity above does not include multiphonon absorption, and where possible we will supplement the above result with the measured index of refraction.

\begin{figure}
\includegraphics[width=0.49\textwidth]{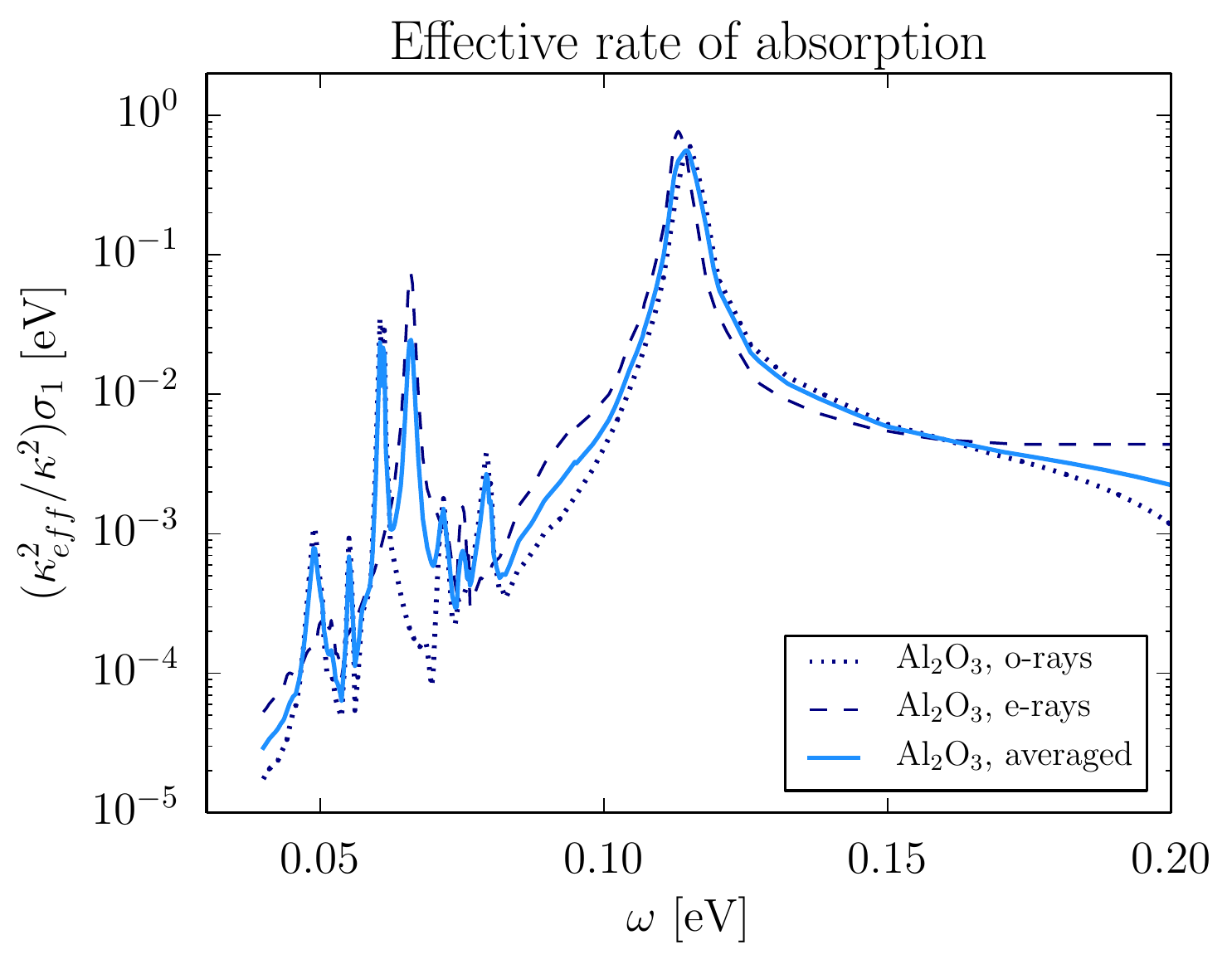}
\includegraphics[width=0.49\textwidth]{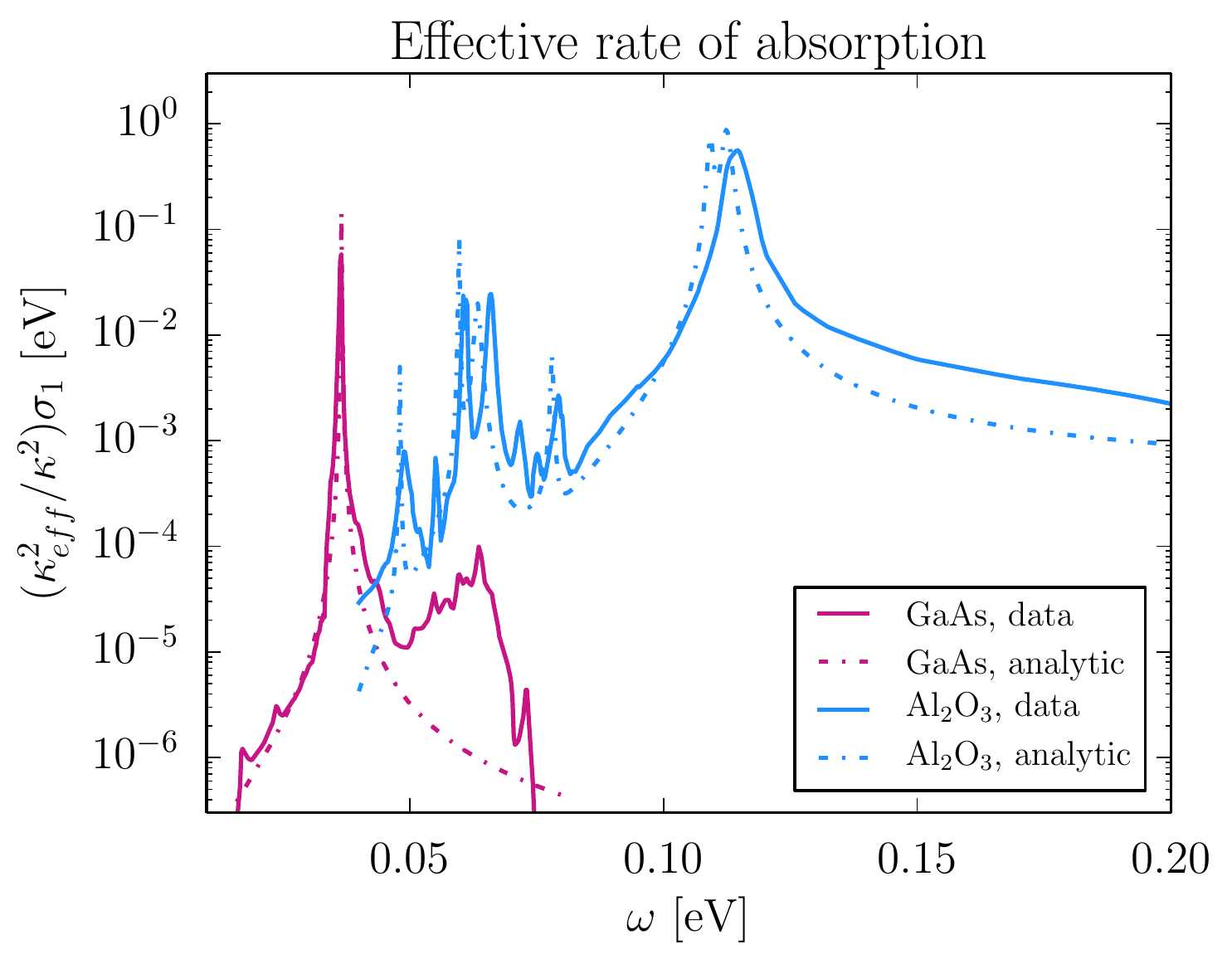}
\caption{ (Left) Effective absorption rate of dark photons into  phonons for sapphire. We show both the absorption into ordinary and extraordinary rays, as well as the weighted average that we expect for a dark photon field. (Right) Comparison of the effective absorption rate obtained from measurements of the optical properties, and that obtained using the analytic approximation in Eq.~\eqref{eq:permittivity}, with best fit parameters quoted in Tab.~\ref{tab:absparams}. \label{fig:absorption_birefringent} }
\end{figure}

Including the in-medium polarization from Eq.~\eqref{eq:inmediumpol} in the Lagrangian and diagonalizing, we obtain a coupling of the dark photon with the EM current given by $ \kappa_{\rm eff} J_{\rm EM}$, where  the effective in-medium kinetic mixing parameter is
\begin{align}
	\kappa_{\rm eff}^2 = \frac{\kappa^2 m_{A'}^4}{\left[m_{A'}^2 - \mbox{Re}~\Pi(\omega) \right]^2 + \mbox{Im}~\Pi(\omega)^2} = \frac{\kappa^2}{|\hat \epsilon(\omega) |^2}
\end{align}
where we took $\omega=m_{A'}$ in the second step.
The dark photon absorption rate per unit target mass is then determined in terms of the photon absorption, and can be written as 
\begin{align}
	R = \frac{1}{\rho_T} \frac{\rho_{\rm DM} }{m_{A'}}  \kappa_{\rm eff}^2 \sigma_{\rm 1} 
\end{align}
where $\rho_T$ is the target density. In Ref.~\cite{Knapen:2017ekk}, we applied the above result to GaAs.  The phonon absorption is temperature dependent, so we have selected low-temperature results whenever available.  For the absorption into phonons ($m_{A'} < $eV), we used calculations of the zero-temperature  absorption coefficient $\alpha$ into single and multiple phonons from Ref.~\cite{PhysRevB.70.245209}, where $\alpha =  \sigma_1/n$,  and we use Eq.~\eqref{eq:permittivity} to determine $n$. It can be seen in the right panel of Fig.~\ref{fig:absorption_birefringent} that using only Eq.~\eqref{eq:permittivity} misses a large portion of the absorption, due to multiphonons.   Fig.~\ref{fig:absorption_birefringent}  also shows that the peak of the absorption is actually at $\omega_{\rm LO}$, even though the photon absorption is peaked at $\omega_{\rm TO}$. This is due to the relatively suppressed $\kappa_{\rm eff}$ at $\omega_{\rm TO}$.  For eV and greater masses, we used room-temperature data on $\hat n$ from Ref.~\cite{opticalconstantsGaAs}. 

The absorption of dark photons in sapphire differs from that of GaAs because sapphire is a birefringent material, meaning that the complex index of refraction depends on the polarization of the vector field relative to the optical axis (the crystal axis or $c$-axis in sapphire). In the optical phonon regime, this should not be too surprising: as discussed in the previous sections, there is significant anisotropy in the phonon dipole moments and energies for modes parallel or perpendicular to the $c$-axis.  Data on the index of refraction is typically quoted separately for ordinary rays ($\vec E \perp c$-axis) and for extraordinary rays ($\vec E || c$-axis), with substantially different resonance structures for the two polarizations. Sapphire exhibits uniaxial birefringence, such that all polarizations perpendicular to the $c$-axis have the same index of refraction.

For dark photons as the DM, we expect the field to have a random polarization with respect to its $k$-vector and to the orientation of the $c$-axis. In particular, the coherence time for the dark photon field is $\sim 1/(m_{A'} v^2) \lesssim 1 \mu$s for the masses considered here and with $v \sim 10^{-3}$, and so the polarization will change randomly on a time scale much faster than the rotation of the crystal, for instance. As such, we simply take the average of the absorption rate for polarizations perpendicular and parallel to the $c$-axis,
\begin{align}
	R_{\sapphire} = \frac{1}{\rho_T} \frac{\rho_{\rm DM} }{m_{A'}} \left( \frac{1}{3} \kappa_{\rm eff, e}^2 \sigma_{\rm 1,e} + \frac{2}{3} \kappa_{\rm eff, o}^2 \sigma_{\rm 1,o} \right)
\end{align}
where the subscripts indicate the ordinary (o) and extraordinary (e) directions, respectively.

In the left panel of Fig.~\ref{fig:absorption_birefringent}, we show the effective absorption rate  $\kappa_{\rm eff}^2 \sigma_{\rm 1} /\kappa^2$ for both polarizations in sapphire, as well as the weighted average we use in computing the sensitivity. The data is obtained from Ref.~\cite{opticalconstantsSapphire}, which compiled measurements at room temperature. Similarly to GaAs, while the strongest absorption into photons is at the TO frequencies, we actually find strong dark photon absorption peaks at the LO frequencies due to the in-medium $\kappa_{\rm eff}$. In particular, we find the strongest absorption at the mode with $\omega_{\rm LO} \approx 110$ meV, which we identified earlier as having the largest dipole moment. 

 In the right panel of Fig.~\ref{fig:absorption_birefringent}, we compare the room temperature data with the result using  Eq.~\eqref{eq:permittivity} and best fit parameters measured at 77 Kelvin from Ref.~\cite{Gervais}. The parameters we used are reported in Tab.~\ref{tab:absparams}. It can be seen that the bulk of the absorption is described by the broad resonances in single optical phonon production, and there is good agreement in the two approximations. Ideally, one would obtain data at even lower temperatures, but we did not find any in the literature. We expect that reducing the temperature further would lead to reduced phonon widths by an $O(1)$ factor, and thus somewhat narrower peaks. Depending on the details of the eventual experimental setup, it may be possible to measure the low temperature absorption rate during a calibration run.

\begin{table}
\begin{tabular}{|c|c|c|c|}\hline
$\omega_{\rm LO}$&   $\gamma_{\rm LO}$  & $\omega_{\rm TO}$ & $\gamma_{\rm TO}$\\\hline
906.6 & 16 & 633.6 & 3.8 \\
629.5 & 4.4 & 569 & 3.2\\
481.7 & 1.4 & 439.1 & 1.5\\
387.6 & 1.4 & 385 & 1.4\\
\hline
\end{tabular}
\quad
\begin{tabular}{|c|c|c|c|}\hline
$\omega_{\rm LO}$&   $\gamma_{\rm LO}$  & $\omega_{\rm TO}$ & $\gamma_{\rm TO}$\\\hline
881.1 & 16* & 582.4 & 3.2* \\
510.9 & 1.4* & 397.5 & 1.5*\\
\hline
\end{tabular}
\caption{Values used in Eq.~\eqref{eq:permittivity} for the ordinary (left) and extraordinary (right) optical response of sapphire. Frequencies are from Ref.~\cite{PhysRevB.61.8187} with error bars of less than 0.5\%. Widths are from Ref.~\cite{Gervais} at 77K, and only reported for the o-ray case; for the e-rays, we adopt the same values as in the o-ray case for similar phonon frequency. All values are quoted in units of 1/cm.}
\label{tab:absparams}
\end{table}

\begin{figure}
\includegraphics[width=0.8\textwidth]{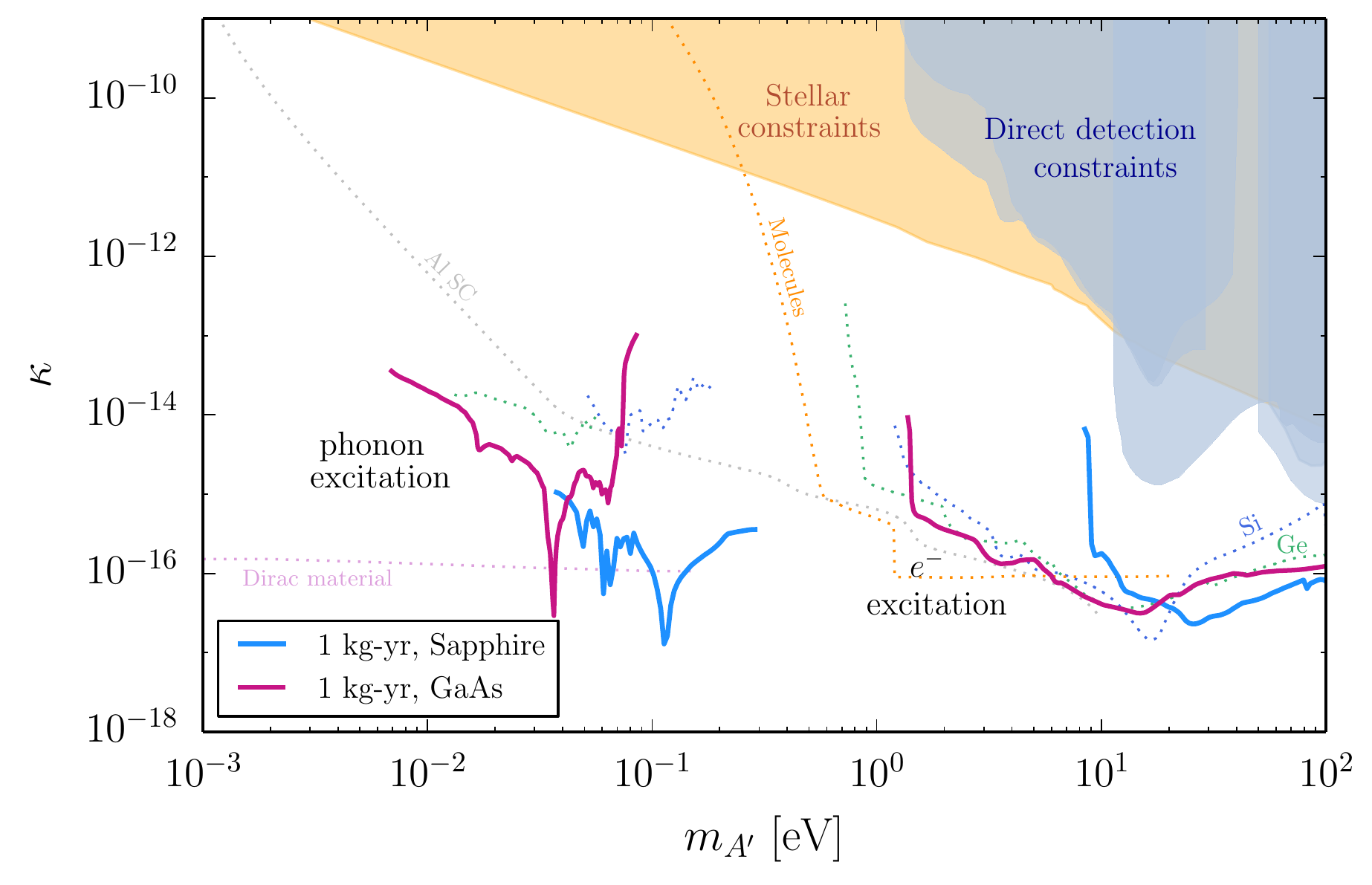}
\caption{ Reach of GaAs and sapphire to dark photon dark matter in terms of kinetic mixing $\kappa$ and mass $m_{A'}$, assuming kg-year exposure. For $m_{A'} < $ eV, the dark matter is absorbed into single and multi-phonon excitations. 
For $m_{A'} > $ eV, the dark matter is absorbed into electron excitations. Also shown are existing direct detection constraints from DAMIC~\cite{Aguilar-Arevalo:2016zop}, SuperCDMS~\cite{Agnese:2018col}, Xenon10~\cite{An:2014twa,Bloch:2016sjj}, and Xenon100~\cite{Hochberg:2016sqx,Bloch:2016sjj} (shaded blue) and constraints on emitting dark photons in the Sun~\cite{An:2013yfc,An:2013yua}. The dotted lines are projections from Al superconductors~\cite{Hochberg:2016ajh}, Ge and Si  semiconductors~\cite{Hochberg:2016sqx}, Dirac materials~\cite{Hochberg:2017wce} and molecules~\cite{Arvanitaki:2017nhi}. See Ref.~\cite{Bloch:2016sjj}  for absorption on GaAs for \mbox{$m_{A'} > $ eV}. Molecular magnets \cite{Bunting:2017net} have a reach in the $\kappa\sim 10^{-17} - 10^{-15}$ range for $10^{-2}\,\text{eV}\lesssim m_{A'}\lesssim 10$ eV.
 \label{fig:absorption} }
\end{figure}

Fig.~\ref{fig:absorption} shows the resulting sensitivity to dark photon DM, parameterized in terms of the vacuum kinetic mixing $\kappa$. We again assume kg-year exposure and zero background, and find that polar materials provide an excellent broadband target in the mass range of few meV up to 0.1 eV via the multiphonon signal.

%%%%%%%%%%%%%%%%%%%%%%%%%%%%%%%%%%%
\section{Conclusion}
\label{sec:conclusions}
%%%%%%%%%%%%%%%%%%%%%%%%%%%%%%%%%%%
Except for the simplest of crystals, most materials have gapped lattice vibrations (optical phonons) with energies between 10 meV and 100 meV. This matches the typical kinetic energy of DM in the Galaxy for masses between the $\sim 10$ keV warm DM limit and up to 1 MeV, allowing for single optical phonons to be excited in DM collisions with the crystal. We used Density Functional Theory (DFT) methods to compute the rate for DM to create an optical phonon in GaAs or sapphire in the zero temperature limit. Both crystals are examples of polar materials, where the optical phonon modes give rise to long range electric fields in the crystal. This implies a  coupling to any DM candidate that scatters through an ultralight dark photon mediator, which is a challenging scenario for other direct detection proposals targeting sub-MeV DM such as superconductors \cite{Hochberg:2015fth} or superfluid helium \cite{Knapen:2017ekk}. 

In previous work~\cite{Knapen:2017ekk}, we studied the example of GaAs with an analytic treatment. Here we go significantly beyond the earlier work in several ways: we validated the analytic treatment for GaAs using DFT methods, we extended the calculations to the more complex but potentially more promising example of sapphire, and we studied the directional dependence of the scattering rate in sapphire. In particular, sapphire has higher energy optical phonon modes that can be more readily accessed in an experiment, and the crystal anisotropy leads to a sizable directional dependence, which is manifest as a modulation in rate over a sidereal day. This directional dependence is much smaller in GaAs due to the more isotropic nature of the crystal. The dependence of the modulation pattern and amplitude on the target material suggests that if a signal were to be observed, one could employ a number of different polar material targets to extract details on the DM model and further confirm its cosmic origin.

We analyzed sub-MeV DM scattering via both a dark photon (vector) mediator and a scalar mediator. For the dark photon mediator, the scattering occurs dominantly into optical phonon modes, and the resulting reach and daily modulation are shown in Fig.~\ref{fig:darkphotonlimit} and Fig.~\ref{fig:darkphotondirection} respectively. In the case of the scalar mediator, the best sensitivity can be obtained if acoustic phonon modes are accessible; the reach and modulation are shown in Fig.~\ref{fig:neutronreach} and Fig.~\ref{fig:neutrondirection}. The modulation pattern and which phonons are excited thus depend strongly on the DM model, and a definitive observation of the modulation could in principle be used to infer the DM mass and mediator spin. For the scalar mediator, we studied the example where the mediator couples to nuclei, but our analysis also applies to the scenario where the scalar mediator couples to electrons. This is because the scattering into phonons is really a scattering off of the nucleus plus the inner-shell electrons rather than just the nucleus, so that one can estimate the rate by substituting the atomic mass numbers in Eq.~\eqref{eq:phononformfac} with the number of bound electrons in each atom. The results of this procedure are summarized in Appendix~\ref{app:electrons}. 

Polar materials are also sensitive to the scenario where the DM is a boson with mass below $\sim$~eV, where the DM could be absorbed into single or multi-phonon excitations. Fig.~\ref{fig:absorption} shows the reach for dark photon DM, for which the absorption rate can be related to the measured optical conductivity of the material. We expect that polar materials could also be sensitive to the absorption of scalar/pseudoscalar DM with a coupling to nucleons and/or electrons, but the absorption rate on optical phonons should be subject to the same destructive interference that we found for scalar-mediated scattering (Sec.~\ref{sec:nucleonint}). This implies that the multiphonon absorption could increase in importance, as compared to the case where a dark photon is absorbed. We reserve this computation for future work as it requires knowledge of anharmonic phonon interactions.  

In Tab.~\ref{tab:expsummary} we provide  a summary of the target materials that so far have been proposed for sub-MeV dark matter scattering and sub-eV dark matter absorption, and their sensitivity to different models. Single element semiconductors such as Ge and Si have a similar phonon spectrum as polar materials and therefore have sensitivity to the same models, with the exception of dark photon mediated scattering. The optical phonons in Ge and Si crystals do not give rise to long-range dipole fields, so dark photon mediated interactions cannot excite a single optical phonon in the long-wavelength limit (multiphonon excitations are still possible, and have been considered for dark photon DM absorption~\cite{Hochberg:2016sqx}). We conclude that polar materials can test a wide range of models for sub-MeV dark matter, with the added advantage of a directional dependence in the scattering rate for certain materials. That polar materials are readily available and well-understood crystals also makes them an exciting prospect for experimental realization.

\newcolumntype{C}[1]{>{\centering\let\newline\\\arraybackslash\hspace{0pt}}m{#1}}
\begin{table}
\resizebox{\textwidth}{!}{
{\footnotesize
\begin{tabular}{l|C{1.95cm}C{1.8cm}C{1.8cm}|C{1.8cm}C{1.8cm}C{1.8cm}|}\cline{2-7}
&\multicolumn{3}{c|}{Scattering}&\multicolumn{3}{c|}{Absorption}\\\cline{2-7}
& Dark photon& Electron & Nucleon & Dark photon & Electron & Nucleon \\\hline
\multicolumn{1}{|l|}{ Superconductor~\cite{Hochberg:2015pha,Hochberg:2015fth,Hochberg:2016ajh}}			&  			&\checkmark 	&  			&\checkmark 	&\checkmark	&\\
\multicolumn{1}{|l|}{Superfluid He \cite{Schutz:2016tid,Knapen:2016cue} }								   	&   			&(\checkmark) 	& \checkmark	& 			& 			&\\
\multicolumn{1}{|l|}{Dirac Materials \cite{Hochberg:2017wce} }											&\checkmark	&\checkmark 	& 			& \checkmark 	&\checkmark	&\\
\multicolumn{1}{|l|}{Polar Materials (this work)}														&\checkmark  	&\checkmark  &\checkmark 	&\checkmark  	&(\checkmark)	&(\checkmark)\\ \hline
\end{tabular}
}
}
\caption{Summary table of experimental proposals probing scattering (absorption) of sub-MeV (sub-eV) dark matter, and their sensitivity to various models. ``Electron'' and ``Nucleon'' refer to a scalar coupling to electrons and nuclei respectively. $(\checkmark)$ refers to cases where sensitivity is expected, but no  calculation has been performed at this time. In addition, molecular magnets \cite{Bunting:2017net} have been shown to have good reach to dark photon absorption, and may be sensitive to scattering and/or scalar absorption processes.  
 \label{tab:expsummary}}
\end{table}

\section*{Acknowledgments}

We thank Matt Pyle for collaboration on related work and for useful discussions. We also thank Florian Altvater, Jonah Haber, Rafael Lang, Mikhail Malkov, Jeffrey Neaton and Tom Melia for useful discussions.   SK, SG and KZ are supported by the DoE under contract DE-AC02-05CH11231, and SK is also supported in part by the National Science Foundation (NSF) under grants No.~PHY-1316783 and No.~PHY-1002399. This work was performed in part at the Aspen Center for Physics, which is supported by National Science Foundation grant PHY-1607611 and at the Kavli Institute for Theoretical Physics, supported in part by the National Science Foundation under Grant No.~NSF PHY-1748958. It also used resources of the National Energy Research Scientific Computing Center and the Molecular Foundry, which are supported by the Office of Science of the DoE under Contract No.~DE-AC02-05CH11231.

\appendix

\section{Phonon eigenmodes\label{app:phonons}}

Density functional theory (DFT) \cite{DFT, DFT_Kohm&Sham} is the workhorse of modern computational materials physics~\cite{Martinbook:2004}. It is an \textit{ab initio} method which requires only the location of the atoms in a crystal and a potential describing the ions to find solutions to the many-body Schrodinger equation. It is routinely applied to calculate a broad range of chemical and physical properties of materials ranging from electronic and phonon band structures to binding energies and magnetic properties. DFT's power lies in its versatility in addressing several areas in quantum materials while maintaining chemical and structural specificity that is not possible in tight-binding and other analytical methods. In its most basic form, DFT calculates the total energy of the system under consideration. From this, many related properties -- such as forces and response functions -- can be calculated by taking derivatives of the total energy and by perturbative methods. 

To calculate the phonon eigenmodes,  $\nu$, for a particular crystal, we require solutions to the eigenvalue equation:
\begin{equation}
\sum_{j'}\bfD_{\bfq,j,j'}\cdot\bfe_{\nu,j',\bfq}= \omega_{\nu,\bfq}^2 \bfe_{\nu,j,\bfq}
\end{equation}
with the dynamical matrix $\bfD_{\bfq,j,j'}$ given by
\begin{equation}
\bfD_{\bfq,j,j'}=\sum_{\bfl'}\frac{\mathbfcal{V}^{(2)}_{0,j,\bfl',j'}}{\sqrt{m_j m_{j'}}} e^{i\bfq\cdot(\bfr^0_{j'}+\bfl'-\bfr^0_{j})}
\end{equation}
and $\mathbfcal{V}^{(2)}_{\bfl,j,\bfl',j'}$ are the force constants to be calculated, as shown in Section III.

In this work we use the frozen-phonon method to calculate the force constants and the corresponding dynamical matrix. This method displaces each atom in the unit cell and calculates the resulting forces on the other atoms using DFT. From a combination of symmetry-inequivalent displacements, the full force-constant matrix can be built up using DFT calculations. A post-processing software package,  \textsf{phonopy} \cite{phonopy}, is then used to solve the eigenvalue problem for  $\omega_{\nu,\bfq}$ and $\bfe_{\nu,j,\bfq}$.

\subsection{Computational Details for DFT}

Our density functional theory calculations were performed with the projector augmented-wave (PAW) method~\cite{Bloechl:1994} as implemented in the VASP code~\cite{VASP1}. All calculations were performed using the Perdew-Becke-Ernzerhof (PBE) parametrization of the generalized gradient approximation (GGA)~\cite{PBE1}. The wavefunctions were expanded using plane waves with an energy cutoff of 600 eV, and used a Monkhorst-Pack~\cite{Monkhorst_Pack} $k$-point sampling mesh of 12x12x4 for 10-atom calculations, 6x6x4 for 30-atom calculations and 4x4x4 for 90-atom (tripled unit cell) calculations. We performed a full relaxation of the lattice constants and internal coordinates of the structure until the forces were converged to 0.01eV/\AA. The phonon calculations and modulations of the phonon modes were performed using the frozen-phonon method as implemented in the  \textsf{phonopy} \cite{phonopy} software. 

\subsection{Crystallographic properties of GaAs and Al$_{2}$O$_{3}$ \label{app:crystallographic}}

GaAs and Al$_{2}$O$_{3}$ adopt the zincblende (space group \textit{F-43m}) and sapphire (space group \textit{R-3c}) structures respectively, with the conventional unit cells shown. The cubic lattice of GaAs is equivalent in all three crystallographic directions, with all Ga and As atoms in the cell being equivalent. The primitive unit cell in this case is made up of two atoms -- one Ga and one As. However, sapphire's rhombohedral unit cell has inequivalent in-plane and out-of-plane crystal axes. The primitive unit cell of Al$_{2}$O$_{3}$ has two copies of five atoms -- two Al and three O. These differing Al and O occupy inequivalent symmetry positions in the unit cell and thus have different surrounding chemical environments. Owing to this, the Born effective charges for each of these five atoms can differ since they will have different responses to external perturbations. The calculated Born effective charges for Al$_{2}$O$_{3}$ for the inequivalent atoms are
\begin{align}
\begin{array}{ll}
\mathbf{Z^{*}}_{\text{Al(1)}}=\left(\!\begin{array}{ccc}2.98&0.034&\\-0.034& 2.98&\\ &&2.951 \end{array}\!\!\right)\qquad \qquad
&\mathbf{Z^{*}}_{\text{Al(2)}}=\left(\!\begin{array}{ccc}2.98&-0.034&\\0.034& 2.98&\\ &&2.951 \end{array}\!\!\right) \\
\mathbf{Z^{*}}_{\text{O(1)}}=\left(\!\begin{array}{ccc}-1.937&-0.086&0.23\\-0.086& -2.037&-0.133\\0.314&-0.181&-1.967 \end{array}\!\!\right)\qquad \qquad
&\mathbf{Z^{*}}_{\text{O(2)}}=\left(\!\begin{array}{ccc}-2.087&&\\& -1.887&0.266\\ &0.363&-1.967 \end{array}\!\!\right) \\ 
\mathbf{Z^{*}}_{\text{O(3)}}=\left(\!\begin{array}{ccc}-1.937&0.086&-0.23\\0.086& -2.037&-0.133\\ -0.314&-0.181&-1.967 \end{array}\!\!\right).
\end{array}
\label{eq:bornsaphinequiv}
\end{align}

\begin{figure}
\includegraphics[width=0.75\textwidth]{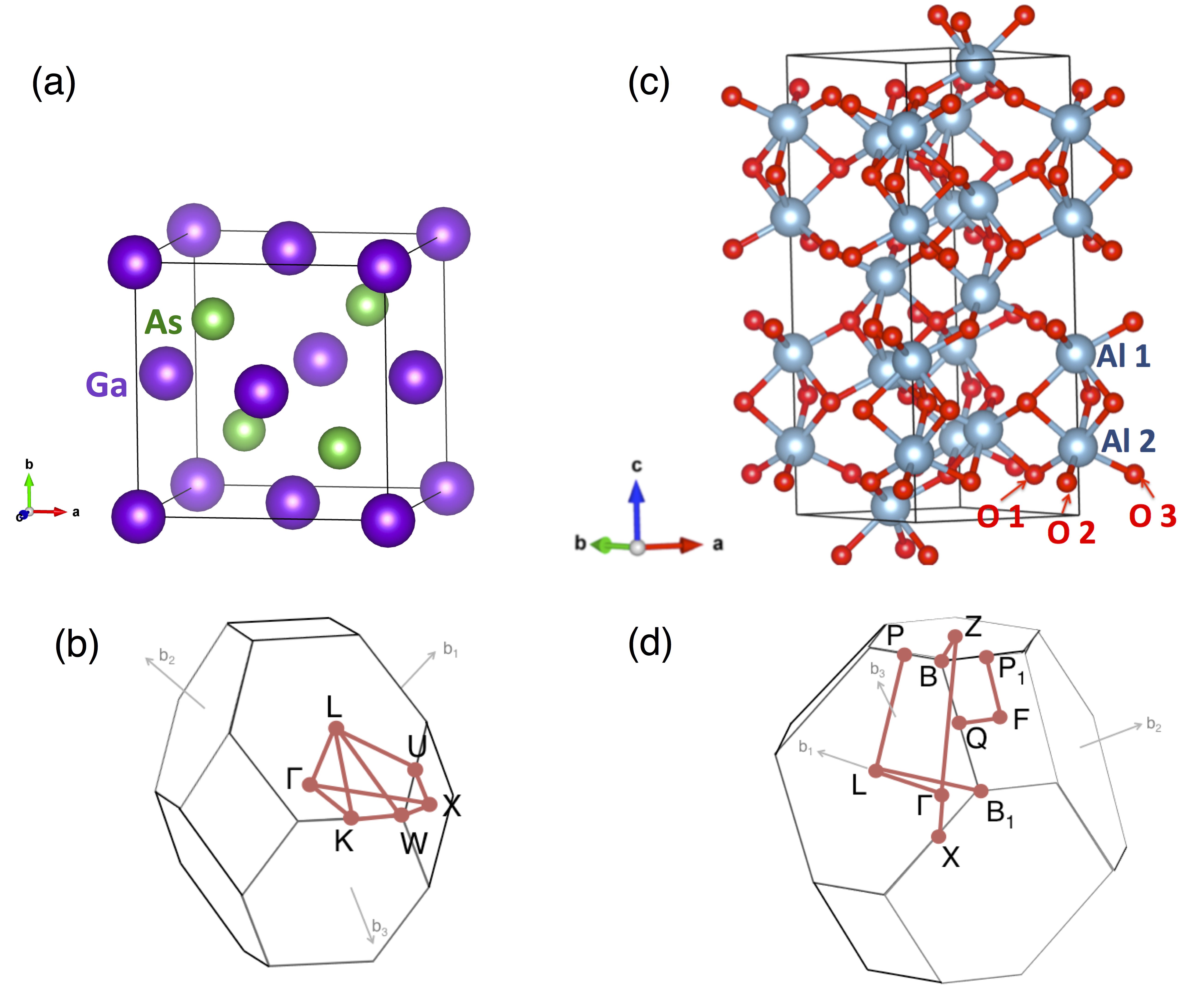}
\caption{\label{crystal_conventional}Conventional unit cells and the primitive first Brillouin zones for zincblende GaAs (a), (b) and sapphire Al$_{2}$O$_{3}$ (c), (d). For GaAs, the primitive unit cell comprises two atoms -- one Ga and one As. However, for sapphire, the primitive unit cell comprises two copies of five atoms -- two inequivalent Al, and three inequivalent O. These inequivalent Al and O atoms are labelled in (c). The Brillouin zones are labelled by the high-symmetry points as is given by convention \cite{Setyawan/Curtarolo:2010}.}\end{figure}

\section{Fr\"ohlich Hamiltonian\label{app:frolich}}

For the derivation of the Fr\"ohlich matrix element we largely follow the discussion in \cite{PhysRevLett.115.176401,YuCardona}. We start with a regular lattice of $\Ncells$ cells, with each cell containing a point charge $Q$ at location $\bftau$ relative to the center of the cell. The Poisson equation for the potential $\phi(\bfr,\bftau)$ for this configuration is:
\begin{equation}\label{eq:poisson}
\nabla\cdot \bfeps_\infty \cdot \nabla \phi(\bfr,\mathbf{\tau})=-\sum_{\bfl} \left[ Q \delta (\bfr-\bftau-\bfl) - \frac{Q}{ \unitcell} \right]. 
\end{equation}
with $\bfl$ the lattice vectors, such that $\phi(\bfr+\bfl,\bftau)=\phi(\bfr,\bftau)$. We have included also a background average charge in the unit cell volume $\unitcell$, to ensure that the system is neutral at long distances. $\bfeps_\infty$ is the high frequency dielectric matrix, which describes the (fast) response of the electrons in the presence of the charge displacement. The solution is 
\begin{equation}
\phi(\bfr,\mathbf{\tau})=\frac{Q}{\Ncells \unitcell} \sum_{\bfl} \sum_{\bfq} \sum_{\bfG\neq-\bfq} \frac{1}{(\bfq+\bfG)\cdot\bfeps_\infty\cdot(\bfq+\bfG)}e^{i(\bfq+\bfG)\cdot (-\bfr+\bftau+\bfl)}
\end{equation}
where $\Ncells\times \unitcell$ is the volume of the entire lattice. The $\bfq$ form a regular, $\Ncells$-point discretization of the first \brill\ zone, the $\bfG$ are  the reciprocal lattice vectors, and we have dropped any constant contributions. 

We now allow for a dipole $\mathbf{p}$ for every atomic displacement in the cell. For each cell, let us denote the equilibrium position for an ion relative to the origin of the primitive cell as $\bfr^0_j$, where $j$ labels atoms in the cell. The potential induced by the displacement of a single ion is defined as $ \phi_{\text{dip}}(\bfr)=\lim_{{\bfu} \to 0} \phi(\bfr,\bfr^0_j + \bfu_j )-\phi(\bfr, \bfr^0_j)$. 
Placing the dipoles in each cell of the lattice, we find 
\begin{equation}\label{eq:dipolepot}
\phi_{\text{dip}}(\bfr)=\frac{i}{\Ncells \unitcell}\sum_{\bfl} \sum_j \sum_{\bfq}\sum_{\bfG\neq-\bfq}\frac{\bfp_{\bfl,j}\cdot(\bfq+\bfG)}{(\bfq+\bfG)\cdot\bfeps_\infty\cdot(\bfq+\bfG)}e^{i(\bfq+\bfG)\cdot (-\bfr+ \bfr^0_j   + \bfl)}.
\end{equation}
where $\bfp_{\bfl,j}$ is the dipole moment for atom $j$ in the cell specified by the lattice vector $\bfl$. 
The dipole is given by the displacement of each ion from its equilibrium position in the cell, multiplied by its Born effective charge,
\begin{align}
\mathbf{p}_{\bfl,j}=&e \bfZ^\ast_j \cdot \bfu_{j,\bfl}(0)\\
=&e\sum_{\nu} \sum_{\bfG\neq-\bfq}\sum_{\bfq}\frac{1}{\sqrt{2 N m_j \omega_{\nu,\bfq}}}\left(\bfZ^\ast_j \cdot \mathbf{e}_{\nu,j,\bfq}\, \hat a_{\nu,\bfq}\, e^{i(\bfq+\bfG)\cdot(\bfl+\bfr^0_j)}+ \text{h.c.}  \right)
\end{align}
where we used the displacement operator in \eqref{eq:displacementquant}. 

To obtain the Hamiltonian for DM with effective charge $e'$, we multiply the potential in \eqref{eq:dipolepot} with $e'$. Using the completeness relation $\sum_{\bfl} e^{i(\bfq-\bfq')\cdot \bfl}=N\delta_{\bfq',\bfq}$, we find that for emission of a single phonon,
\begin{equation}
H=\frac{i e e'}{\unitcell}\sum_{j,\nu}\sum_{\bfG\neq-\bfq}\sum_\bfq\frac{1}{\sqrt{2 N m_j \omega_{\nu,\bfq}}}\frac{(\bfq+\bfG)\cdot\bfZ^\ast_j \cdot \bfe_{j,\nu,\bfq}^*}{(\bfq+\bfG)\cdot\bfeps_\infty\cdot(\bfq+\bfG)} \hat a_{\nu,\bfq}^\dagger  e^{-i(\bfq + \bfG) \cdot \bfr  }
+\text{h.c} 
\end{equation}
The incoming and outgoing DM states can be modeled by plane waves $ \bra \bfp_i -(\bfq + \bfG)  |  $ and $|\bfp_i\ket$, such that the transition matrix element $ \bra \bfp_i-\bfq - \bfG | H| \bfp_i  \ket$ is 
\begin{equation}\label{eq:frolichapp}
\mathcal{M}_{{\bfq + \bfG,\nu}}= \frac{i e e'}{\unitcell}\sum_{j}\frac{1}{\sqrt{2 N m_j \omega_{\nu,\bfq}}}\frac{(\bfq+\bfG)\cdot\bfZ^\ast_j \cdot \bfe^*_{j,\nu,\bfq}}{(\bfq+\bfG)\cdot\bfeps_\infty \cdot(\bfq+\bfG)}. 
\end{equation}
The expression for electronic transitions is identical, except that the appropriate in-medium wave functions must be used instead of plane waves. Since $\Ncells$ is formally infinite, the matrix element in \eqref{eq:frolichapp} appears to go to zero. However, in Fermi's golden rule the squared matrix element is always be evaluated as a sum over $\bfq$, which diverges as well for $N\rightarrow\infty$. We can thus go to continuum limit by taking
\begin{equation}
\sum_\bfq |\mathcal{M}_{{\bfq + \bfG,\nu}}|^2 \rightarrow  \Ncells \unitcell \int_{\text{BZ}}\!\frac{d^3\bfq}{(2\pi)^3} |\mathcal{M}_{{\bfq + \bfG,\nu}}|^2
\end{equation}
where the integral runs over the \brill\ zone. Since we work in the continuum limit for the calculations in Sec.~\ref{sec:darkphoton}, it is convenient to absorb the $\sqrt{\Ncells \unitcell}$ factor directly into the matrix element, which then becomes 
\begin{equation}\label{eq:frolichappcont}
\mathcal{M}_{{\bfq + \bfG,\nu}}= i e e'\sum_{j}\frac{1}{\sqrt{2 \unitcell m_j \omega_{\nu,\bfq}}}\frac{(\bfq+\bfG)\cdot\bfZ^\ast_j \cdot \bfe^*_{j,\nu,\bfq}}{(\bfq+\bfG)\cdot\bfeps_\infty \cdot(\bfq+\bfG)} 
\end{equation}
 which manifestly independent of the number of cells in the lattice. 
 
In the isotropic, long-wavelength limit we can drop the dependence on the reciprocal lattice vectors ${\bf G}$. Taking a 2-atom unit cell such as for GaAs, the expression reduces to 
\begin{equation}
	\label{eq:frolichisotempapp}
	\mathcal{M}^{\text{iso}}_{{\bfq}}\approx i\frac{ee'}{\epsilon_\infty}\frac{|Z^\ast|}{\sqrt{2 \unitcell \dipmu \omega_{LO}}}\frac{1}{|\bfq|}
\end{equation}
where $\omega_{LO}$ is the frequency of the optical phonon and $\dipmu\equiv(1/m_1+1/m_2)^{-1}$ is the reduced mass. Here we used that the eigenvectors are normalized within the unit cell (see condition above Eq.~\ref{eq:dynmatrix}), so $|{\bf e}_j| = 1/\sqrt{2}$ for a 2-atom unit cell, and that $\tfrac{1}{\sqrt{\dipmu}} \approx \tfrac{1}{\sqrt{2}}(\tfrac{1}{\sqrt{m_1}}+\tfrac{1}{\sqrt{m_2}})$, which is valid if $m_1+m_2\gg |m_1-m_2|$. With the identity
\begin{equation}\label{eq:born}
eZ^\ast=\Bigg[\unitcell\mu\left(\frac{1}{\epsilon_\infty}-\frac{1}{\epsilon_0}\right)\Bigg]^{1/2}\epsilon_{\infty}\omega_{LO}
\end{equation}
\eqref{eq:frolichisotempapp} then reproduces Eq.~\eqref{eq:frolichiso}.

It now only remains to derive \eqref{eq:born}. Following Ref.~\cite{YuCardona}, we consider a harmonic oscillator with reduced mass $\dipmu$, charge $Z^\ast$, and natural oscillation frequency $\omega_{TO}$ (this will be identified as the frequency of the TO modes, hence the notation). When the oscillator is driven by an electric field  with amplitude $\bfE_0$ and frequency $\omega$, the amplitude of the oscillations is given by
\begin{equation}\label{phononsol}
\mathbf{u}_0=\frac{eZ^\ast \mathbf{E}_0}{\dipmu(\omega_{TO}^2-\omega^2)}.
\end{equation}
The macroscopic polarization vector is $\bfP=eZ^* \bfu_0/\unitcell$, where the $1/\unitcell$ is merely the number density of the oscillators. The displacement vector of the system is then
\begin{equation}
 \bfD = \epsilon_\infty \bfE + \bfP= \epsilon \bfE
\end{equation}
with the frequency dependent dielectric function:
\begin{equation}\label{eq:dieelect}
\epsilon(\omega)= \epsilon_\infty + \frac{ e^2Z^{\ast2}}{\dipmu\unitcell(\omega_{TO}^2-\omega^2)}.
 \end{equation}
The $\epsilon_\infty$ term is again the contribution from the valence electrons, while the second term is the contribution from the oscillators. At high frequencies the ions are too slow to respond and only the electron contribution remains. Gauss' law demands that $\bfk \cdot \bfD=0$, which is trivially satisfied for the transverse modes. For the longitudinal mode $\bfk \parallel\bfE$, this requires that $\epsilon(\omega)=0$, which is satisfied at the frequency $\omega= \omega_{LO}$ with \begin{equation}\label{eq:omegaL}
 \omega_{LO}^2=\omega_{TO}^2+\frac{e^2Z^{\ast2}}{\epsilon_\infty\dipmu\unitcell}
 \end{equation}
 One may interpret the additional term as the self-energy correction to the LO mode from the back-reaction of its induced electric field. Combining \eqref{eq:dieelect} and \eqref{eq:omegaL} yields the Lyddane-Sachs-Teller relation
 \begin{equation}
\frac{\epsilon_0}{\epsilon_\infty}=\frac{\omega_{LO}^2}{\omega_{TO}^2},
\end{equation}
with $\epsilon_0\equiv \epsilon(0)$.
Combining this with \eqref{eq:omegaL} results in \eqref{eq:born}.

%%%%%%%%%%%%%%%%%%%%%%%%%%%%%%%%%%%
\section{Nucleon-scattering structure factor\label{app:neutronscat}}
%%%%%%%%%%%%%%%%%%%%%%%%%%%%%%%%%%%

In this Appendix, we present the derivation of the dynamic structure factor for DM scattering in a lattice at zero temperature. We follow closely the discussion presented in Ref.~\cite{Schober2014}, which reviews scattering of cold neutrons in a lattice.
 To compute the structure factor for hard sphere scattering, we treat the crystal as a regular, periodic lattice with $\Ncells$ cells and $\Nunit$ atoms in a unit cell, for a total of $\Ncells \times \Nunit$ atoms in the lattice. Summing the potential of the individual scattering centers gives the total potential 
\begin{equation}\label{eq:neutronpotential}
\mathcal{V}(\bfr )=\sum_{J=1}^{\Nlattice} \mathcal{V}_J(\bfr _J-\bfr )=\frac{2\pi b_X}{m_X}\sum_{J=1}^{\Nlattice} A_J\delta(\bfr _J-\bfr )
\end{equation}
where $J$ sums over all the atoms in the lattice, $b_X$ is the DM-nucleon scattering length, and $A_J$ is the mass number of the nucleus $J$. 
In Fourier space, the potential is
\begin{equation}
\mathcal{V}(\bfq )= \frac{2\pi b_X }{m_X}\sum_J^{\Nlattice} A_J e^{i \bfq \cdot \bfr_J}.
\end{equation}
We then define the structure function by 
\begin{equation}\label{skodef}
S(\bfq,\omega)\equiv \frac{1}{\Ncells} \sum_{\lambda_i,\lambda_f} p(\lambda_i)\left|\sum_J^{\Nlattice} A_J\bra \lambda_f |e^{i \bfq\cdot \bfr _J}|\lambda_i \ket\right|^2 \delta (E_{\lambda_f}-E_{\lambda_i} -\omega)
\end{equation}
with $\lambda_{i,f}$ the initial and final states, and $p(\lambda_i)$ is the thermal distribution over the initial states. Since we envision a very cold target, we only consider the ground state in the sum of the initial states, setting $\lambda_i=\lambda_0$. We have normalized $S(\bfq,\omega)$ such that it is an intrinsic quantity under $\Ncells\rightarrow\infty$. 
With this definition, the rate from Fermi's golden rule is
\begin{equation}\label{eq:appneutronrate}
\Gamma =(2\pi)\left(\frac{2\pi b_X}{m_X}\right)^2 \frac{1}{\unitcell}\int_{\text{BZ}}\!\frac{d^3\bfq}{(2\pi)^3}\, S(\bfq,\omega)
\end{equation}
where we treated the incoming and outgoing DM particle as plane waves. The integral is over the \brill\ zone and $\unitcell$ is the volume of the primitive unit cell. 

To compute the structure function, first we note that the squared matrix element in \eqref{skodef} can be rewritten as a single correlation function as follows:
\begin{align}
\Bigg|\sum_J^{\Nlattice} A_J\bra &\lambda_f |e^{i \bfq\cdot \bfr_J}|\lambda_0 \ket\Bigg|^2 \delta (E_{\lambda_f}-E_{\lambda_0} -\omega)\\
&=\sum_{J,J'} A_J A_{J'}
\bra \lambda_f | e^{i\bfq\cdot\bfr _J}|\lambda_0\ket\bra \lambda_0 | e^{-i\bfq\cdot\bfr _{J'}}|\lambda_f\ket \delta (E_{\lambda_f}-E_{\lambda_0} -\omega)\\
&=\frac{1}{2\pi}\sum_{J,J'} A_J A_{J'}\int_{-\infty}^{+\infty}\!\!\!\!\!\!dt\,
\bra \lambda_0 | e^{-i\bfq\cdot\bfr _{J'}} |\lambda_f\ket\bra \lambda_f | e^{i E_{\lambda_f}t}e^{i\bfq\cdot\bfr_J} e^{-i E_{\lambda_0}t}|\lambda_0\ket e^{-i\omega t}
\intertext{Since $\lambda_{0,f}$ are eigenstates of the Hamiltonian, we can replace the $E_{\lambda_0,\lambda_f}$ with the operator  $\bfH$:}
&=\frac{1}{2\pi}\sum_{J,J'} A_J A_{J'}\int_{-\infty}^{+\infty}\!\!\!\!\!\!dt\,
\bra \lambda_0 | e^{-i\bfq\cdot\bfr _{J'}} |\lambda_f\ket\bra \lambda_f | e^{i \mathbf{H}t}e^{i\bfq\cdot\bfr _J} e^{-i \mathbf{H}t}|\lambda_0\ket e^{-i\omega t}\\
&=\frac{1}{2\pi}\sum_{J,J'} A_J A_{J'}\int_{-\infty}^{+\infty}\!\!\!\!\!\!dt\,
\bra \lambda_0 | e^{-i\bfq\cdot\bfr _{J'}(0)} |\lambda_f\ket\bra \lambda_f | e^{i\bfq\cdot\bfr_J(t)} |\lambda_0\ket e^{-i\omega t}
\end{align}
In the last step we used the quantum evolution operator on the phase factor, and made the time dependence of the $\bfr_J$ explicit. By applying the sum over the final states from \eqref{skodef}, we can use the completeness of the $|\lambda_{f}\ket$ states to obtain
\begin{equation}
S(\bfq,\omega)= \frac{1}{2\pi \Ncells} \sum_{J,J'} A_J A_{J'}\int_{-\infty}^{+\infty}\!\!\!\!\!\!dt\,
\bra \lambda_0 | e^{-i\bfq\cdot\bfr _{J'}(0)} e^{i\bfq\cdot\bfr_J(t)} |\lambda_0\ket e^{-i\omega t}.
\end{equation}
In what follows we will drop the $|\lambda_0\ket$ to facilitate the notation and all expectation values are understood to be with respect to the ground state.

To compute this two-point correlation function, we write the position vectors in terms of the atomic displacements $\bfu$ relative to their equilibrium position, 
\begin{equation}
	\bfu_{j,\bfl} (t)\equiv\bfr _J(t)-\bfr _j^0 -\bfl
\end{equation}
where now $\bfl$ labels the lattice vector for a given primitive cell, and $\bfr_j^0$ are the equilibrium positions of the atoms relative to the origin of the primitive cell. We thus replace the sum over all atoms in lattice (labelled by $J$) with a sum over all lattice vectors $\bfl$ and atoms in a single primitive cell (labelled by $j$). Since the mass numbers $A_J$ are identical within each cell, we can also take $A_J \to A_j$.
Inserting this in the correlation function, 
\begin{align}\label{eq:appinterm1}
\bra e^{-i\bfq\cdot\bfr _{J'}(0)} e^{i\bfq\cdot\bfr _J(t)} \ket=e^{i\bfq\cdot(\bfr ^0_{j}-\bfr ^0_{j'})}e^{i\bfq\cdot(\bfl-\bfl')} \bra e^{-i\bfq\cdot\bfu_{j',\bfl'}(0)} e^{i\bfq\cdot\bfu_{j,\bfl}(t)} \ket.
\end{align}

We wish to expand this in the displacements, and keep only the leading correlation function. We can do so by applying the Baker-Campbell-Hausdorff identity and truncate at leading order. Concretely, for two operators $\mathbf{A}=i\bfq\cdot \bfu_{j,\bfl}$ and $\mathbf{B}=-i\bfq\cdot \bfu_{j',\bfl'}$ we have
\begin{equation}\label{eq:haussdorf}
e^{\mathbf{A}}e^{\mathbf{B}}\approx e^{\mathbf{A}+{\mathbf{B}}+\frac{1}{2} [\mathbf{A},\mathbf{B}]}.  
\end{equation}
Since we are in the small displacement (harmonic) approximation, the operators $\bfu_{j,\bfl}$ can be written as a linear combination of creation and annihilation operators. The commutator in \eqref{eq:haussdorf} is therefore proportional to the identity operator and we can pull it outside of the expectation value: 
\begin{align}
\left\langle e^{\mathbf{A}}e^{\mathbf{B}} \right\rangle &\approx \left\langle e^{\mathbf{A}+{\mathbf{B}}+\frac{1}{2} [\mathbf{A},\mathbf{B}]} \right\rangle\\
&= e^{\frac{1}{2} [\mathbf{A},\mathbf{B}]} \left\langle e^{\mathbf{A}+{\mathbf{B}}} \right\rangle
\intertext{Next we use the Bloch identity, $\bra e^A \ket=e^{\frac{1}{2}\bra A^2\ket}$, which only applies to linear combinations of creation and annihilation operators,}
&= e^{\frac{1}{2} [\mathbf{A},\mathbf{B}]}  e^{\frac{1}{2}\bra (\mathbf{A}+\mathbf{B})^2\ket} \\
&= e^{\frac{1}{2}\bra \mathbf{A}^2+\mathbf{B}^2+2\mathbf{A}\mathbf{B}\ket}.
\end{align}
In the last step we brought commutator back into the expectation value, again using that it is proportional to the identity as long as $\mathbf{A}$ and $\mathbf{B}$ are linear combinations of creation and annihilation operators. 

Applying the above formula to \eqref{eq:appinterm1}, we find
\begin{align}
\bra  e^{-i\bfq\cdot\bfu_{j',\bfl'}(0)} e^{i\bfq\cdot\bfu_{j,\bfl}(t)} \ket&=e^{-\frac{1}{2}\bra (\bfq\cdot \bfu_{j',\bfl'})^2\ket}e^{-\frac{1}{2}\bra (\bfq\cdot \bfu_{j,\bfl})^2\ket}e^{\bra \bfq\cdot \bfu_{j',\bfl'}(0)\bfq\cdot \bfu_{j,\bfl}(t)\ket}\\
&\approx e^{-\frac{1}{2}\bra (\bfq\cdot \bfu_{j',\bfl'})^2\ket}e^{-\frac{1}{2}\bra (\bfq\cdot \bfu_{j,\bfl})^2\ket} \bra \bfq\cdot \bfu_{j',\bfl'}(0)\bfq\cdot \bfu_{j,\bfl}(t)\ket.
\end{align}
where in the second line we expand the exponential to leading order and drop the constant piece that does not contribute to scattering.
The two exponentials in front are the Debye-Waller factors, defined by
\begin{equation}\label{eq:debeyewaller}
W_j (\bfq) \equiv\frac{1}{2}\bra (\bfq\cdot \bfu_{j})^2\ket.
\end{equation}
where we dropped the $\bfl$ index due to translation invariance over the lattice vectors. From \eqref{eq:debeyewaller}, we see that the Debye-Waller factor measures the average motion of atom $j$ relative to the momentum transfer. 

Putting the above results together, the structure function is then
\begin{equation}
S(\bfq,\omega)= \frac{1}{2\pi \Ncells} \sum_{j,j',\bfl,\bfl'} A_j A_{j'}e^{i\bfq\cdot(\bfr ^0_{j}-\bfr ^0_{j'})}e^{i\bfq\cdot(\bfl-\bfl')}e^{-W_j(\bfq) }e^{-W_{j'}(\bfq)}\int_{-\infty}^{+\infty}\!\!\!\!\!\!dt\,
\bra \bfq\cdot \bfu_{j',\bfl'}(0)\bfq\cdot \bfu_{j,\bfl}(t)\ket e^{-i\omega t}.
\end{equation}
To further simplify the sums, one can use the invariance of the two point function under lattice translations, which permits the replacement $\sum_{\mathbf{l},\mathbf{l}'} e^{i\bfq\cdot(\mathbf{l}-\mathbf{l}')} \to \Ncells \sum_{\mathbf{l}} e^{i\bfq\cdot\mathbf{l}}$:
\begin{equation}
S(\bfq,\omega)= \frac{1}{2\pi} \sum_{j,j',\bfl} A_j A_{j'}e^{i\bfq\cdot(\bfr ^0_{j}-\bfr ^0_{j'})}e^{i\bfq\cdot\mathbf{l}}e^{-W_j(\bfq) }e^{-W_{j'}(\bfq)}\int_{-\infty}^{+\infty}\!\!\!\!\!\!dt\,
\bra \bfq\cdot \bfu_{j',0}(0)\bfq\cdot \bfu_{j,\bfl}(t)\ket e^{-i\omega t}.
\end{equation}

It remains to compute the correlation function and the Debye-Waller functions in terms of the phonon eigenvectors and dispersion relations. To this end, we decompose the displacement operators in creation and annihilation operators, as in \eqref{eq:displacementquant} 
\begin{align}\label{displacement2}
\bfu_{j,\bfl}(t) = \sum_{\nu}^{3 \Nunit}\sum_{\bfk}  \sqrt{\frac{1}{2 \Ncells m_j \omega_{\nu,\bfk}}} \left(\mathbf{e}_{\nu,j,\bfk} \hat a_{\nu,\bfk} e^{i\bfk\cdot(\mathbf{l}+\bfr^0_j)-i \omega_{\nu,\bfk}t}+\mathbf{e}^\ast_{\nu,j,\bfk} \hat a^\dagger_{\nu,\bfk} e^{-i\bfk\cdot(\mathbf{l}+\bfr^0_j)+i \omega_{\nu,\bfk}t}  \right)
\end{align}
where the index $\nu$ runs over all $3n$ phonon modes and $\bfk$ is a regular, $\Ncells$-point discretization of the first \brill\ zone. The $1/\sqrt{\Ncells}$ factor implies that $S(\bfq,\omega)$ is an intrinsic quantity, as mentioned below \eqref{skodef}. Inserting this in the two-point function, we can trivially perform the Wick contractions, at least in the zero temperature limit. (For the finite temperature result we refer to section 9.12 of \cite{Schober2014}.) This results in
\begin{align}
\bra (\bfq\cdot \bfu_{j',0})(0)(\bfq\cdot \bfu_{j,\bfl})(t)\ket&=\frac{1}{2\Ncells\sqrt{m_j m_{j'}}}\sum_{\nu,\bfk} \frac{1}{\omega_{\nu,\bfk}}(\bfq\cdot\mathbf{e}_{\nu,j',\bfk})(\bfq\cdot\mathbf{e}^\ast_{\nu,j,\bfk})e^{i \omega_{\nu,\bfk}t}e^{-i\bfk\cdot\mathbf{l}}e^{i\bfk\cdot(\bfr_{j'}^0-\bfr_j^0)} .
\end{align}
The Debye-Waller function is just the special case where $j=j'$, $\mathbf{l}=0$ and $t=0$:
\begin{align}
W_j (\bfq) &=\frac{1}{2}\bra (\bfq\cdot \bfu_{j,0})^2\ket=\frac{1}{4 \Ncells m_j}\sum_{\nu,\bfk}\frac{1}{\omega_{\nu,\bfk}}|\bfq\cdot\mathbf{e}_{\nu,j,\bfk}|^2
\end{align}

Putting everything back together, we find
\begin{align}
S(\bfq,\omega)=& \frac{1}{2\Ncells} \sum_{j,j',\bfl} A_j A_{j'}e^{i\bfq\cdot(\bfr ^0_{j}-\bfr ^0_{j'})}e^{i\bfq\cdot\mathbf{l}}e^{-W_j(\bfq) }e^{-W_{j'}(\bfq)}\frac{1}{\sqrt{m_j m_{j'}}}
\\&
\times \sum_{\nu}^{3 \Nunit}\sum_{\bfk} \frac{1}{\omega_{\nu,\bfk}}(\bfq\cdot\mathbf{e}_{\nu,j',\bfk})(\bfq\cdot\mathbf{e}^\ast_{\nu,j,\bfk})e^{-i\bfk\cdot\mathbf{l}} e^{i\bfk\cdot(\bfr_{j'}^0-\bfr_j^0)} \delta( \omega_{\nu,\bfk}-\omega)
\end{align}
where we used $\int^{+\infty}_{-\infty}\!\!dt\,e^{i (\omega_{\nu,\bfk}-\omega)t}=2\pi \delta (\omega_{\nu,\bfk}-\omega)$. With the identity $\sum_{\mathbf{l}} e^{i\bfq\cdot\mathbf{l}}= \Ncells \sum_{\mathbf{G}}\delta_{\bfq,\bfG}$ with $\mathbf{G}$ the reciprocal lattice vectors, this finally reduces to 
\begin{equation}
S(\bfq,\omega)=\frac{1}{2}\sum_{\bfG,\bfk,\nu}\frac{1}{\omega_{\nu,\bfk}}\left|F_{\nu}(\bfq,\bfk)\right|^2\delta_{\bfk-\bfq,\bfG}\delta( \omega_{\nu,\bfk}-\omega)
\end{equation}
with the phonon form factor
\begin{equation}
F_\nu(\bfq,\bfk)\equiv\sum_{j} \frac{A_j}{\sqrt{m_j}}e^{-W_j(\bfq)}\bfq\cdot \mathbf{e}_{\nu,j,\bfk} e^{i (\bfq-\bfk)\cdot \bfr^0_j}
\end{equation}
Both energy and crystal momentum conservation are now manifest in these expressions. For scattering with sub-MeV dark matter, the momentum transfer is typically smaller than the size of the \brill\ zone, such that we can neglect the sum over the reciprocal lattice vectors and set $\bfG=0$. In this limit, the structure factor further simplifies to
\begin{equation}
S^{\text{low}}(\bfq,\omega)\approx\frac{1}{2}\sum_{\nu}\frac{1}{\omega_{\nu,\bfq}}\left|F^{\text{low}}_{\nu}(\bfq)\right|^2\delta( \omega_{\nu,\bfk}-\omega)
\end{equation}
with
\begin{equation}
F^{\text{low}}_\nu(\bfq)\equiv\sum_{j} \frac{A_j}{\sqrt{m_j}}e^{-W_j(\bfq)}\bfq\cdot  \mathbf{e}_{\nu,j,\bfq}.
\end{equation}
Note that this expression differs from the one in \cite{Knapen:2017ekk} by a phase factor, since a different convention was used for the phonon eigenvectors.

%%%%%%%%%%%%%%%%%%%%%%%%%%%%%%%%%%%
 \section{DM-electron scattering \label{app:electrons}}
%%%%%%%%%%%%%%%%%%%%%%%%%%%%%%%%%%%
In this appendix we comment on the reach for models where the DM couples to electrons through a scalar mediator. Such models tend to be extremely constrained by stellar cooling and Big Bang Nucleosynthesis bounds for $m_X\lesssim1$ MeV~\cite{Green:2017ybv,Knapen:2017xzo}, and at the moment we are not aware of models which can achieve $\bar \sigma_e\gtrsim10^{-45}\,\text{cm}^2$. While it is likely difficult for near future experiments to access such low cross sections, we briefly discuss the reach for the sake of completeness. 

Because the displacements involved in a phonon excitation correspond to displacements of the nucleus and tightly-bound inner shell electrons, a DM-electron coupling also results in an effective DM-phonon coupling, analogous to the discussion in Sec.~\ref{sec:nucleonint} for DM-nucleon couplings. The difference is that we must replace the mass number of the atom in the form factor \eqref{eq:phononformfac} with the number of core electrons for each atom. Ga and As both have 28 core electrons, while O and Al have 2 and 10 core electrons, respectively.\footnote{As matter of convention we treat fully-filled shells as core electrons, explicitly this designates Al: $1s^{2}$, $2s^{2}$, $2p^{6}$, O: $1s^{2}$, Ga: $1s^{2}$, $2s^{2}$, $2p^{6}$, $3s^{2}$, $3p^{6}$, $3d^{10}$ and As: $1s^{2}$, $2s^{2}$, $2p^{6}$, $3s^{2}$, $3p^{6}$, $3d^{10}$ as core electrons.} Note that the form factors for coherently scattering off the electrons in the atom are constant for $|\bfq|\lesssim 1$ keV \cite{doi:10.1063/1.556027}, and we can neglect their effect for the DM mass range of interest. For higher DM masses, these form factors are expected to suppress the rate.

\begin{figure}[h]
\includegraphics[width=0.49\textwidth]{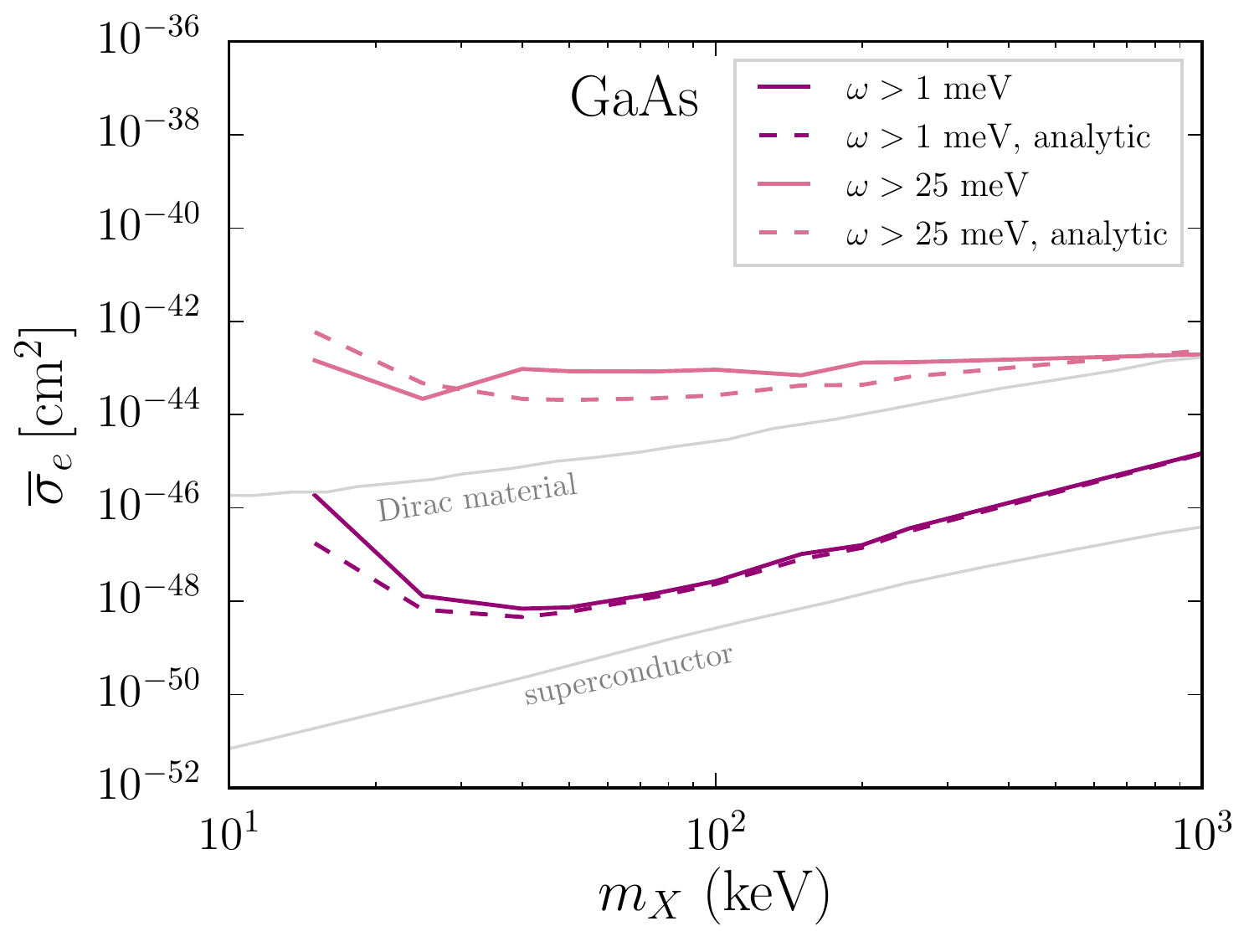}\hfill
\includegraphics[width=0.49\textwidth]{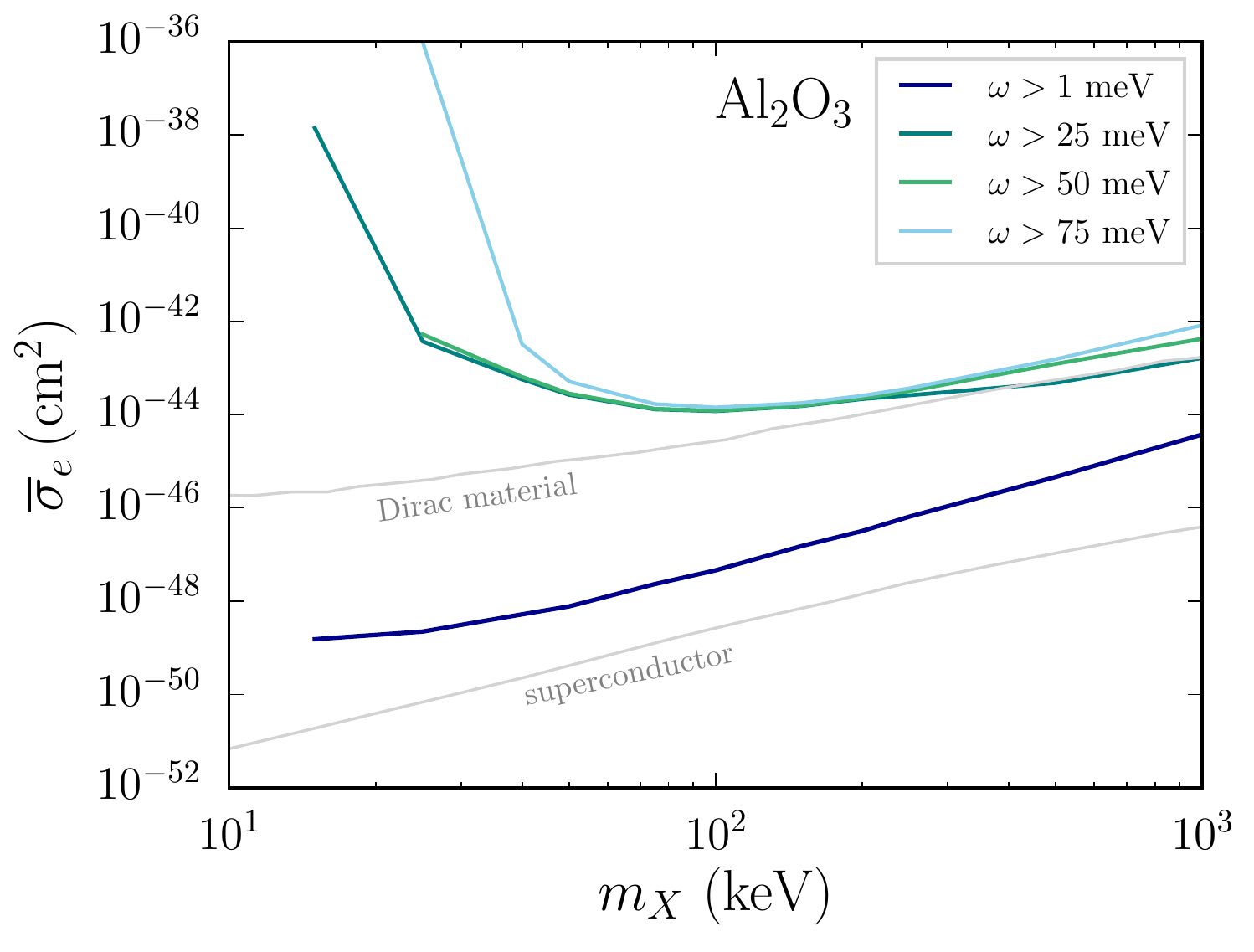}
\caption{The projected reach for scattering through a scalar mediator coupling to electrons, for GaAs (left) and $\text{Al}_2\text{O}_3$ (right) with a kg-year exposure and different experimental thresholds. The solid lines show the reach using the numerically computed phonon modes, while the dashed lines use the analytic approximation in \eqref{eq:formfactor_approx}. Also shown are projections for Dirac materials \cite{Hochberg:2017wce} and superconductors \cite{Hochberg:2015fth}. 
\label{fig:electronreach}}
\end{figure}

The results are shown in Fig.~\ref{fig:electronreach} for a massless scalar mediator for a kg$\times$year exposure, where we plot
\begin{equation}
\bar \sigma_e \equiv \frac{y_e^2 y_X^2}{4\pi}\frac{\mu_{eX}^2}{(\alpha m_e)^4} ,
\end{equation}
where $y_e$ ($y_X$) is the electron-mediator (DM-mediator) coupling, $\mu_{eX}$ is the DM-electron reduced mass, and $\alpha$ is the fine structure constant. If only the optical branches are accessible, we find a reach that is competitive with that of  Dirac material targets, in which the DM can create an electron excitation with $\sim$ meV threshold. In the optimistic case where the acoustic modes can also be resolved, polar materials could have a reach approaching that of a superconducting target.

%%%%%%%%%%%%%%%%%%%%%%%%%%%%%%%%%%%
 \section{Statistical power of daily modulation signal \label{app:statistics}}
%%%%%%%%%%%%%%%%%%%%%%%%%%%%%%%%%%%

To estimate the discriminating power of the daily modulation, we calculate how many events are needed to distinguish the scenarios where (i) all observed events are due to a hypothetical, non-modulating background and (ii) all events are due to a modulating signal, as predicted in Sec.~\ref{sec:darkphoton}.  

\begin{figure}[h]
    \centering
    \begin{minipage}{0.48\textwidth}
        \centering
        \includegraphics[height=6cm]{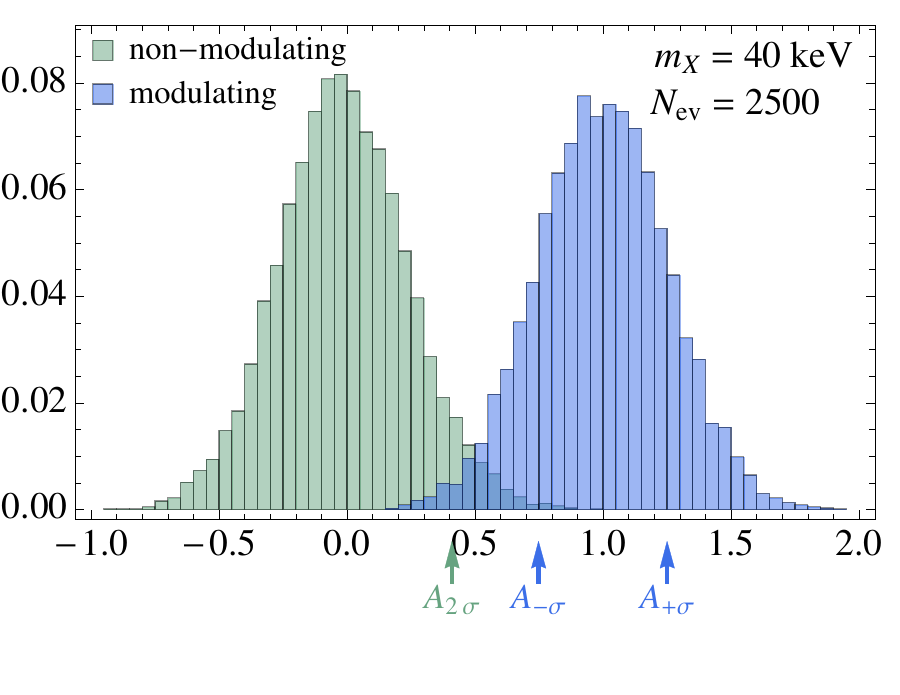} 
        \caption{Distributions of the modulation amplitude, assuming 2500 expected events.  \label{fig:modhistogram}}
    \end{minipage}\hfill
    \begin{minipage}{0.49\textwidth}
        \centering
        \includegraphics[height=6cm]{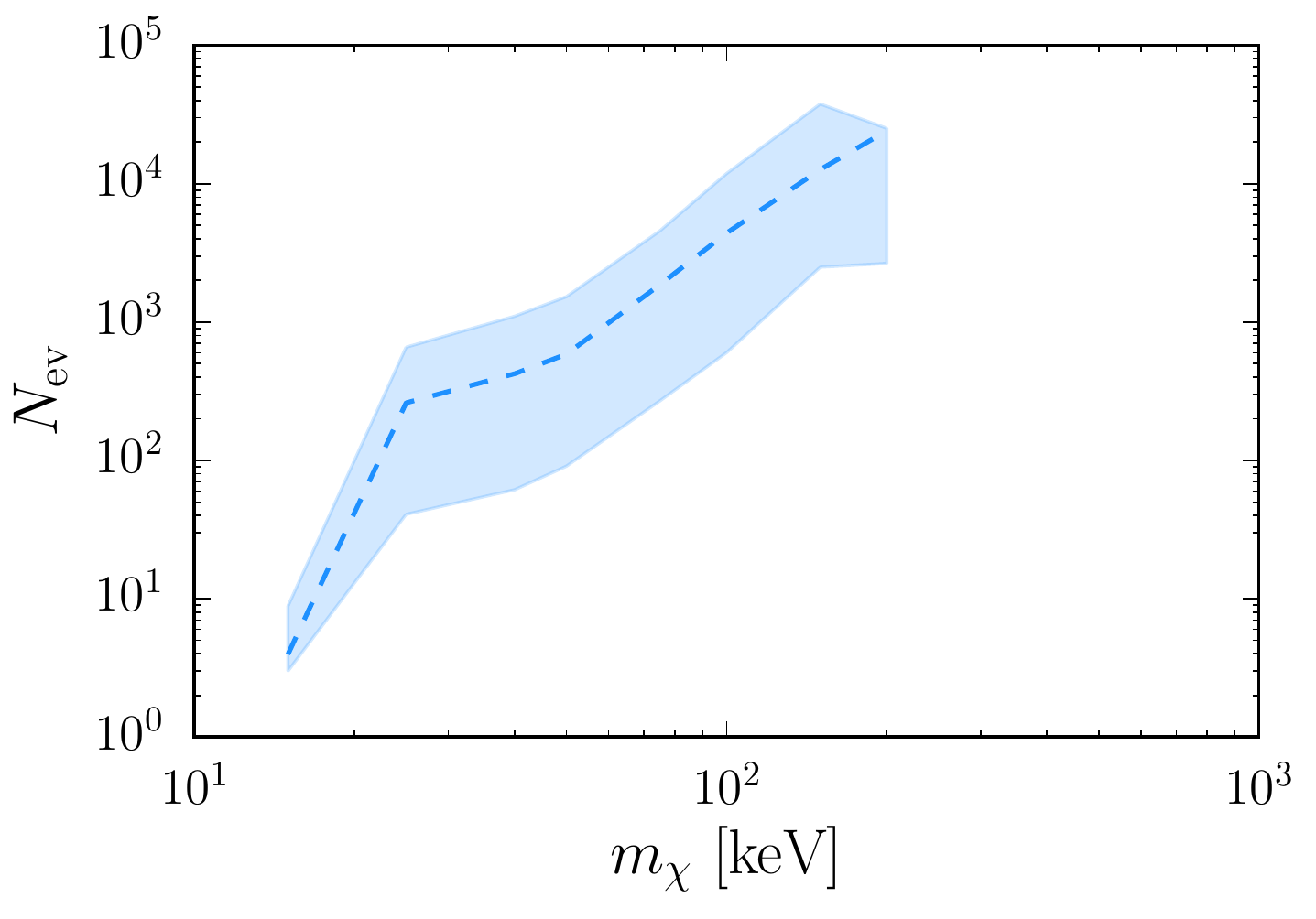}
        \caption{Expected number of events needed for $2\sigma$ observation of the daily modulation.  \label{fig:expected_events}}
    \end{minipage}
\end{figure}

For a given mass point $m_X$ and an expected number of events $N_{\text{ev}}$, we generate simulated datasets for both scenarios above. The number of events in each dataset is Poisson distributed with average $N_{\text{ev}}$, and for the modulating sample the probability distribution in $t$ is given by the computations in Sec.~\ref{sec:darkphoton}. For each dataset, we then perform a fit to the modulation, allowing for both a constant component and a modulating component with amplitude $A$, fixing the template for that $m_X$. Denote the modulation amplitude as $A^{\text{non-mod}}$ for the datasets that are purely background, and $A^{\text{mod}}$ for the datasets that are purely signal.  Repeating this procedure for many datasets, we generate the expected distribution in the modulation amplitude, shown in Fig.~\ref{fig:modhistogram} for an example set of parameters. By construction, $\langle A^{\text{mod}}\rangle=1$, and $\langle A^{\text{nod-mod}}\rangle=0$.

For each $N_{\text{ev}}$, we then compute the 2$\sigma$ upper value (95\% quantile) on $A$ for the non-modulating data ($A^{\text{non-mod}}_{2\sigma}$, indicated by the green arrow in Fig.~\ref{fig:modhistogram}). Interpolating in $N_{\text{ev}}$, we can then find the number of events needed such that $A^{\text{non-mod}}_{2\sigma}$ is below the expected amplitude for the modulating sample, in other words $A^{\text{non-mod}}_{2\sigma}<1$. This gives the number of events needed so that in 50\% of the purely signal datasets, we can reject the background hypothesis at 2$\sigma$.

Similarly, we can obtain $\pm \sigma$ quantiles about the mean expectation for the modulating signal ($A^{\text{mod}}_{\pm\sigma}$, indicated by the blue arrows in Fig.~\ref{fig:modhistogram}). The $\pm\sigma$ bands are then obtained by demanding that   $A^{\text{non-mod}}_{2\sigma}<A^{\text{mod}}_{\pm\sigma}$. The results of this procedure are shown in Fig.~\ref{fig:expected_events} as a function of $m_X$, and translated in terms of cross section in Fig.~\ref{fig:darkphotonlimit} (blue shaded band). 

\FloatBarrier

\bibliography{sapphire.bib,lightdm.bib,sinead.bib}

\end{document}